\documentclass{aa}

\usepackage{graphicx}
\usepackage{natbib}
\usepackage{savesym}
\usepackage{txfonts}
\savesymbol{iint}
\savesymbol{iiint}
\savesymbol{iiiint}
\savesymbol{idotsint}
\usepackage{amsmath}
\restoresymbol{AMS}{iint}
\restoresymbol{AMS}{iiint}
\restoresymbol{AMS}{iiiint}
\restoresymbol{AMS}{idotsint}

  \usepackage{longtable,lscape}

\usepackage{aas_macros}
\usepackage{dsfont}
\usepackage{fixltx2e}
\begin{document} 
\title{An absorption-selected survey of neutral gas in the Milky Way halo}
\titlerunning{An absorption-selected survey of neutral gas in the Milky Way halo}

\subtitle{New results based on a large sample of \ion{Ca}{ii}, \ion{Na}{i}, and \ion{H}{i} spectra towards QSOs}

\author{	N. Ben Bekhti\inst{1}      
\and        B. Winkel\inst{2}
\and        P. Richter\inst{3}\inst{4}
\and        J. Kerp\inst{1}
\and        U. Klein\inst{1}
\and  	    M. T. Murphy\inst{5}
} 
\authorrunning{Ben Bekhti et al.}
\offprints{N. Ben Bekhti} 
\institute{Argelander-Institut f\"ur Astronomie, 
Universit\"{a}t Bonn, Auf dem H\"{u}gel 71, 53121 Bonn, Germany\\
\email{nbekhti@astro.uni-bonn.de}
\and Max-Planck-Institut f\"{u}r Radioastronomie, Auf dem H\"{u}gel 69, 53121 Bonn, Germany%\\
\and Institut f\"{u}r Physik und Astronomie, Universit\"{a}t Potsdam, Haus 28, Karl-Liebknecht-Str. 24/25 14476, Potsdam, Germany%\\
\and Leibniz-Institut f\"ur Astrophysik Potsdam (AIP), An der Sternwarte 16, 14482 Potsdam, Germany%\\  
\and  Centre for Astrophysics \& Supercomputing, Swinburne
University of Technology, Hawthorn, Victoria 3122, Australia 
}

\date{Received xxxx; accepted month xxx}

\abstract{}{We aim at analysing systematically the distribution and physical properties of neutral and mildly ionised gas in the Milky Way halo, based on a large absorption-selected data set.}
{Multi-wavelength studies were performed combining optical absorption line data of \ion{Ca}{ii} and \ion{Na}{i} with follow-up \ion{H}{i} 21-cm emission line observations along 408 sight lines towards low- and high-redshift QSOs. We made use of archival optical spectra obtained with UVES/VLT. \ion{H}{i} data were extracted from the Effelsberg-Bonn \ion{H}{i} survey and the Galactic All-Sky survey. For selected sight lines we obtained deeper follow-up observations using the Effelsberg 100-m telescope.}
{\ion{Ca}{ii} (\ion{Na}{i}) halo absorbers at intermediate and high radial velocities are present in $40-55$\% ($20-35$\%) of the sightlines, depending on the column density threshold chosen.
Many halo absorbers show multi-component absorption lines, indicating the presence of sub-structure. In 65\% of the cases, absorption is associated with \ion{H}{i} 21-cm emission.  The \ion{Ca}{ii} (\ion{Na}{i}) column density distribution function follows a power-law with a slope of $\beta \approx -2.2$ ($-1.4$).}
{Our absorption-selected survey confirms our previous results that the Milky Way halo is filled with a large number of neutral gas structures whose high column density tail represents the population of common \ion{H}{i} high- and intermediate-velocity clouds seen in 21-cm observations. 
We find that \ion{Ca}{ii}/\ion{Na}{i} column density ratios in the halo absorbers are typically smaller than those in the Milky Way disc, in the gas in the Magellanic Clouds, and in damped Lyman\,$\alpha$ systems. The small ratios (prominent in particular in high-velocity components) indicate a lower level of Ca depletion onto dust grains in Milky Way halo absorbers compared to gas in discs and inner regions of galaxies.}

\maketitle 

\section{Introduction}
High-resolution emission and absorption measurements in different wavelength regimes have demonstrated that spiral galaxies at low and high redshifts are surrounded by extended gaseous haloes \citep[e.g.,][ and references therein]{savage_massa87, majewski04, fraternalietal_07, boomsmaetal08,steidel2010,bouche2012,rudie2012}. Sensitive observations suggest that up to $30\%$ of the neutral hydrogen of a spiral galaxy could be located in the halo \citep[e.g.,][]{Fraternalietal_01, Oosterlooetal_07}. The origin and nature of this extra-planar gas in the halo is still uncertain, because the halo is continuously fuelled with gas by different mechanisms that are connected with the on-going formation and evolution of galaxies \citep[outflows, galaxy merging and gas accretion from the intergalactic medium, e.g.][]{shapirofield76, Fraternali04, kaufmann06, sancisi08}.

Understanding the nature of extra-planar gas in galaxy haloes is important to characterise the infall of gaseous material feeding star formation and the outflow of gas as part of galactic winds. Since haloes represent the interface region between the dense galactic discs and the surrounding intergalactic medium (IGM) the study of matter exchange between galaxies and their environment is an important aspect in our understanding of the IGM.

A particularly powerful method for studying the properties and nature of extra-planar gaseous structures in the haloes of galaxies is the analysis of absorption lines in the spectra of distant quasars (QSOs). This can best be done in the optical and ultraviolet, since most of the ion transitions of interest for QSO absorption spectroscopy are located in these wavelength regimes. For a recent review we refer to \citet{richterp06}. QSO absorption spectroscopy provides information about the physical properties, kinematics, and spatial distribution of gas in galaxy haloes over a large range of column densities. The analysis of different ions like \ion{Mg}{ii}, \ion{C}{iv}, \ion{O}{i}, \ion{Si}{ii}, and \ion{C}{ii} \citep[e.g.,][]{charltonetal00, dingetal03, masieroetal05, boucheetal06, richteretal09} along sight lines that pass through the halo of the Milky Way and the inner and outer regions of other galaxies reveal different types of absorbers (e.g., discs absorbers, infalling and outflowing gas, tidal structures). All of these have a specific range of physical properties, such as the degree of ionisation and the metallicity. 

The most prominent gaseous structures in the Milky Way halo are the intermediate- and high-velocity clouds \citep[IVCs, HVCs;][] {mulleroortraimond63}, which show radial velocities that are inconsistent with a simple model of a Galactic disc rotation. They possibly represent one group of local analogues of extra-planar gas features seen in the vicinity of other galaxies. IVCs and HVCs represent multi-phase structures that can contain regions of partly molecular gas \citep[H$_2$ absorption,][]{richtersembachetal01, wakker06}, cold and warm neutral gas seen in 21-cm emission \citep{Bruenskerp05, benbekhtietal08}, warm-ionised plasma as traced by H$\alpha$ emission \citep{putmanetal_03, hill09} and UV absorption \citep{Luetal_94}, and highly-ionised plasma seen in \ion{O}{vi} absorption \citep{sembach03, foxetal06} and in X-ray emission \citep[][and references therein]{kerp99,deboer04}.

Distance estimates of IVCs and HVCs around the Milky Way \citep[e.g.,][]{Sembachetal91, vanWoerden99, wakker01, thom2006, wakker_york_howketal07, wakkeryorkwilhelmetal08} indicate that most IVCs are relatively nearby objects at distances of $d< 2$~kpc, while HVCs are more distant clouds, located in the halo of the Milky Way at $5<d<50$~kpc.

Metal abundances of several IVCs/HVCs have been determined by absorption-line measurements \citep[see reviews by][]{wakker01, richterp06}. Results show metallicities varying between $\sim 0.1$ and $\sim 1.0$ solar. This wide range of metallicities suggests that not all IVCs and HVCs share a common origin. The chemical composition of some IVCs/HVCs with nearly solar metallicities can be explained by the galactic fountain model. There are, however, IVCs and HVCs with metal abundances clearly below solar. The Magellanic Stream, for example, has abundances similar to that of the Small Magellanic Cloud ($\sim 0.3$~solar). There are also HVCs with abundances less than $~0.3$~solar \citep[e.g., Complex C,][] {wakkeretal99,richtersembachetal01b}, but still larger than one would expect for primordial gas. The conclusion is that the gas has already been processed, but at different levels depending on the origin.

In our previous paper \citep[][BB08 hereafter]{benbekhtietal08} we systematically analysed \ion{Ca}{ii} absorption and \ion{H}{i} emission data of gas in the Milky Way halo towards about 100 QSOs and we generated a first absorption-selected halo-cloud sample. We found that next to the massive IVC/HVC complexes, there exists a population of low-column density absorbers with \ion{H}{i} column densities of $N_\mathrm{HI} < 10^{19}$\,cm$^{-2}$ \citep[][ BB08]{richterwestmeierbruens05} that are traced by optical \ion{Ca}{ii}   absorption. We further showed that optical halo absorbers have a substantial area-filling factor of about $30\%$ \citep[see also][]{richteretal10}, suggesting that low-column density neutral gas clumps represent an important absorber population that is important for our understanding of the gas distribution in the Milky Way halo.

In this follow-up paper we extend our previous study of extra-planar gas structures in the Milky Way halo. We almost quadrupled the size of our sample to a total of 408 extragalactic sight lines through the halo. This large data set allows us to perform a more solid statistical study of the absorbing structures. Furthermore, we now have a sufficiently large data base to provide statistical analyses independently for IVCs and HVCs and, even more important, to incorporate \ion{Na}{i} into our investigation. The sky distribution of the sight lines is very homogeneous (though the Galactic plane and the region $\delta>30\degr$ is not traced). This is an important point, since the MW halo is known to contain gaseous structures at large angular scales (i.e., the 21-cm IVC/HVC complexes), which could bias our statistics if the analysis was based on a small or very inhomogeneous sky coverage.

For all the positions where \ion{Ca}{ii} and/or \ion{Na}{i} halo absorption is detected we searched for counterparts in \ion{H}{i} 21-cm emission. We performed deep pointed observations using the 100-m telescope at Effelsberg for selected sightlines. For the remaining lines of sight we make use of the new large-area \ion{H}{i} 21-cm surveys GASS \citep[Galactic All-Sky survey,][]{McClure_Griffithsetal09, kalberla10} and EBHIS \citep[Effelsberg-Bonn HI survey,][]{winkel10a, Kerp11}. Compared to BB08, this leads to a substantially increased number of \ion{H}{i} spectra available for follow-up analyses. The large-area surveys also provide the opportunity to search for \ion{H}{i} structures in the direct vicinity of the absorption sight lines. The study of these as well as of follow-up \ion{H}{i} aperture synthesis observations is under-way and will be presented in a subsequent paper.

In Section\,\ref{secdata} of our paper we describe the data acquisition and reduction as well as the sample of extra-planar clouds. In Section\,\ref{secresults}  the results of the statistical analysis are presented along with a discussion of the various findings. Finally, Section\,\ref{secsummary} summarises our work.

\section{Data and cloud catalogue}\label{secdata}
\subsection{UVES data}\label{uves_data}

For our large absorption-selected data set we made use of the publicly available ESO data archive.\footnote{http://archive.eso.org/wdb/wdb/eso/uves/form} We use 408 optical spectra of low- and high-redshift QSOs that were observed between 1999 and 2004 with the Ultraviolet and Visual Echelle Spectrograph (UVES) at the ESO Very Large Telescope (VLT). A detailed description of the UVES instrument is given by \citet{dekkeretal.00}. The raw data were reduced and normalised as part of the UVES Spectral Quasar Absorption Database (SQUAD; PI: Michael T. Murphy) using a modified version of the UVES pipeline. For a more detailed description of the reduction steps see BB08 and \citet{richteretal10}.

The reduced data set provides spectra with a resolution of $R \approx 40000-60000$, corresponding to a spectral resolution of $\delta v\approx6.6\,\mathrm{km\,s}^{-1}$ (FWHM). The pixel separation, $\Delta v\approx 2.5\,\mathrm{km\,s}^{-1}$, is small enough to ensure full Nyquist sampling.

The spectral features were analysed via Voigt-profile fitting using the \textsc{Fitlyman} package in \textsc{Midas} \citep{fontanaandbalester95}, which, among other parameters, delivers column densities and Doppler parameters ($b$-values). The latter are automatically de-convolved by \textsc{Fitlyman} to account for the  instrumental (spectral) response function. Therefore, $b$-values obtained from the fit are usually much smaller than the apparent width of the absorption lines in the spectra. As a consequence, in many cases the obtained $b$-values are much smaller than the instrumental resolution, i.e., the profiles are unresolved. Therefore, they should be treated with care --- they provide only an estimate of the true Doppler widths under the assumption that the absorption is due to a \textit{single component}.

\begin{figure} 
\includegraphics[width=0.5\textwidth,clip=]{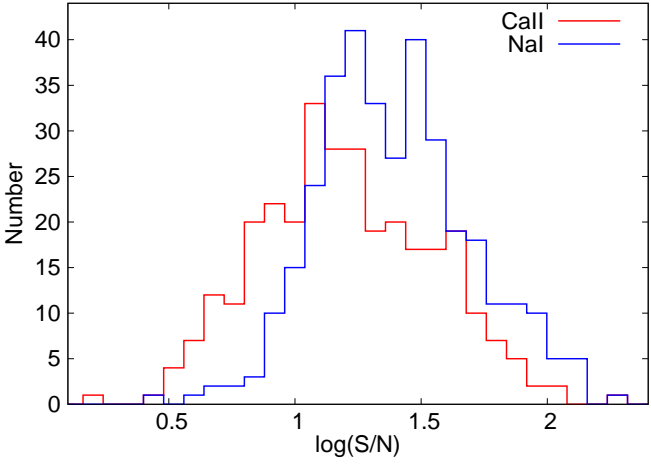}
\caption{Distribution of $\mathrm{S/N}_\mathrm{p}$ values for the QSO absorption spectra in our sample.}
\label{figsnrhistograms}
\end{figure}

We use the \ion{Ca}{ii}\,$\lambda3935, 3967$ and \ion{Na}{i}\,$\lambda5892,5898$ doublets which were fitted simultaneously with \textsc{Fitlyman}. For the normalised spectra the signal-to-noise ratio, $\mathrm{S/N}_\mathrm{p}=1/\mathrm{RMS}$, per pixel was computed using \textsc{Midas}. It is important to note that the $\mathrm{S/N}_\mathrm{p}$ values vary significantly for the different spectra and are also dependent on the wavelength. In Fig.\,\ref{figsnrhistograms} we show the distribution calculated for the spectral ranges near the two \ion{Ca}{ii} ($\sim4000$\,\AA{}) and \ion{Na}{i} ($\sim6000$\,\AA{}) lines. The median value for \ion{Ca}{ii} is 15.5, while for \ion{Na}{i} we find a median of 23.4. Unfortunately, in several cases one (or rarely both) desired spectral ranges of the 408 available data sets were corrupted. In total, our sample includes 306 (345) spectra suited to search for \ion{Ca}{ii} (\ion{Na}{i}) absorption in the Milky Way halo.

\subsubsection{Column density detection limits and sample completeness}\label{subsubsecsamplecompleteness}

From the $\mathrm{S/N}_\mathrm{p}$ values we can compute a (theoretical) detection limit for \ion{Ca}{ii} and \ion{Na}{i} for each spectrum. We consider a line as detected if the equivalent width, $W_\lambda$, of the stronger line of each doublet is at least six times the equivalent width limit, $W_\lambda^\mathrm{lim}(1\sigma)$, for that spectrum.

The $W_\lambda^\mathrm{lim}$ values can be computed analytically if one makes the following simplifications. Firstly, for lower limits it is safe to assume that the absorption line profiles are Gaussian-like. Secondly, we use a typical expected line width of $w_\ion{Ca}{ii}=5.5\,\mathrm{km\,s}^{-1}$ (FWHM, corresponding to a $b$-value of $3.3\,\mathrm{km\,s}^{-1}$) and $w_\ion{Na}{i}=3.5\,\mathrm{km\,s}^{-1}$ (FWHM, corresponding to a $b$-value of $2.1\,\mathrm{km\,s}^{-1}$)\footnote{$w=2\sqrt{2\ln2}\,b\approx1.66 \,b$.}; see also Section\,\ref{subsec:bvalues}. These values have to be convolved with the instrumental resolution to obtain the `apparent' line width $\mathcal{D}v=\sqrt{w^2+\delta v^2}$ in the spectra, i.e., $\mathcal{D}v_\ion{Ca}{ii}\approx8.6\,\mathrm{km\,s}^{-1}$ and $\mathcal{D}v_\ion{Na}{i}\approx7.5\,\mathrm{km\,s}^{-1}$, respectively.

After (hypothetical) smoothing of the spectra until the spectral resolution matches the apparent line widths a detection would be made in just one physical resolution unit (not pixel, as these are correlated due to the slight oversampling). Due to the smoothing the noise will decrease,
\begin{equation}
\mathrm{RMS}^\mathrm{sm}=\mathrm{RMS}\sqrt{\frac{\delta v}{\mathcal{D}v}}\,.
\end{equation}
Since, the lines are assumed to be Gaussian-shaped, one can easily calculate the integrated intensity,
\begin{equation}
W=I_0\sqrt{2\pi}\,\frac{\mathcal{D}v}{\sqrt{8\ln2}}\,,
\end{equation}
of a profile where $I_0$ is the amplitude of the Gaussian. To calculate the limit we set $I_0=\mathrm{RMS}^\mathrm{sm}$ and obtain
\begin{equation}
\begin{split}
W_\lambda^\mathrm{lim}(1\sigma)&=\mathrm{RMS}\sqrt{2\pi \frac{\mathcal{D}v}{\sqrt{8\ln2}} \frac{\delta v}{\sqrt{8\ln2}}}\quad [\mathrm{km\,s}^{-1}]\\
&=\frac{\mathrm{RMS}}{\sqrt{8\ln2}}\frac{\lambda_0}{c}\sqrt{2\pi\, \mathcal{D}v\,\delta v}\quad [\AA{}] \,.
\end{split}
\end{equation}
This can be easily converted to a column density detection limit, via
\begin{equation}
N_\mathrm{limit}(1\sigma)=\frac{W_\lambda\, \mathrm{m}_\mathrm{e}c^2}{\pi\, \mathrm{e}^2\,\lambda^2\,f}\,,
\end{equation}
where $f$ is the oscillator strength of the transition considered. Note, that the pixel resolution $\Delta v$ does not enter the detection limits, but only the true (instrumental) resolution $\delta v$.

\begin{figure} 
\includegraphics[width=0.45\textwidth,clip=,bb=105 54 352 299]{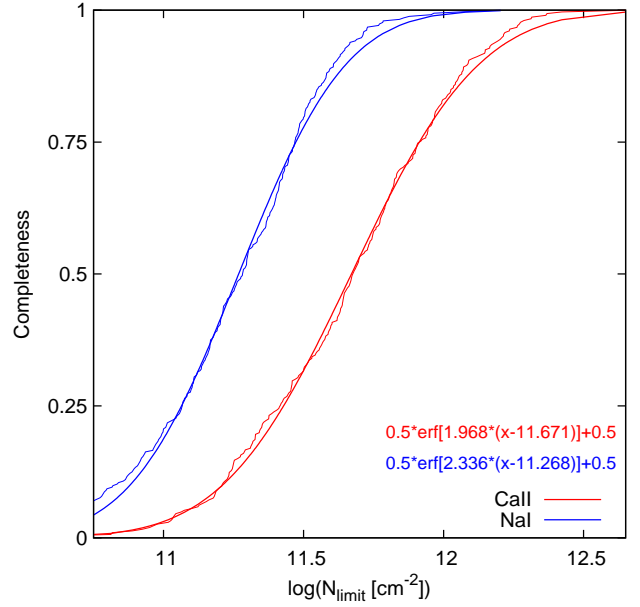}
\caption{Completeness,i.e., the ratio of the sight lines with a given detection limit compared to the total number of sight lines in the sample. To parametrise the curves for later use we fitted error functions to the data.}
\label{figcompleteness}
\end{figure}

The column density limit values allow us to explore the completeness of the sample. In Fig.\,\ref{figcompleteness} we show the (cumulative) size of the \ion{Ca}{ii} and \ion{Na}{i} samples if certain $N_\mathrm{limit}$ thresholds are applied. Both curves (thin solid lines) were normalised to unity and can be parametrised using an error function of the form

\begin{equation}
C\left(\log N_\mathrm{limit}; a,b\right)=\frac{1}{2} \mathrm{erf}\left[a(\log(N_\mathrm{limit}-b)\right]+ \frac{1}{2}\,.
\end{equation}

This function $C$ can be understood as a completeness function, as it returns the fraction of sight lines that can potentially reveal an absorber having a column density of $N\geq N_\mathrm{limit}$. Note that the $N_\mathrm{limit}$ values are purely theoretical and likely need to be calibrated (i.e., shifting the parameter $b$) to match the true effective limits. However, the shape $a$ of the true completeness function matches the values determined from  Fig.\,\ref{figcompleteness}.

\begin{figure*} 
\includegraphics[width=0.98\textwidth,clip=,bb=94 531 928 917]{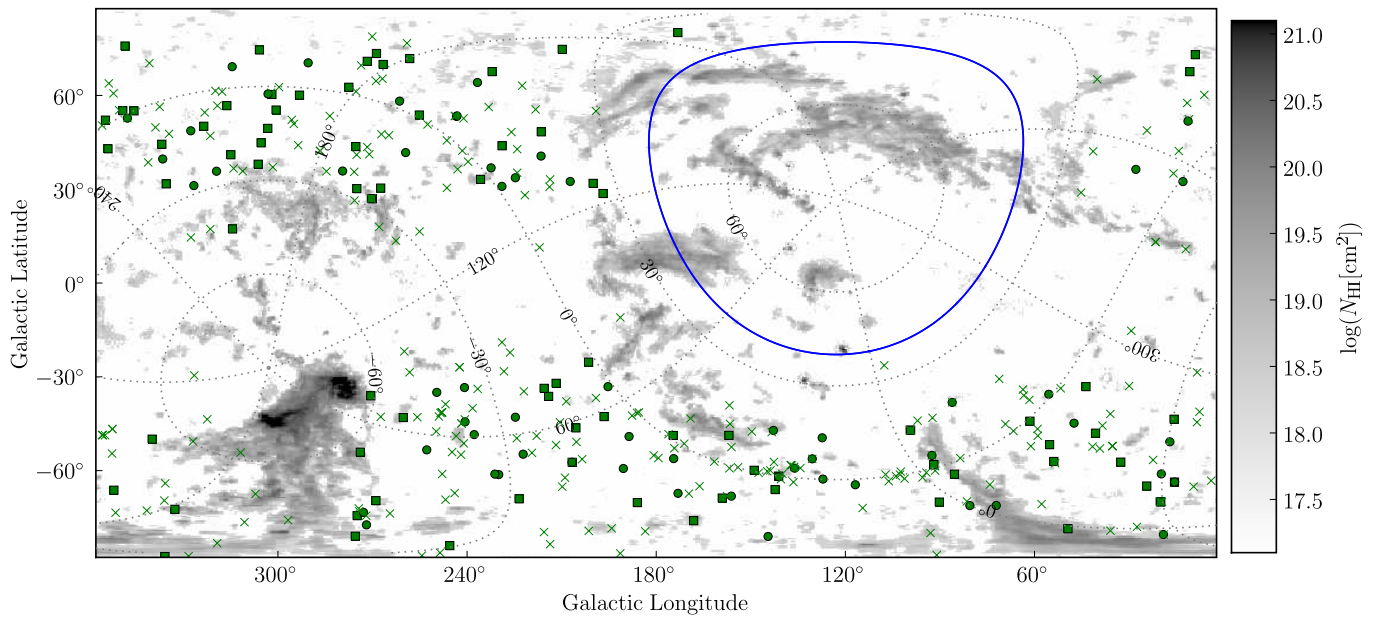}
\caption{HVC-all-sky map derived from the data of the LAB Survey (Kalberla et al. 2005) using a simple model of Galactic rotation kindly provided by T. Westmeier \citep[for details see][]{westmeier07}. The different symbols mark the locations
of 408 QSO sight lines that were observed with UVES. Along 126 (75) lines of sight we detect \ion{Ca}{ii} (\ion{Na}{i}) absorption components (marked with circles and boxes). The boxes mark the positions where we found corresponding \ion{H}{i} emission. The crosses show non-detections. The blue line marks the region not accessible with the VLT ($\delta\gtrsim40^\circ$). Furthermore, the Galactic plane is not covered by our sample.}
\label{fig_hvc_allskymap} 
\end{figure*} 

\subsection{EBHIS and GASS data} 

For our follow-up analysis we used data from the Effelsberg--Bonn \ion{H}{i} Survey \citep[EBHIS;][]{winkel10a,Kerp11} and the Galactic All-Sky Survey \citep[GASS;][]{McClure_Griffithsetal09,kalberla10}, which are the most sensitive, highest angular resolution, large-scale surveys of Galactic \ion{H}{i} emission ever made in the northern and southern sky. The former is currently undertaken with the 100-m telescope at Effelsberg, while the latter (using the 64-m Parkes telescope) was recently finished.
Unfortunately, as the EBHIS is not yet completed for the full northern hemisphere, data was not available for every sight line.

The data reduction pipeline used for the GASS data is described in detail in \citet{kalberla10}. The angular resolution of the final data cubes is $15\farcm6$, leading to an RMS level of $57\,\mathrm{mK}$ per spectral channel ($\Delta v=0.8\,\mathrm{km\,s}^{-1}$). For EBHIS the data reduction scheme presented in \citet{winkel10a} was used. EBHIS has a slightly higher nominal noise level of $\lesssim90\,\mathrm{mK}$ ($\Delta v=1.2\,\mathrm{km\,s}^{-1}$), but due to the better angular resolution of $10\farcm1$ the resulting column density detection limit (after angular smoothing to the Parkes beam) is almost identical: $N^\mathrm{limit}_\ion{H}{i}=4.1\times10^{18}\,\mathrm{cm}^{-2}$ (GASS) and  $N^\mathrm{limit}_\ion{H}{i}=5.9\times10^{18}\,\mathrm{cm}^{-2}$ (EBHIS), calculated for a Gaussian-like emission line with a width of $20\,\mathrm{km\,s}^{-1}$ (FWHM). All data sets were corrected for stray-radiation using the method of \citet{kalberla80} previously applied to the Leiden/Argentine/Bonn survey \citep[LAB;][]{kalberlaetal05}.

We also performed pointed observations towards 17 sources having $\delta>-30\degr$ with the 100-m telescope at Effelsberg. The integration times for these pointed measurements were between 10 and 20\,min, leading to an RMS noise level of about $30\ldots50\,\mathrm{mK}$ ($N^\mathrm{limit}_\ion{H}{i}=2\ldots4 \times10^{18}\,\mathrm{cm}^{-2}$).

\subsection{The sample}

Fig.\,\ref{fig_hvc_allskymap} shows the population and distribution of high-velocity \ion{H}{i} gas across the entire sky, based on the data of the LAB (grey scale). There are several coherent and extended HVC complexes covering the northern and southern sky, some of which are spanning tens of degrees like the Magellanic Stream, which is thought to have been stripped off the Magellanic Clouds by ram-pressure and/or tidal forces. New \ion{H}{i} data obtained with EBHIS and GASS revealed that complexes like the Galactic Center negative (GCN) are strongly fragmented into several warm and compact clumps \citep{winkel11} if observed with higher resolution. Furthermore, numerous isolated and compact HVCs can be seen all over the sky \citep{braunburton99, Putmanetal02, deheijbraunburton02}.

The symbols in Fig.\,\ref{fig_hvc_allskymap} mark the positions of the 408 sight lines that were observed with UVES (the blue line marks the region not accessible with the VLT). Along 126 lines of sight we detect 226 \ion{Ca}{ii} absorption components at intermediate or high velocities, as indicated by the filled circles and boxes. For \ion{Na}{i} we found 96 components along 75 sight lines. \ion{H}{i} spectra were available for 133 lines of sight (EBHIS: 28, GASS: 88, pointed Effelsberg: 17). For 65\% (EBHIS: 80\%, GASS: 60\%, pointed Effelsberg: 70\%) of these we detect \ion{H}{i} emission connected to the absorption features.

The conditional probability, $p(\ion{Na}{i}\,\vert\,\ion{Ca}{ii})$, to find \ion{Na}{i} if \ion{Ca}{ii} was detected is about 50\%, while $p(\ion{Ca}{ii}\,\vert\,\ion{Na}{i})$ is 90\%, i.e. in just 6 of all cases we found only  \ion{Na}{i} but no  \ion{Ca}{ii} along a sight line.

\begin{figure*}[!tp]
\centering
\includegraphics[width=0.44\textwidth,clip=,bb=3 14 414 581]{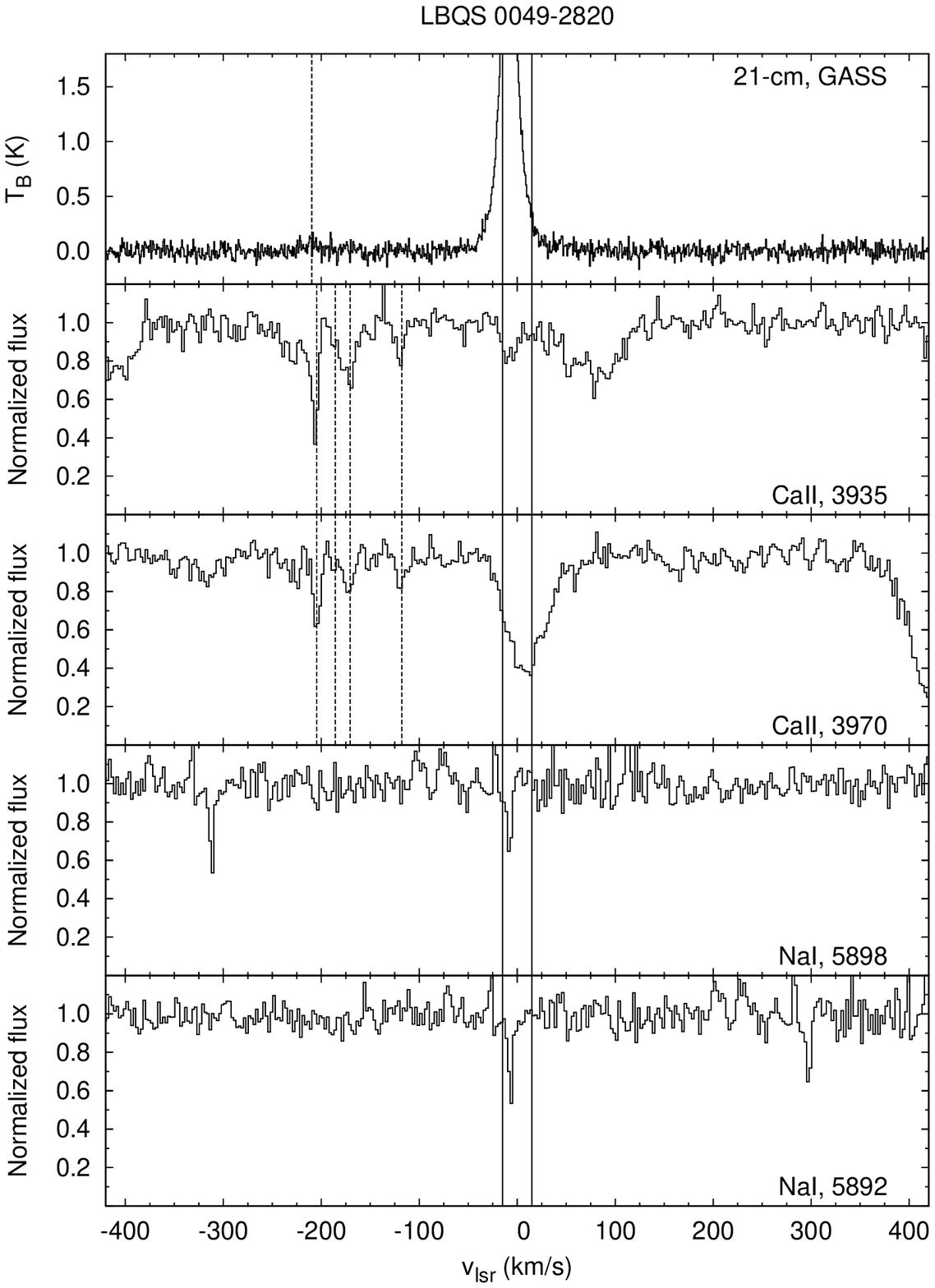}\hspace{5ex}
\includegraphics[width=0.44\textwidth,clip=,bb=3 14 414 581]{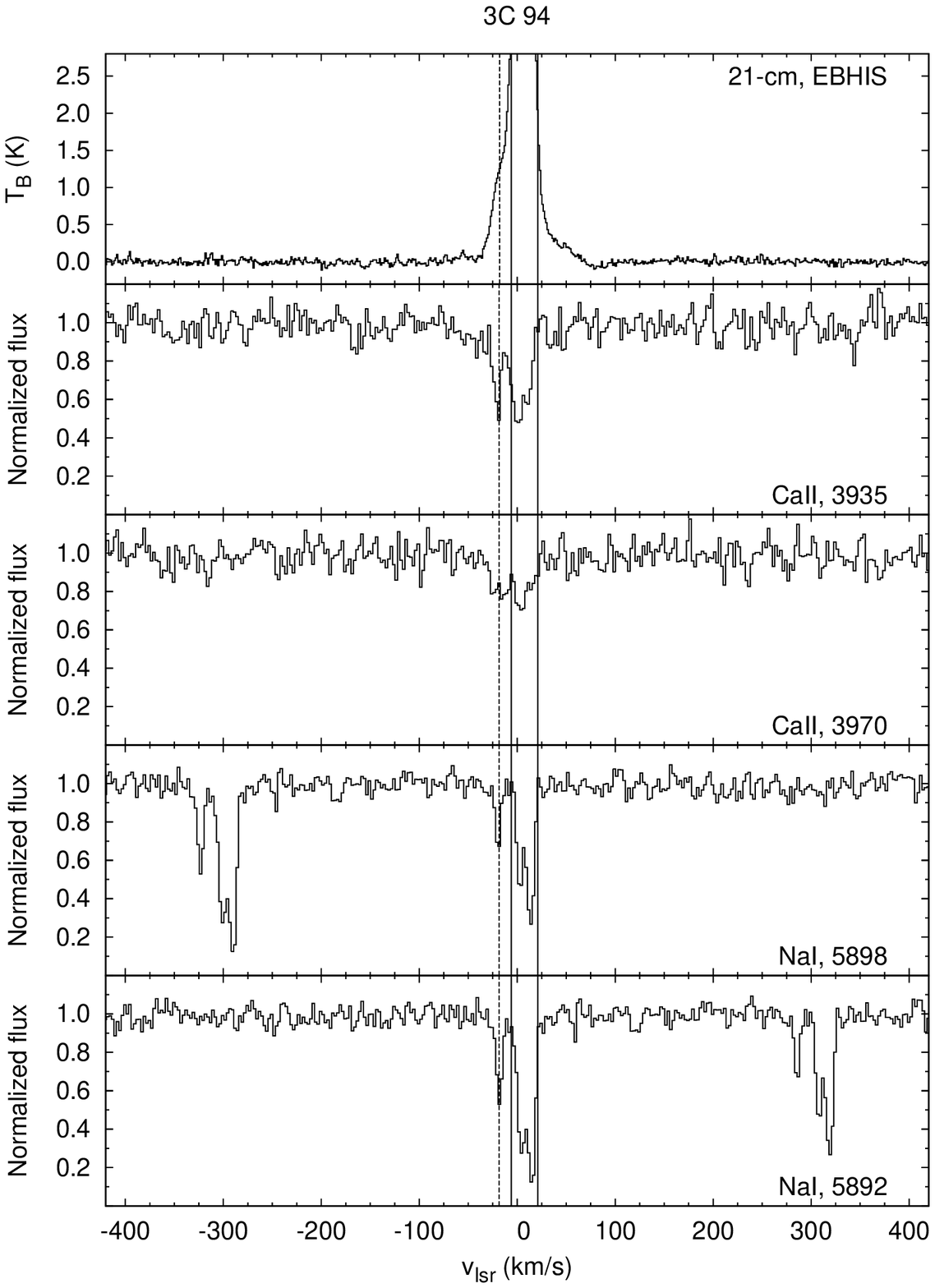} \\[1ex]
\includegraphics[width=0.44\textwidth,clip=,bb=3 14 414 581]{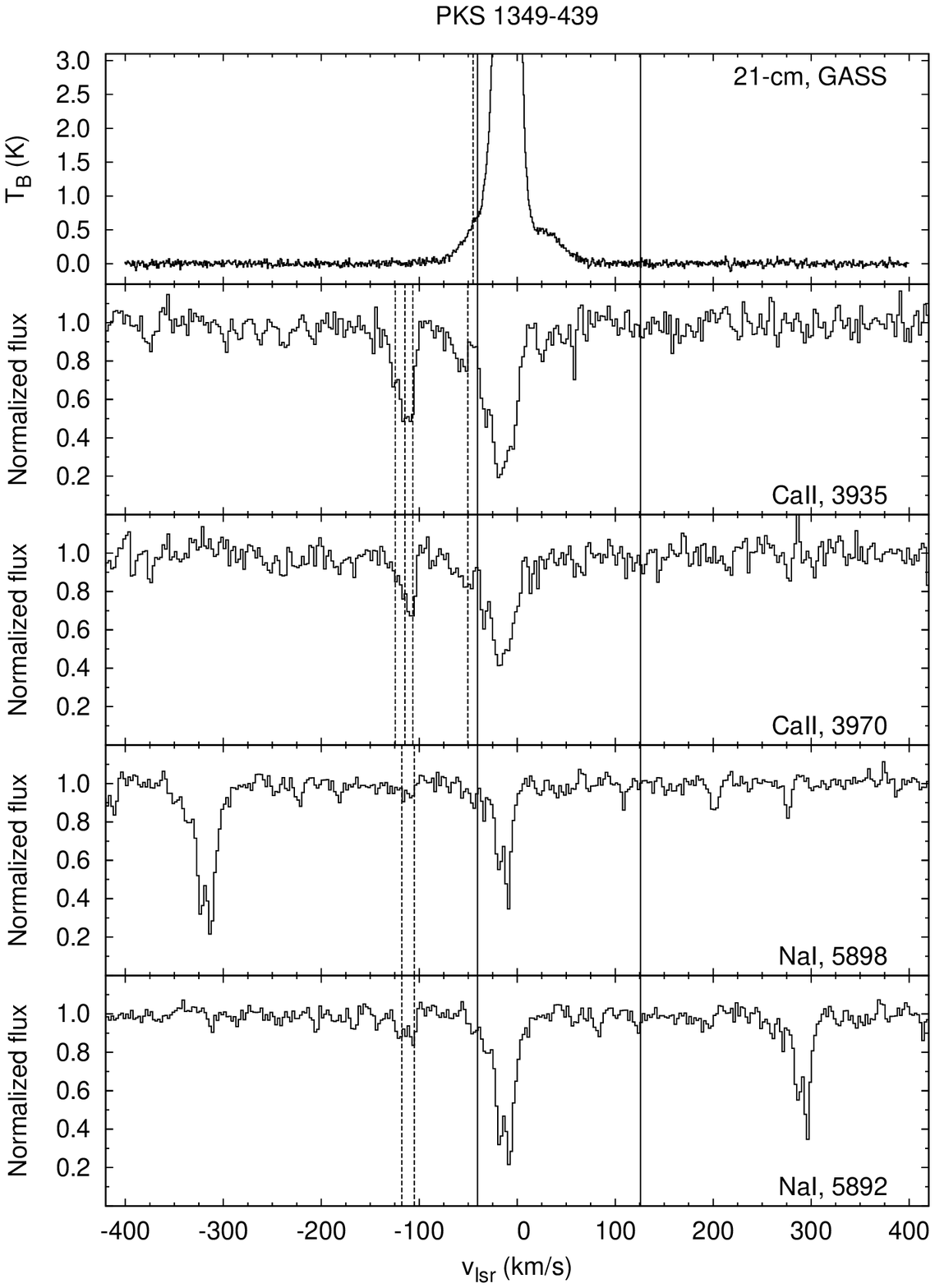}\hspace{5ex}
\includegraphics[width=0.44\textwidth,clip=,bb=3 14 414 581]{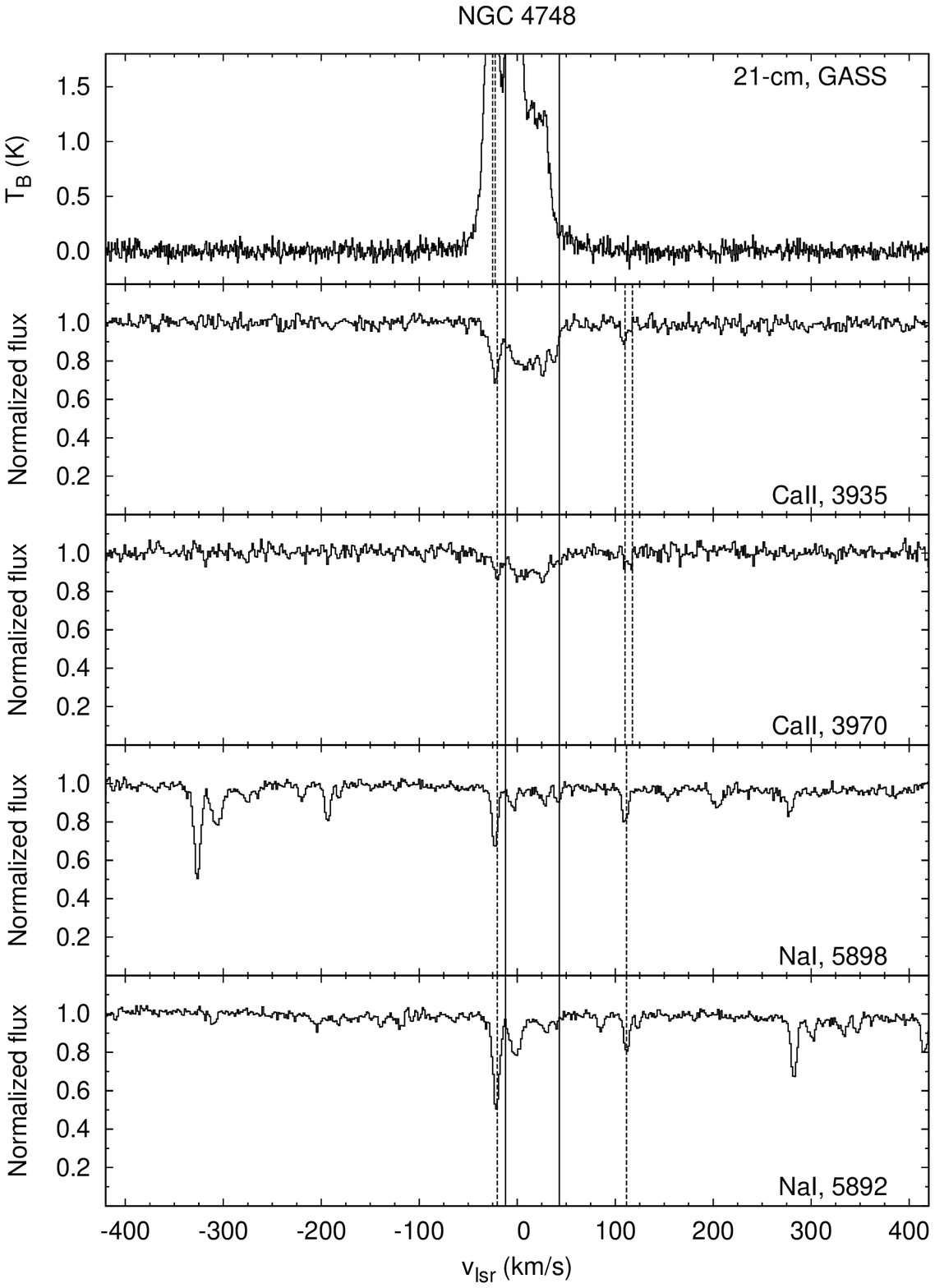}
\caption{Example \ion{Ca}{ii} and \ion{Na}{i} absorption and \ion{H}{i} emission spectra in the direction of the quasars LBQS\,0049$-$2820, 3C\,94, PKS\,1349$-$439 and NGC\,4748 obtained with UVES, EBHIS, and GASS. Detected components are indicated by dashed lines. The solid lines mark the $v_\mathrm{lsr}$ velocity range expected for Milky Way gas according to a model; see text.}
\label{fig_spectra}
\end{figure*}

Figure~\ref{fig_spectra} shows four example spectra of the sample with optical absorption of \ion{Ca}{ii} and \ion{Na}{i}, along with the corresponding \ion{H}{i} emission profiles observed with GASS and EBHIS, respectively. Most of the emission and absorption near zero velocities can be attributed to the local Galactic disc. To decide whether the gas in a given direction participates in Galactic disc rotation or not, we used a kinematic Milky Way model of the warm neutral medium \citep[WNM,][]{kalberla03, kalberlaetal07} and the concept of the deviation velocity, $v_\mathrm{dev}$, that is the difference of the radial velocity and the terminal velocity of the Milky Way disc as introduced by \citet{wakker91}. The model of the WNM disc gas predicts for each line of sight the radial velocity, the density, and the so-called $f$-layer as a function of distance, (out to 50\,kpc). The $f$-layer, $f_z$, is defined as
\begin{equation}
f_z=(z-z_\mathrm{disc})/w_\mathrm{disc}
\end{equation}
with $z_\mathrm{disc}$ the mean offset of the WNM from $z=0$, i.e., $z_\mathrm{disc}$ describes the Milky Way warp, and $w_\mathrm{disc}$ the exponential scale height of the WNM. The latter accounts for the flaring of the disc. Its local value is 400\,pc. For each line of sight we used the model to find the distance at which $f_z>4$ and extracted the associated radial velocities. They should resemble rather conservative limits on typical velocities of the WNM. Of course local phenomena cannot be predicted by the model, so in few cases we might wrongly classify a cloud as being located in the halo when it is in fact disc material (and vice versa).   

Supporting our previous results, $35\%$ of the detected halo \ion{Ca}{ii} and $20\%$ of \ion{Na}{i} halo absorbers show multiple intermediate- and high-velocity components, indicating the presence of sub-structure (see Section \ref{subsec:substructure}). Typical \ion{H}{i} column densities of the gas structures lie in the range of $2 \times 10^{18}\,\mathrm{cm}^{-2}$ (detected in the pointed Effelsberg observations) up to $2 \times 10^{20}\,\mathrm{cm}^{-2}$. Typical \ion{H}{i} line widths vary from $\Delta v_\mathrm{FWHM}=4$~km\,s$^{-1}$ to $25$~km\,s$^{-1}$, although the distribution is very wide-spread and several detections have more than $25$~km\,s$^{-1}$, which can possibly explained by substructure (compare Section\,\ref{subsec:bvalues}). These values convert to upper kinetic temperature limits ranging from 100\,K up to about $10^4$\,K, as expected for CNM and WNM, respectively. Measured column densities and $b$-values for all sight lines are summarised in Table~\ref{qsotable}.

The directions as well as the velocities of $45\%$ ($40\%$) of the halo \ion{Ca}{ii} (\ion{Na}{i}) absorption components indicate a possible association with known HVC or IVC complexes (see Section\,\ref{HVC_IVC_complexes}). 
The other components, in contrast, do not appear to be associated with any known HVC or IVC complex. Confirming our previous results, optical absorption spectra allow us to trace the full column density range of neutral gas structures in the Galactic halo. The distribution of neutral or weakly ionised gas obviously is more complex than indicated by \ion{H}{i} 21-cm observations alone.

Five sight lines (QSO\,B1448$-$232, HE\,0001$-$2340, QSO\,B0450$-$1310B, [HB89]\,1331$+$170 and HS\,0810$+$2554) with prominent halo \ion{Ca}{ii} and \ion{Na}{i} absorption lines were previously analysed with data from the Very Large Array (VLA) and the Westerbork Radio Synthesis Telescope (WSRT).
The high-resolution \ion{H}{i} data resolve the IVCs/HVCs into several compact, cold clumps \citep{richterwestmeierbruens05, benbekhtietal_09}. Additional lines of sight have been re-observed in a similar manner and the results will be presented in a forthcoming paper.

\section{Results of the statistical analysis}\label{secresults}

The large data sample allows us to perform a solid statistical analysis of the optical halo absorption components. In contrast to BB08 we are now able to include \ion{Na}{i} into the analysis as well.

\subsection{Column densities} 
\begin{figure}
\centering
\includegraphics[width=0.46\textwidth,clip=,bb=60 60 395 291]{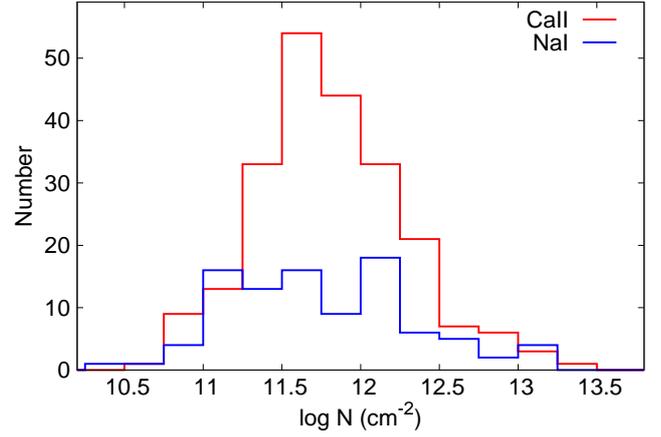}
\caption{Observed column density distributions of the \ion{Ca}{ii} and \ion{Na}{i} absorbers.}
\label{fig_Anzahl_vs_N}
\end{figure}

Figure~\ref{fig_Anzahl_vs_N} shows a histogram of the \textit{observed} column densities of the \ion{Ca}{ii} and \ion{Na}{i} halo absorbers. The column densities range from $\log(N_\ion{Ca}{ii}/\mathrm{cm}^{-2})=10.5 \ldots 13.5$ to $\log(N_\ion{Na}{i}/\mathrm{cm}^{-2})=10 \ldots 13.3$. Interpretation of the numbers is very difficult, because our sample is very inhomogeneous in terms of sensitivity. The peak of the histogram at $\log (N)/\mathrm{cm}^{-2}\sim 11.5$ is mainly due to selections effects.

\begin{figure}
\centering
\includegraphics[width=0.48\textwidth]{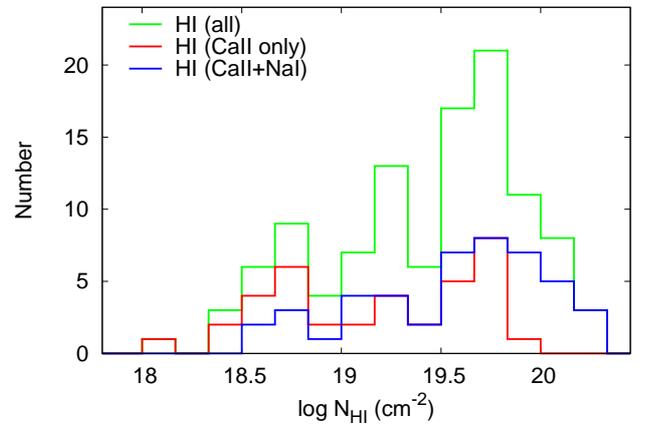}
\caption{Distribution of all \ion{H}{i} column densities (green line). The blue line displays the number of absorbers were both, \ion{Ca}{ii} and \ion{Na}{i}, is present, while the red line is for components showing only \ion{Ca}{ii}. }
\label{fig_HI_EBHIS+GASS}
\end{figure}
\begin{figure}
\centering
\includegraphics[width=0.48\textwidth]{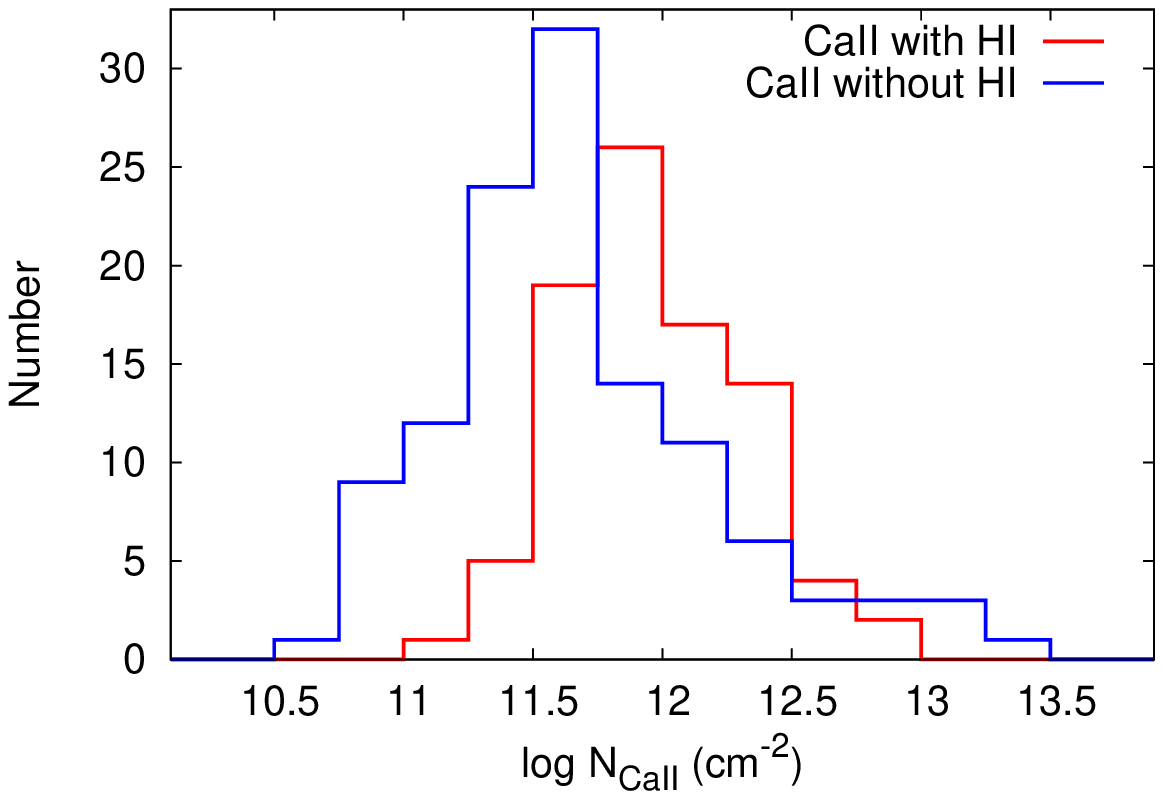}\\[-2ex] 
\includegraphics[width=0.48\textwidth]{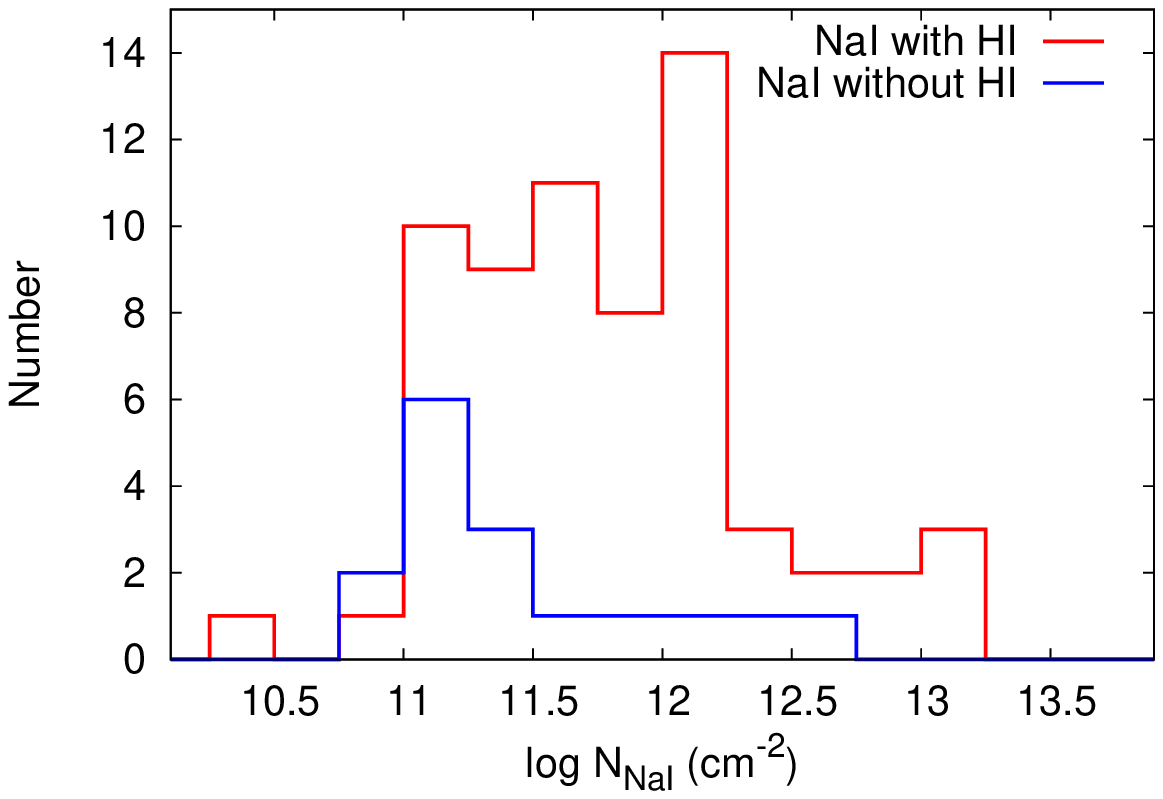}
\caption{Number of \ion{Ca}{ii} and \ion{Na}{i} absorbers with (red lines) and without (blue lines) corresponding \ion{H}{i} emission.}
\label{fig_number_CaIIandNaI_withandwithout_HI}
\end{figure}

Figure~\ref{fig_HI_EBHIS+GASS} shows the distribution of all \ion{H}{i} column densities in our halo sample observed with Effelsberg, EBHIS and GASS (green line). The blue line displays the histogram of $\log N_\ion{H}{i}$ for cases where both \ion{Ca}{ii} and \ion{Na}{i} were also detected, while the red line is associated to those components where only \ion{Ca}{ii} was detected. The probability to find higher \ion{H}{i} column densities (above $\log(N_\ion{H}{i}/\mathrm{cm}^{-2})\gtrsim19.8$) in our sample is slightly higher if both \ion{Ca}{ii} and \ion{Na}{i} are present, and not only \ion{Ca}{ii}.

This can be seen more clearly in Fig.\,\ref{fig_number_CaIIandNaI_withandwithout_HI} which displays the number of sight lines with \ion{Ca}{ii} (upper panel) and \ion{Na}{i} (lower panel) halo absorption with (red lines, A1) and without (blue lines, A2) corresponding \ion{H}{i} emission. We find \ion{H}{i} in almost all cases where \ion{Na}{i} was previously detected. Apparently, with our \ion{H}{i} and \ion{Na}{i} data we are (detection-)limited to structures of similar gas column densities, while for \ion{Ca}{ii} the limiting column density is about 0.5\,dex lower. Hence, \ion{Ca}{ii} can be considered a much better probe to study the low-column density gas phase in the Galactic halo.

\begin{figure}
\centering
\includegraphics[width=0.46\textwidth,clip=,bb=60 60 395 291]{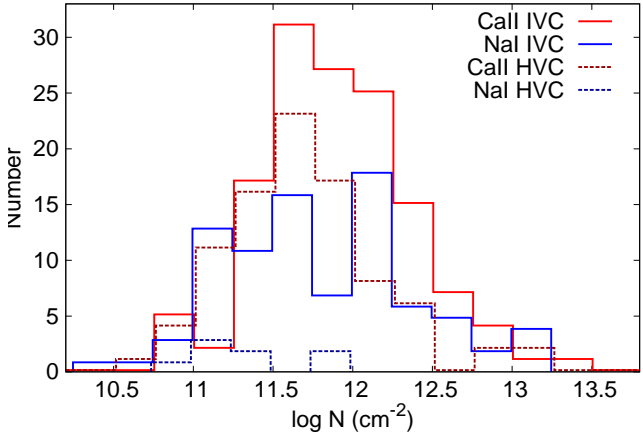}\hfill
\caption{Column densities of the absorbers separated by their deviation velocity. IVCs have $\vert v_\mathrm{dev}\vert<50\,\mathrm{km\,s}^{-1}$, while HVCs have $\vert v_\mathrm{dev}\vert\geq 50\,\mathrm{km\,s}^{-1}$.}
\label{fig_logN_ivc_hvc}
\end{figure}

In Fig.\,\ref{fig_logN_ivc_hvc} we show column density histograms separately for IVC and HVC gas. To decide whether a cloud is considered as high- or intermediate velocity we use the deviation velocity defined by \citet{wakker91}, which is the velocity distance to the closer terminal velocity of MW disc gas. Based on \citet{wakker91} we classify absorbers with $\vert v_\mathrm{dev}\vert\geq 50\,\mathrm{km\,s}^{-1}$ as HVCs. While for \ion{Ca}{ii} about one third of the absorbers originates from high-velocity gas, only a tiny fraction of \ion{Na}{i} detections stems from HVCs. The latter also have very low column densities, while the \ion{Ca}{ii} is similar for IVCs and HVCs; only the total number of absorbers is different.

\subsubsection{Column density distribution functions}\label{subseccdd}
\begin{figure} 
\centering
\includegraphics[width=0.48\textwidth,clip=,bb=51 60 391 291]{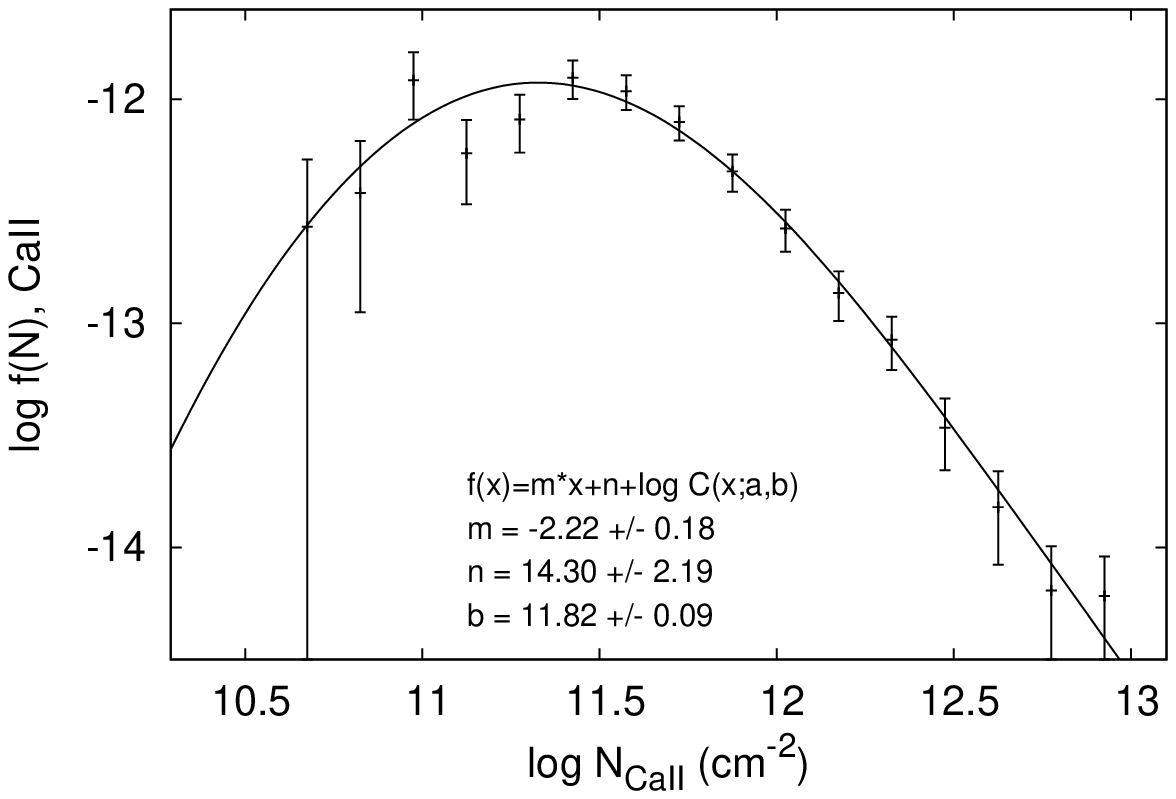}\\[1ex]%\hfill
\includegraphics[width=0.48\textwidth,clip=,bb=51 60 391 291]{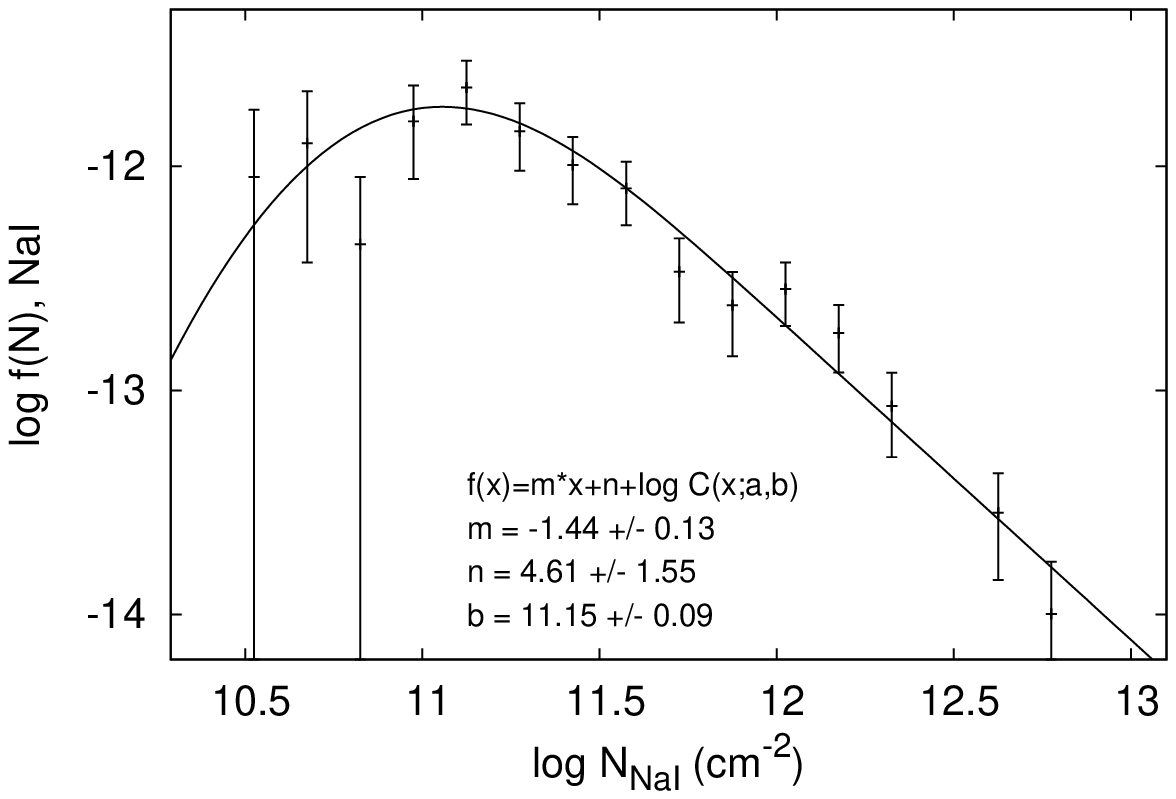}
\caption{The \ion{Ca}{ii} and \ion{Na}{i} column density distributions $f(N)$ (using logarithmic binning). The solid lines represent a power-law fit under consideration of the completeness function of our sample. From $\chi^2$-fitting we obtain a slope of $\beta=-2.2 \pm 0.2$ for \ion{Ca}{ii} and  $\beta=-1.4 \pm 0.1$ for \ion{Na}{i}.}
\label{fig_num_vs_columndensity} 
\end{figure}

Fig.\,\ref{fig_num_vs_columndensity} shows the \ion{Ca}{ii} and \ion{Na}{i} column density distribution (CDD) functions, 
\begin{equation}
f(N)=\frac{1}{M}\frac{m(N)}{\Delta N},\qquad M=\int_{N_\mathrm{min}}^{N_\mathrm{max}} f(N)\,\mathrm{d}N 
\end{equation}
the normalised number of absorbers $m(N)$ per column density interval $[N, N+ \Delta N]$ \citep[][BB08]{churchillvogtcharlton03}.

Due to the increasing incompleteness of our sample the CDD drops off towards smaller column densities. While one usually circumvents this problem by using only the rightmost values for fitting the slope \citep[e.g., BB08, ][]{richteretal10} we incorporate the inferred sample completeness function (see Section\,\ref{subsubsecsamplecompleteness}) into fitting of the CDD. This has two major advantages. First, one does not need to introduce an arbitrary threshold to the CDD values used for the fit and second, one can use all of the data points, thus increasing the statistical significance of the fit. In order to include the completeness into the fit, one just has to multiply the power-law fit function with the completeness function, $C(\log N; a,b)$. For convenience, we convert to the log--log regime where, instead of fitting a linear function to the rightmost data points (as in BB08), we use
\begin{equation}
f(\log N)=\beta \log N+n+\log C(\log N; a,b)\label{eqcddfitfunction}
\end{equation}
to parametrise the CDD. Here, $\beta$ is the slope of the underlying power-law, $n$ is a normalisation constant. In Fig.\,\ref{fig_num_vs_columndensity} the result of a $\chi^2$-fit is shown. Note that only the offset $b$ of the completeness function was fitted, while the shape parameter $a$ was kept constant (i.e., $a_\ion{Ca}{ii}=1.968\,\mathrm{cm}^2$, $a_\ion{Na}{i}=2.336\,\mathrm{cm}^2$), the values of which were determined in Section\,\ref{subsubsecsamplecompleteness}. The resulting offset values of $b_\ion{Ca}{ii}=11.8\,\mathrm{cm}^{-2}$, $b_\ion{Na}{i}=11.2\,\mathrm{cm}^{-2}$ are not completely matching the theoretical predictions, but are similar. The difference could be explained with blending of lines, an effect which decreases the completeness of our sample, but has not been incorporated into the $\mathrm{S/N}_\mathrm{p}$-statistics. Furthermore, the detected lines have of course different line widths, while for the detection limits a constant width was assumed for all spectra.

We obtain power-law slopes of $\beta_\ion{Ca}{ii}=-2.2 \pm 0.2$ and  $\beta_\ion{Na}{i}=-1.4 \pm 0.1$. The former is steeper than found in BB08, but still consistent within $3\sigma$ confidence with the older result. However, the new finding has smaller errors and describes the CDD much better (in terms of reduced $\chi^2$).

\begin{figure}
\centering
\includegraphics[width=0.48\textwidth,clip=,bb=58 60 395 291]{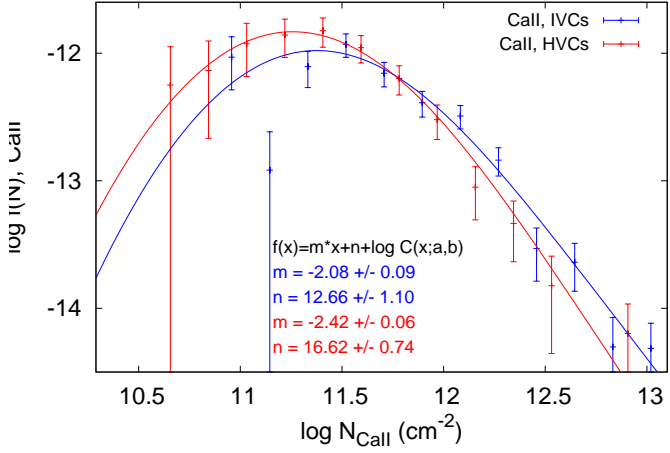}
\caption{The \ion{Ca}{ii} column density distributions $f(N)$ separately for IVC and HVC gas. Note that the completeness function $C(\log N; a,b)$ was kept fixed using the values from Fig.\,\ref{fig_num_vs_columndensity}.}
\label{fig_hvcivccdd}
\end{figure}

Our \ion{Ca}{ii} slope is also steeper than found for extra-galactic samples \citep[e.g.,][$\beta_\ion{Ca}{ii}\approx-1.7$]{richteretal10}, which is, however, not too surprising. In the latter not only pure halo absorbers are measured but also discs (having \ion{H}{i} column densities of up to $10^{22}\,\mathrm{cm}^{-2}$) which thus flattens the CDD. Another effect probably is the depletion of Calcium onto dust grains. As our absorber sample contains a good fraction of HVCs (which are known to have little dust), more Ca will reside in the free form in the high-velocity structures, which populate preferentially the low-column density regime. Especially the latter effect can be seen in Fig.\,\ref{fig_hvcivccdd} where we calculated the CDD separately for IVCs and HVCs. High-velocity gas shows even steeper CDD slopes (though the residual scatter is somewhat larger due to the smaller sample sizes), while the IVC slope is slightly flatter than that of the full sample. Dust depletion is further discussed in Section\,\ref{subsecdustdepletion}.

\begin{figure}
\centering
\includegraphics[width=0.48\textwidth,clip=,bb=58 60 395 291]{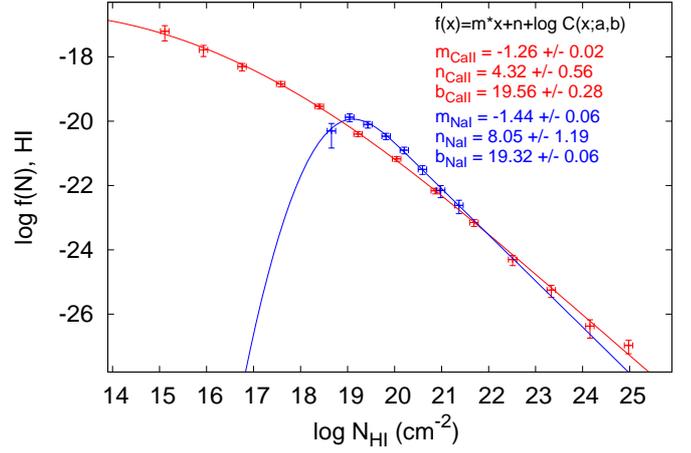}
\caption{Converted \ion{H}{i} column density distribution functions $f(N)$. The \ion{H}{i} column densities were calculated from the \ion{Ca}{ii} and \ion{Na}{i} column densities
using the correlation found by \citet{wakkermathis00}. The solid lines represent power-law fits (again corrected for completeness of the sample), with a slope of $\beta_{\ion{H}{i}[\ion{Ca}{ii}]}=-1.26 \pm 0.02$ and $\beta_{\ion{H}{i}[\ion{Na}{i}]}=-1.44 \pm 0.06$.}
\label{fig_lognum_vs_logHIcolumndensity_UVES}
\end{figure}

% using log A(Z) = log N(Z) - log N(H), and
% log A(CaII) = -0.78*(log N(HI) - 19.5) -7.76
% log A(NaI) = -0.16*(log N(HI) - 19.5) -8.12

Using the correlations
\begin{align}
\log \left(N_\ion{H}{i}/\mathrm{cm}^{-2}\right) &= 4.55 \left[\log \left(N_\ion{Ca}{ii}/\mathrm{cm}^{-2}\right)-7.45\right]
\label{eq_caIIintoHIconversion}\\
\log \left(N_\ion{H}{i}/\mathrm{cm}^{-2}\right) &=  1.19 \left[\log \left(N_\ion{Na}{i}/\mathrm{cm}^{-2}\right)+5\right]
\label{eq_naIintoHIconversion}.
\end{align}
between \ion{Ca}{ii}, \ion{Na}{i} and \ion{H}{i} found by \citet{wakkermathis00}, we can convert the \ion{Ca}{ii} and \ion{Na}{i} column densities into \ion{H}{i} column densities. One should keep in mind though that the conversion formula is affected by uncertainties as discussed in their paper. 
Fig.\,\ref{fig_lognum_vs_logHIcolumndensity_UVES} shows the converted \ion{H}{i} column density distribution functions $f(N)$ calculated from the \ion{Ca}{ii} and \ion{Na}{i} column densities. 
The solid lines represent again the best-fit through the data using Eq.\,(\ref{eqcddfitfunction}). Apparently, the slopes $\beta_{\ion{H}{i}[\ion{Ca}{ii}]}=-1.26 \pm 0.02$ and $\beta_{\ion{H}{i}[\ion{Na}{i}]}=-1.44 \pm 0.06$ inferred from the two species differ. Although both values are still consistent within a $3\sigma$-level, it is likely that the conversion formula of \citet{wakkermathis00} is not perfectly in agreement with our halo-absorber data. Dust depletion might play a role, as in the halo we expect more Calcium to reside in the free form which would flatten the CDD (with respect to \ion{Na}{i}) (compare Section\,\ref{subsecdustdepletion}).  Furthermore, \citet{wakkermathis00} used data having $\log (N_\ion{H}{i}/\mathrm{cm}^{-2})\gtrsim18$, which means that our \ion{Ca}{ii} CDD data points below this threshold are nothing else than extrapolation. Below $\log (N_\ion{H}{i}/\mathrm{cm}^{-2})\sim17$ the conversion formula even leads to unphysically high (i.e., above solar) abundance ratios.

Again, we can compare the (converted) \ion{H}{i} slope with the value of $\beta_{\ion{H}{i}[\ion{Ca}{ii}]}=-1.33 \pm 0.11$ found in BB08. Apparently, the results are consistent within $1\sigma$, but the new finding has much lower error bars.

One can also compare the converted \ion{H}{i} CDD with other samples. \citet{petitjean93} used  low- and high-redshift ($\langle z\rangle=2.8$) QSO absorption line data to obtain a slope of $\beta_\ion{H}{i}=-1.32$ for the $\log (N_\ion{H}{i}/\mathrm{cm}^{-2}) > 16$ regime. This is consistent with our values of $\beta_{\ion{H}{i}[\ion{Ca}{ii}]}=-1.26 \pm 0.02$ and $\beta_{\ion{H}{i}[\ion{Na}{i}]}=-1.44 \pm 0.06$, despite the issues with the conversion formula of \citet{wakkermathis00} which does not account for dust-depletion effects.

Using 21-cm line emission data only, \citet{lockman02} find a distribution of $\beta_\ion{H}{i}\sim-1$ which is not in agreement with our data. However, if beam-smearing plays a role \citep[which is indicated by recent observations, e.g.,][]{winkel11}, their column density detection limit of $\log (N_\ion{H}{i}/\mathrm{cm}^{-2})\sim17.9$ not only needs to be increased, but replaced with a certain selection function that could lead to a steeper CDD.

\subsection{Deviation velocities}\label{Deviation velocities}
\begin{figure*}
\centering
\includegraphics[width=0.49\textwidth, bb=70 98 420 268, clip=]{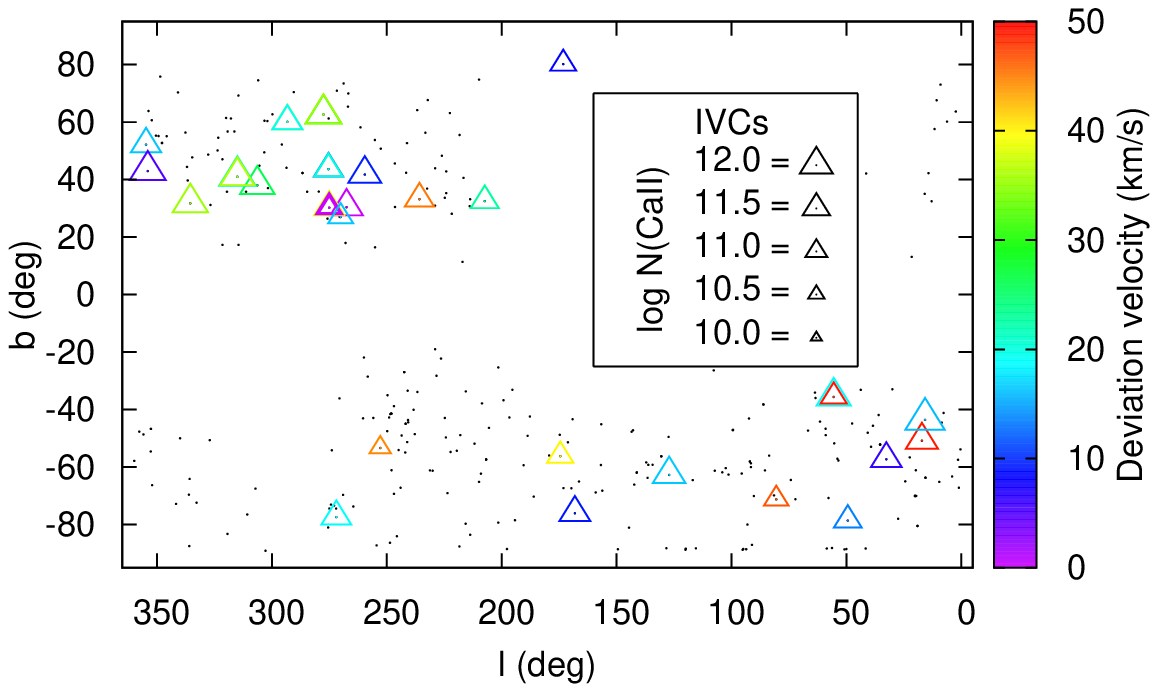} \includegraphics[width=0.49\textwidth, bb=70 98 420 268, clip=]{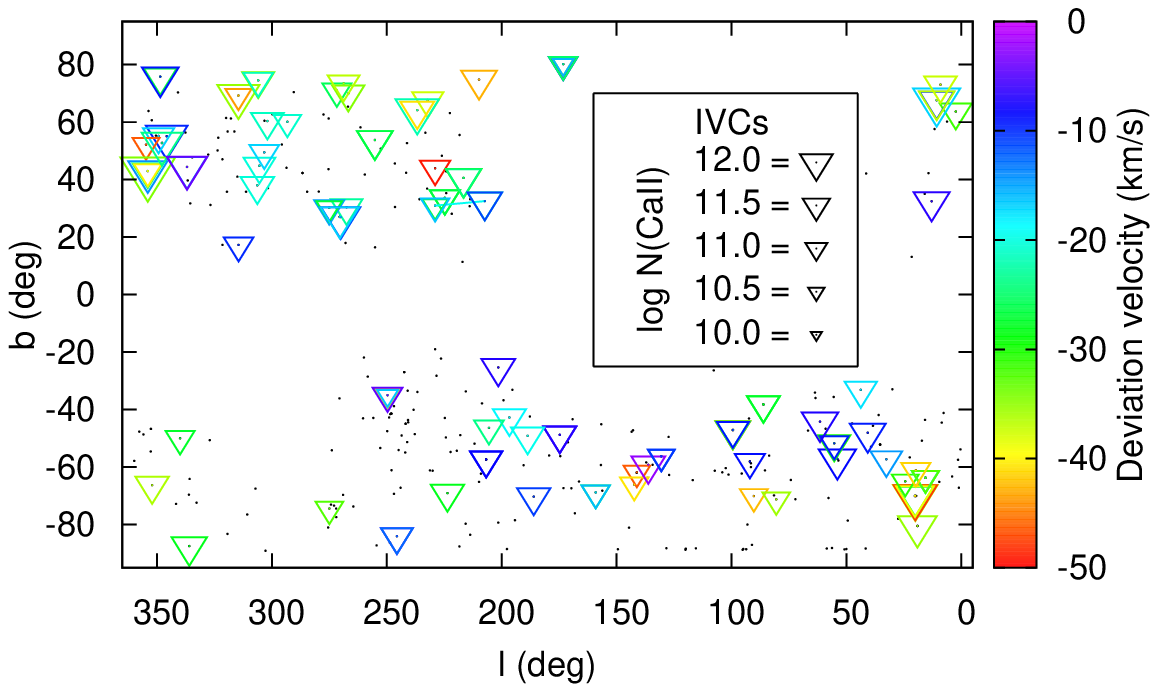}\\[0ex]
\includegraphics[width=0.49\textwidth, bb=70 98 420 268, clip=]{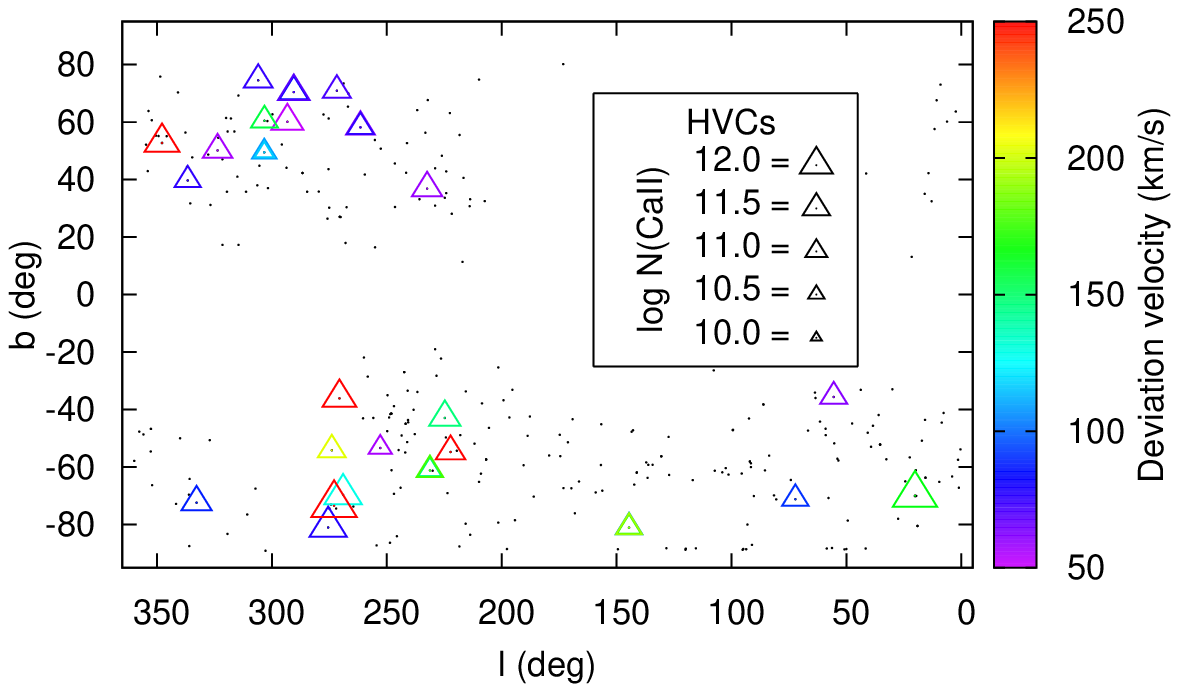} \includegraphics[width=0.49\textwidth, bb=70 98 420 268, clip=]{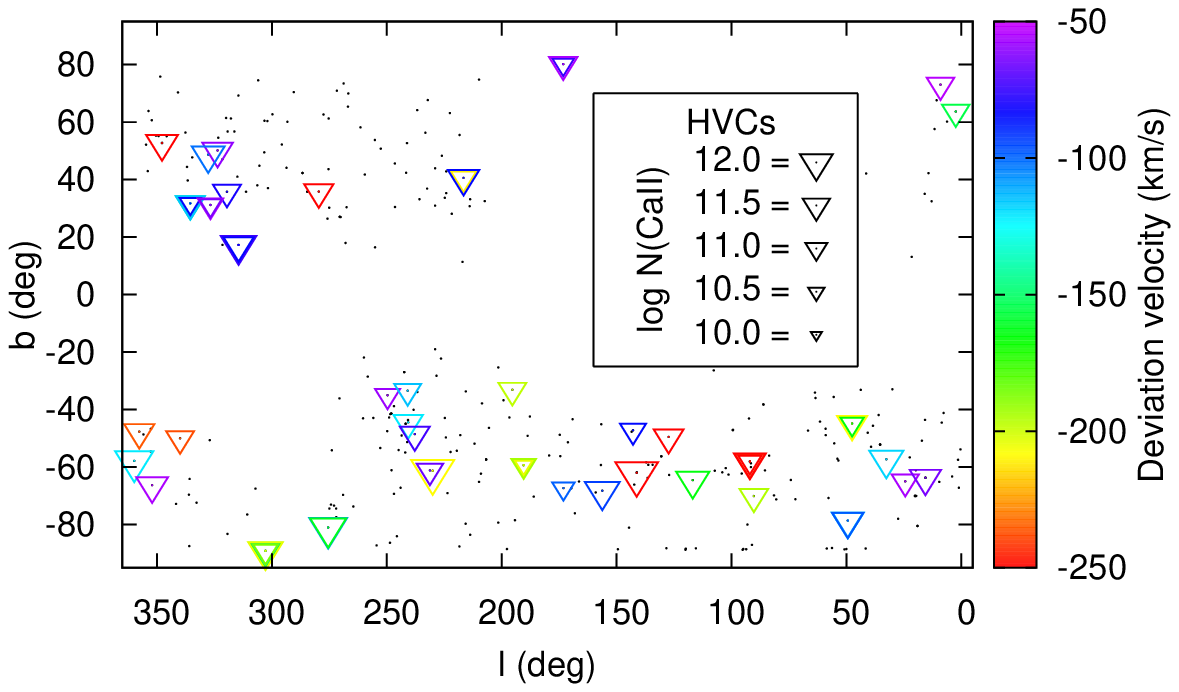}\\[0ex]
\includegraphics[width=0.49\textwidth, bb=70 98 420 268, clip=]{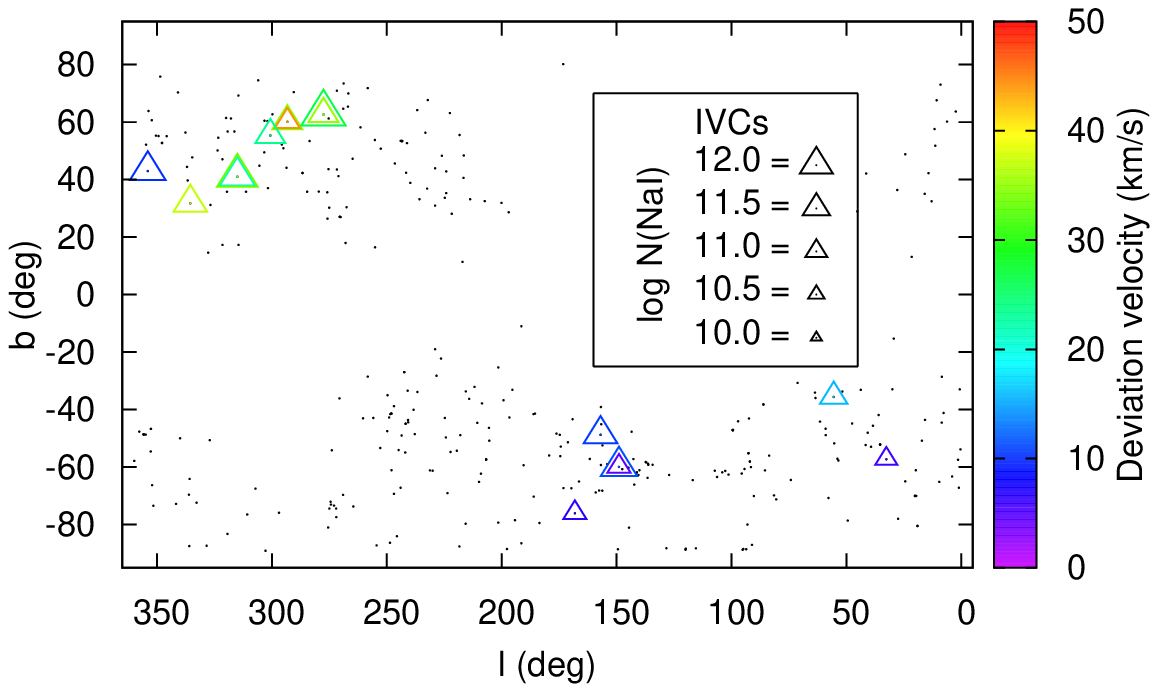} \includegraphics[width=0.49\textwidth, bb=70 98 420 268, clip=]{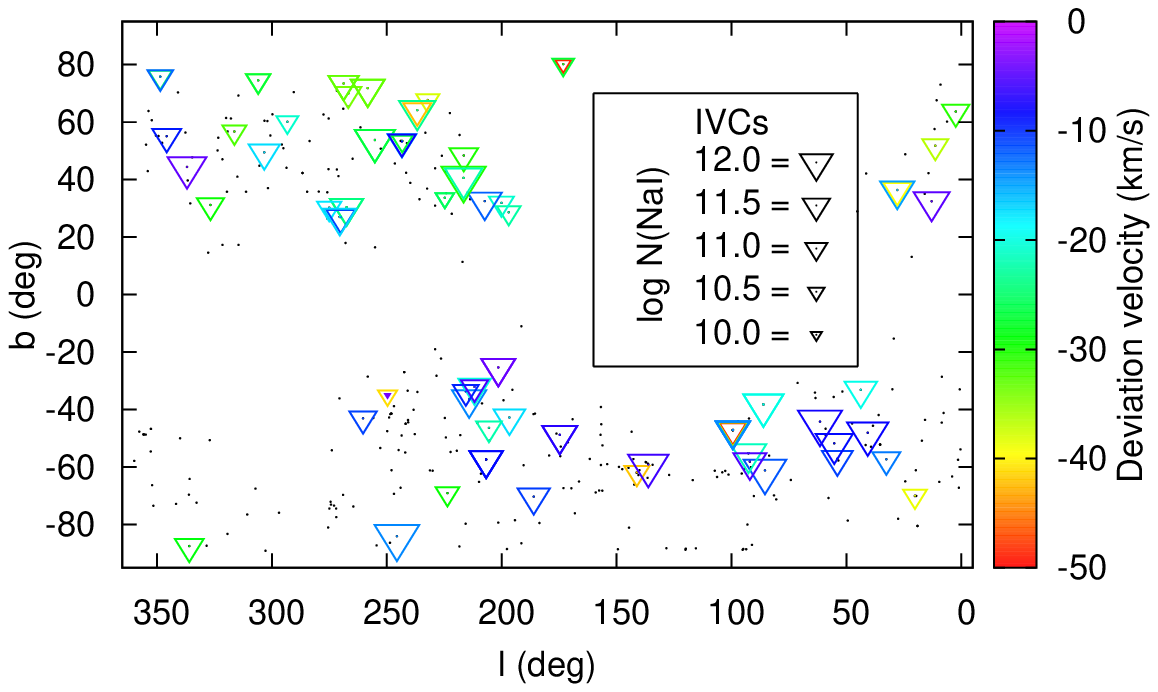}\\[0ex]
\includegraphics[width=0.49\textwidth, bb=70 65 420 268, clip=]{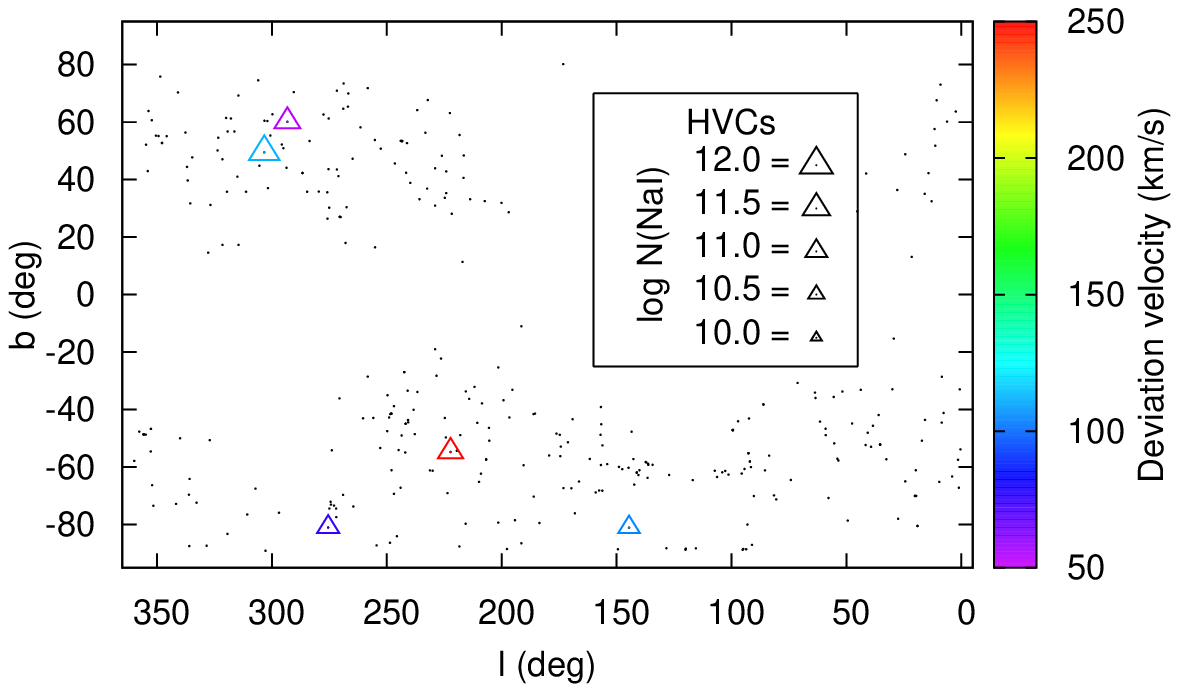} \includegraphics[width=0.49\textwidth, bb=70 65 420 268, clip=]{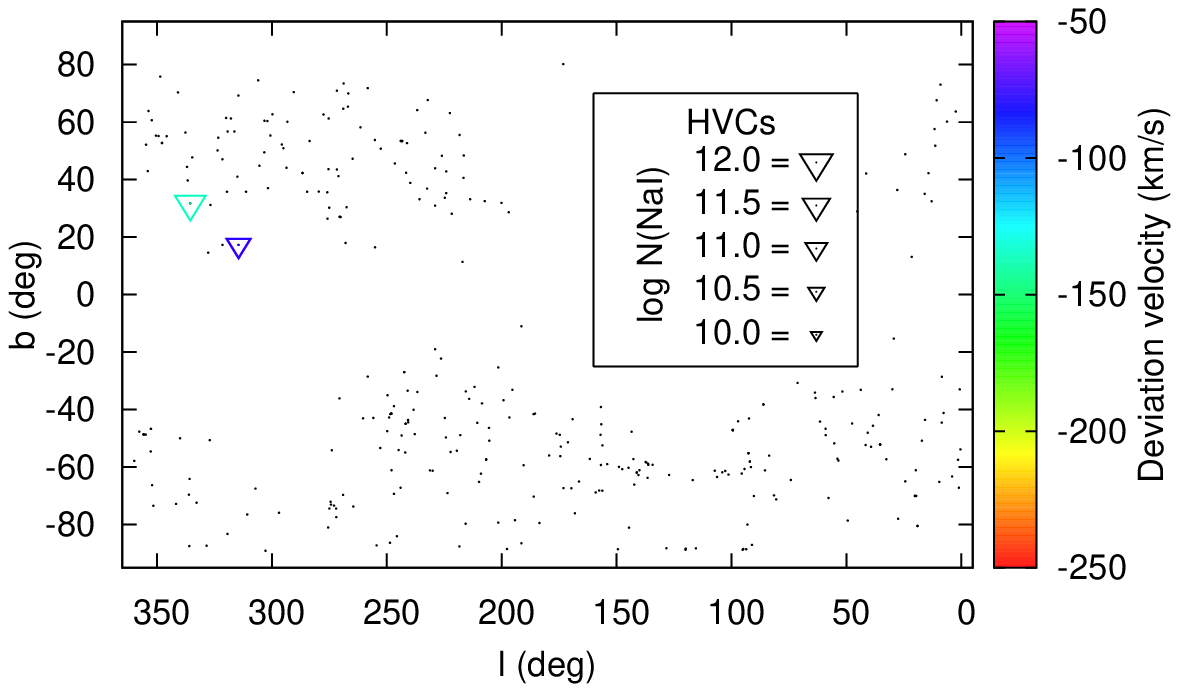}\\[0ex]
\caption{Distribution of deviation velocities of the observed intermediate- and high-velocity \ion{Ca}{ii} and \ion{Na}{i} absorbers versus Galactic longitude and latitude. The dimensions of the triangles are proportional to the logarithm of the \ion{Ca}{ii} and \ion{Na}{i} column densities observed with the UVES instrument. The black dots mark the position of all sight lines for which spectra were available.} 
\label{fig_deviation_velocity_HVC_IVC} 
\end{figure*} 

Fig.\,\ref{fig_deviation_velocity_HVC_IVC} shows the deviation velocities, $v_\mathrm{dev}$, of all observed \ion{Ca}{ii} and \ion{Na}{i} halo absorbers versus galactic longitude ($l$) and latitude ($b$). We use $v_\mathrm{dev}$ to separate Milky-Way disc gas from intermediate- and high-velocity clouds.  The size of the triangles in Fig.\,\ref{fig_deviation_velocity_HVC_IVC} is proportional to the logarithm of the \ion{Ca}{ii} and \ion{Na}{i} column densities observed with UVES. For reference, the triangles in the legend have column densities of $N_\mathrm{CaII, NaI}=1 \times 10^{12}$\,cm$^{-2}$. In these plots, there is no systematic trend visible in the velocity distributions.

\begin{figure} 
\centering
\includegraphics[width=0.48\textwidth,bb=60 60 395 290,clip=]{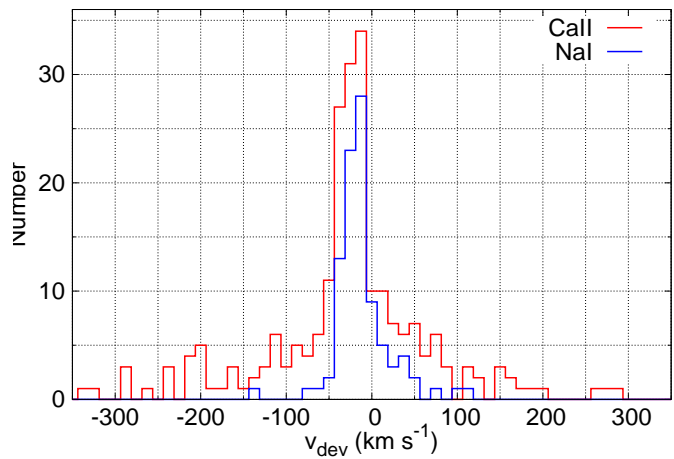}
\caption{Number of \ion{Ca}{ii} and \ion{Na}{i} absorbers versus deviation velocity.}
\label{fig_numvsvdev_IVC_HVC_zusammen} 
\end{figure}

Fig.\,\ref{fig_numvsvdev_IVC_HVC_zusammen} shows a histogram of absorption components as a function of deviation velocity. The figure indicates that IVCs and HVCs trace different cloud populations, as the histograms cannot be described by a single (Gaussian-like) distribution, but rather by a narrow IVC population superposed onto a broad HVC distribution. In addition, we find an excess of negative deviation velocities. It is unlikely that this is due to a selection effect caused by the missing north-eastern part in our sample, as  most of the known complexes in that area have negative deviation velocities \citep{wakker91}. Such an excess was also reported by \citet{blitz99}. They calculated the mean velocity and dispersion for different velocity rest frames. To compare our sample with the data of \citet{blitz99}  we show in Table\,\ref{tabvelocityframes} the median velocity, $\langle v \rangle $, and velocity dispersion, $\langle v^2 \rangle$, of halo \ion{Ca}{ii} and \ion{Na}{ii} absorbers for various rest frames, separated for intermediate and high velocities. As only 6 \ion{Na}{ii} HVCs were detected we omit the latter.

\begin{table}
\caption{Median velocities, $\langle v \rangle $, for different rest frames (deviation velocities, local, Galactic, and Local Group  standard of rest) and velocity dispersions separated for intermediate- and high-velocity clouds. }
\label{tabvelocityframes}
\centering
\begin{tabular}{l c c c c c c }     % 7 columns
\hline\hline
& \multicolumn{4}{c}{IVC}& \multicolumn{2}{c}{HVC}\\
& \multicolumn{2}{c}{\ion{Ca}{ii}}& \multicolumn{2}{c}{\ion{Na}{i}}& \multicolumn{2}{c}{\ion{Ca}{ii}}\\
Frame & $\langle v \rangle $ & $\langle v^2 \rangle $& $\langle v \rangle $ & $\langle v^2 \rangle $& $\langle v \rangle $ & $\langle v^2 \rangle $\\
\hline
DEV & $-17$ & 24 & $-15$ & 20 & $-74$ & 150  \\
LSR & $-20$ & 45 & $-18$ & 22 & $-90$ & 160 \\
GSR & $-42$ & 90 & $-63$ &100 & $-92$ & 150  \\
LGSR & $-66$ & 120 & $-72$ & 130 & $-90$ & 170  \\
\hline
Sample size  & \multicolumn{2}{c}{136}& \multicolumn{2}{c}{88}& \multicolumn{2}{c}{90} \\
\hline
\end{tabular}
\tablefoot{All velocities are given in $\mathrm{km\,s}^{-1}$.}
\end{table}

The values in the tables are again subject to the inhomogeneous coverage of our sample and to some extent also to the different sensitivities of the sight lines. At least the latter effect should be not too large --- the noise levels are distributed more or less randomly on the sky. This is mainly, because the sight lines in our sample were not observed to study the MW halo in the first place.

Not surprisingly, the dispersion of the IVC distributions is lowest in the LSR and deviation velocity frame. As for the full sample, there is a negative-velocity excess. In contrast to the results of \citet{blitz99}, the dispersion of HVCs in our sample is not smaller in the Local Group standard-of-rest frame (LGSR) and GSR. The LSR dispersion value is similar to that of \citet{blitz99} with $\sim 160\,\mathrm{km\,s}^{-1}$. However, we cannot reproduce their much smaller GSR and LGSR dispersion values of $\sim 100\,\mathrm{km\,s}^{-1}$.

Regardless of the rest frame we observe a mean velocity that is significantly below zero (about $-75\ldots-90\,\mathrm{km\,s}^{-1}$ and $-20\,\mathrm{km\,s}^{-1}$ for HVCs and IVCs, respectively). We attribute the excess of negative velocities to a net infall of gas onto the Milky Way disc for both, IVCs and HVCs, which is in agreement with the findings of \citet{marasco11} who used LAB data to find a best-fit model for extra-planar gas in the MW. They report on both, vertical ($v_z=-20\,\mathrm{km\,s}^{-1}$) and radial ($v_R=-30\,\mathrm{km\,s}^{-1}$), infall.

Note that we find 17 sight lines with very high deviation velocities of $\vert v_\mathrm{dev} \vert > 200\, \mathrm{km\,s}^{-1}$. Twelf of these pass the region between the anti-center complex and the Magellanic system, two pass HVC complex Galactic center negative, for both of which high negative deviation velocities are commonly observed. The remaining three sight lines (PKS\,0922$+$14, HE\,1126$-$2259, and QSO\,B1429$-$0053B) cannot be associated to a known HVC structure.

\subsection{Doppler parameters and multi-component structure} \label{subsec:bvalues}

\begin{figure} 
\centering
\includegraphics[width=0.48\textwidth,clip=,bb=42 35 448 583]{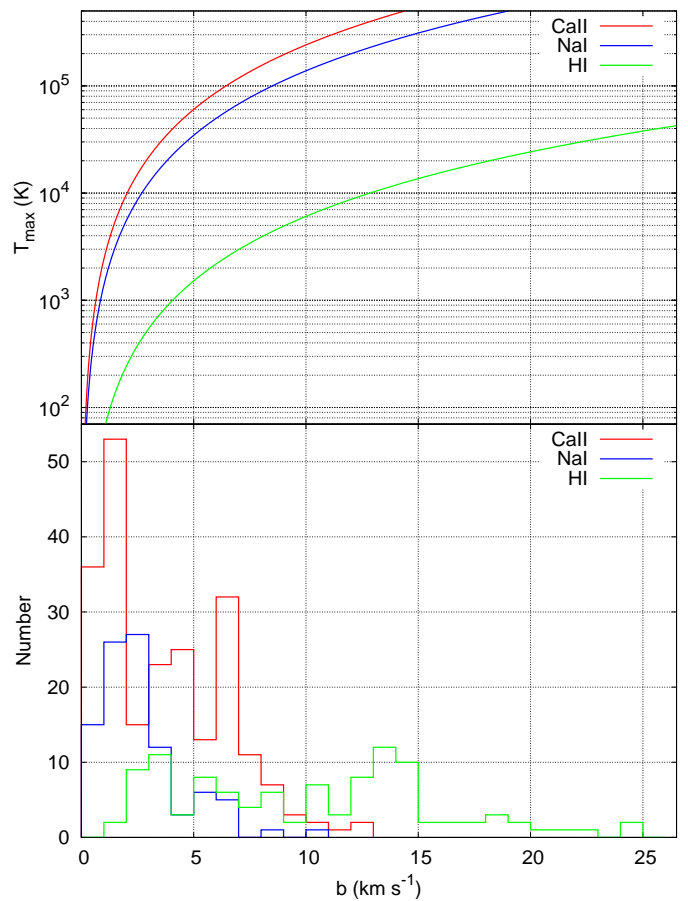}
\caption{Lower panel: Number of \ion{Ca}{ii}, \ion{Na}{i} and \ion{H}{i} detections versus Doppler-parameter ($b$-value). Median values for the Doppler parameter are $b=3.3$\,km\,$^{-1}$ for \ion{Ca}{ii} and $b=2.1$\,km\,$^{-1}$ for \ion{Na}{i}. For neutral atomic hydrogen the $b$-values have an enormous spread, which can most likely be explained with substructure since the spatial resolution of the 21-cm data is very coarse. For convenience we show in the upper panel the equivalent maximal (kinetic) temperatures associated with the $b$-values. }
\label{fig_bvalue} 
\end{figure}

Fig.\,\ref{fig_bvalue} shows the distribution of Doppler parameters ($b$-values) for our halo-absorber sample. The $b$-value distribution of \ion{Ca}{ii} peaks at $b= 1 \ldots 2$\,km\,s$^{-1}$ and the maximum for \ion{Na}{i} lies at $b= 2 \ldots 3$\,km\,s$^{-1}$. In case of \ion{Ca}{ii}, a broad tail extends to $b \approx 13$\,km\,s$^{-1}$. The median values are $b_\mathrm{CaII}=3.3\, \mathrm{km\,s}^{-1}$ and $b_\mathrm{NaI}=2.1\, \mathrm{km\,s}^{-1}$. 

The obtained Doppler widths are mostly smaller than the instrumental resolution of the UVES data. As discussed in Section\,\ref{uves_data} the Voigt profile fitting software \textsc{Fitlyman} applies a de-convolution technique to remove instrumental broadening. Since the spectra are fully Nyquist sampled, i.e. $\Delta v<\delta v/2$, even the narrowest lines are well sampled (after broadening). The uncertainty of the $b$-values is of course higher for very small values. Furthermore, de-convolution provides only an estimate of the true widths, if the absorption was caused by a single component.

From the Doppler-widths one can infer an upper temperature limit, $T_\mathrm{max}$, assuming purely thermal line-broadening, since for an ion with atomic weight $A$ the thermal component of the $b$ value is $b_{\rm therm}=0.129\,(T/A)^{0.5}$. However, non-thermal broadening (e.g., by turbulence, gas flows, or unresolved velocity substructure) often contributes substantially to the observed $b$ values in interstellar absorbers so that $b^2= b_{\rm therm}^2+ b_{\rm non-therm}$ (and $T\ll T_\mathrm{max}$).

For convenience we display in the top panel of Fig.\,\ref{fig_bvalue} the $T^\mathrm{max}$ values as a function of $b$. Apparently, at least half of our \ion{Ca}{ii} and \ion{H}{i} lines and about a third of the \ion{Na}{i} lines have values for $T_\mathrm{max}$ above $10^4\,\mathrm{K}$, which is unrealistically high for neutral gas traced by \ion{Ca}{ii} and \ion{Na}{i}. This strongly hints at the presence of unresolved substructure or other broadening effects adding to the line widths. In case of \ion{H}{i} this is not surprising given the rather low spatial resolution which can easily lead to beam-smearing. For a few sight lines this is confirmed by our previous high-resolution observations using the VLA and the WSRT where we detected small ($\varphi\lesssim3\arcmin$) and cold ($100\leq T^\mathrm{max} \leq 4000\,\mathrm{K}$) clumps \citep{richterwestmeierbruens05, benbekhtietal_09}. In addition, ultraviolet measurements of molecular hydrogen in Galactic extraplanar clouds have shown that small, dense gaseous clumps at sub-pc scale are widespread in the lower halo, in particular in intermediate-velocity clouds \citep{richterdeboeretal99, richtersavagesembachetal03, richterwakkersavagesembach03}.

\subsubsection{Absorption component multiplicity}

\begin{figure}
\centering
\includegraphics[width=0.48\textwidth,clip]{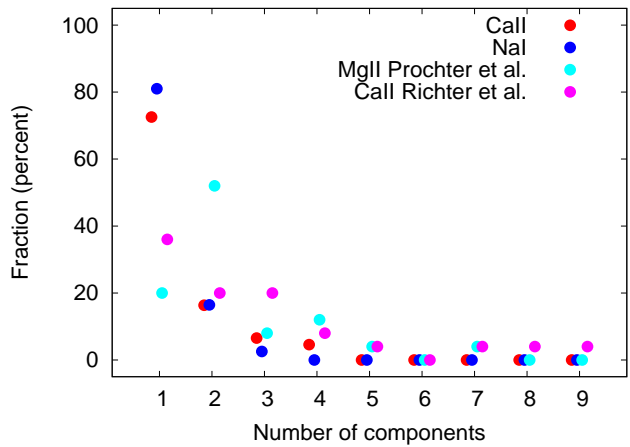}
\caption{Number of \ion{Ca}{ii} and \ion{Na}{i} velocity components per absorption system. The total number of sight lines that show \ion{Ca}{ii} (\ion{Na}{i}) absorption lines is 126 (75), but as some sight lines contain more than one independent system, we normalise with the number of independent systems (153 and 79, respectively). Furthermore, we show the results obtained by \citet{prochteretal06} and \citet{richteretal10} for comparison. Their total number of sight lines is 25 and 23, respectively.}
\label{fig_number_comp} 
\end{figure}

A fraction of the halo \ion{Ca}{ii} and \ion{Na}{i} absorbers are composed of several velocity components. To compare our sample with other studies we show in Fig.\ref{fig_number_comp} the number of velocity components per absorber (normalised to the total number of absorbers in our sample).  Most of our systems show a single absorption component, about 20\% are found in double-component systems, while systems with more than two components are rare in our sample. This is different from the samples studied by \citet{prochteretal06} and \citet{richteretal10}, which both exhibit higher component multiplicity. This is not surprising, however, as these surveys also sample gas discs which have, {\it together} with the surrounding halo gas, a more complex velocity structure compared to halo absorbers alone.

The median group width (of systems with $\geq2$ lines) is $14~\mathrm{km\,s}^{-1}$ for \ion{Ca}{ii} and $15~\mathrm{km\,s}^{-1}$ for \ion{Na}{i}. The small separation in velocity indicates that the multiple (mostly two) components are usually part of a gas structure being physically connected.

\subsection{Equivalent widths}

\begin{figure}
\centering
\includegraphics[width=0.48\textwidth,clip=]{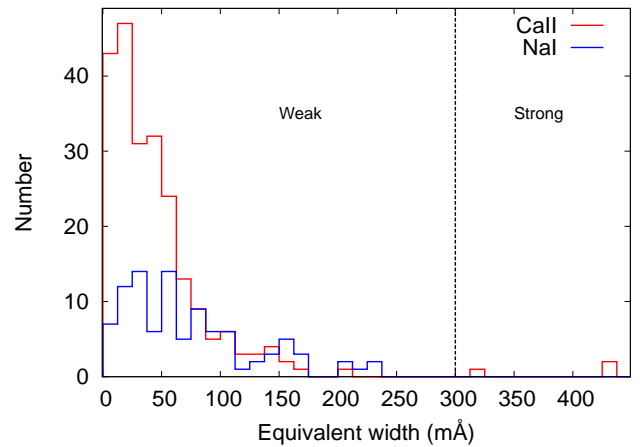}
\caption{Number of \ion{Ca}{ii} and \ion{Na}{i} absorption systems versus their equivalent width. The dashed line represents the equivalent width of $W_{\lambda} = 300$\,km\,s$^{-1}$, which separates weak from strong absorption systems according to \ion{Mg}{ii} systems.}
\label{fig_eqwidth}
\end{figure}

The distribution of the \ion{Ca}{ii} and \ion{Na}{i} equivalent widths $W_{\lambda}$ for the stronger transitions $\lambda3934$ and $\lambda5892$ are shown in Fig.\,\ref{fig_eqwidth}. In our sample the majority of \ion{Ca}{ii} equivalent widths is below $W_{\lambda}^\ion{Ca}{ii} \lesssim 150$\,m\AA{} and for \ion{Na}{i} we find $W_{\lambda}^\ion{Na}{i} \lesssim 250$\,m\AA{}. If we use $W_{\lambda}=300$\,km\,s$^{-1}$ as a separation between weak and strong absorption systems \citep[e.g.,][]{richteretal10}, we find that our sample contains almost exclusively weak systems (note that most of these systems would not be observable with low spectral resolution). Furthermore, the equivalent widths are smaller than in the extra galactic sample of \citet{richteretal10}. Again, this is probably due to the fact that we do not trace discs in our absorption line sample.

\subsection{Area filling factors}
To estimate the area filling factor of the \ion{Ca}{ii}/\ion{Na}{i}-absorbing gas in the halo we compute the fraction of sight lines that show \ion{Ca}{ii}/\ion{Na}{i} absorption with respect to the total number of observed sight lines. This gives a total \ion{Ca}{ii} (\ion{Na}{i}) filling factor of about $40\%$ ($25\%$). For high-velocity absorbers, the total \ion{Ca}{ii} (\ion{Na}{i}) filling factors are about $20\%$ ($2\%$), while for intermediate velocities we find  $30\%$ ($20\%$).

\begin{figure}
\centering
\includegraphics[width=0.48\textwidth,clip=,bb=56 56 363 296]{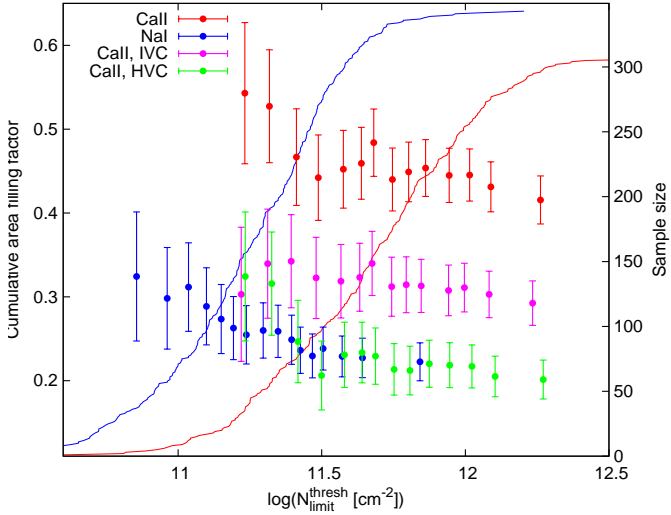}
\caption{Cumulative area filling factors, computed by applying column density thresholds to our sample, i.e., using only those spectra whose (theoretical) column density detection limits are below the threshold, $N_\mathrm{limit}\leq N_\mathrm{limit}^\mathrm{thresh}$. The solid curves show the size of the sub-samples with increasing threshold. }
\label{fig_cumulative_fillingfactors}
\end{figure}

However, as we have seen earlier, the nominal detection limits vary substantially for the different sight lines. To account for that effect we show in Fig.\,\ref{fig_cumulative_fillingfactors} cumulative area filling factors. These were calculated by applying $N_\mathrm{limit}$ thresholds (using all spectra for which $N_\mathrm{limit}\leq N_\mathrm{limit}^\mathrm{thresh}$) to the sample, i.e., the left-most values result from the smallest sub-sample but having the lowest noise in the spectra. The sub-sample grows in size with increasing threshold levels but the mean spectral quality decreases. The right-most values reflect the area filling factor of the full sample.

Not surprisingly, the area filling factor is notably higher for lower $N_\mathrm{limit}^\mathrm{thresh}$. If one only takes into account the highest-quality spectra, which allow for the detection of absorbers with $\log (N_\ion{Ca}{ii}/\mathrm{cm}^{-2})\geq 11.4$ ($\log (N_\ion{Na}{i}/\mathrm{cm}^{-2})\geq 10.9$), the area filling factor is as high as 50\% (35\%). Such an effect was also seen by \citet{wakker91} who applied different brightness temperature thresholds to \ion{H}{i} emission data obtained from an HVC survey. \citet{wakker04} reported on typical \ion{H}{i} area filling factors for HVCs, of order 15\% for $N_\ion{H}{i}>2\times10^{18}\,\mathrm{cm}^{-2}$, and increasing to 30\% for $N_\ion{H}{i}>7\times10^{17}\,\mathrm{cm}^{-2}$.

We have obtained a sufficient number of \ion{Ca}{ii} detections to also compute the cumulative area filling factors of HVCs and IVCs separately (see Fig.\,\ref{fig_cumulative_fillingfactors} green and purple points). 
There might be an increase of the HVC filling factor below $\log (N_\mathrm{limit}^\mathrm{thresh}/\mathrm{cm}^{-2})\lesssim11.4$ but more data would be needed to obtain a statistically significant detection.

\subsection{Possible association with known IVC or HVC complexes}\label{HVC_IVC_complexes}

A total of 102 \ion{Ca}{ii} (45$\%$) and 38 \ion{Na}{i} (40$\%$) components in our sample is possibly associated with known IVC or HVC complexes; see Table\,\ref{qsotable}. Due to the unknown distances of the absorption systems (and many complexes) the only indicator for an association can be the position and velocity of each absorber. We also analysed how many groups of lines are associated. The latter number yields a more reasonable estimate of the correlation with complexes, as a single complex could produce several absorption lines. Also, a simple counting of sight lines containing an IVC/HVC association would completely neglect the possibility of having several groups in a single sight line, some of them eventually associated with a complex, while others are not. In total, we have 138 IVC and 66 HVC groups in our sample, with 83 and 38 groups being associated with known complexes, respectively. This in turn means that there is a substantial fraction of low-column density absorption systems that possibly do not belong to a known HVC/IVC complex.

\begin{figure}
\centering
\includegraphics[width=0.48\textwidth,clip=,bb=60 93 395 291]{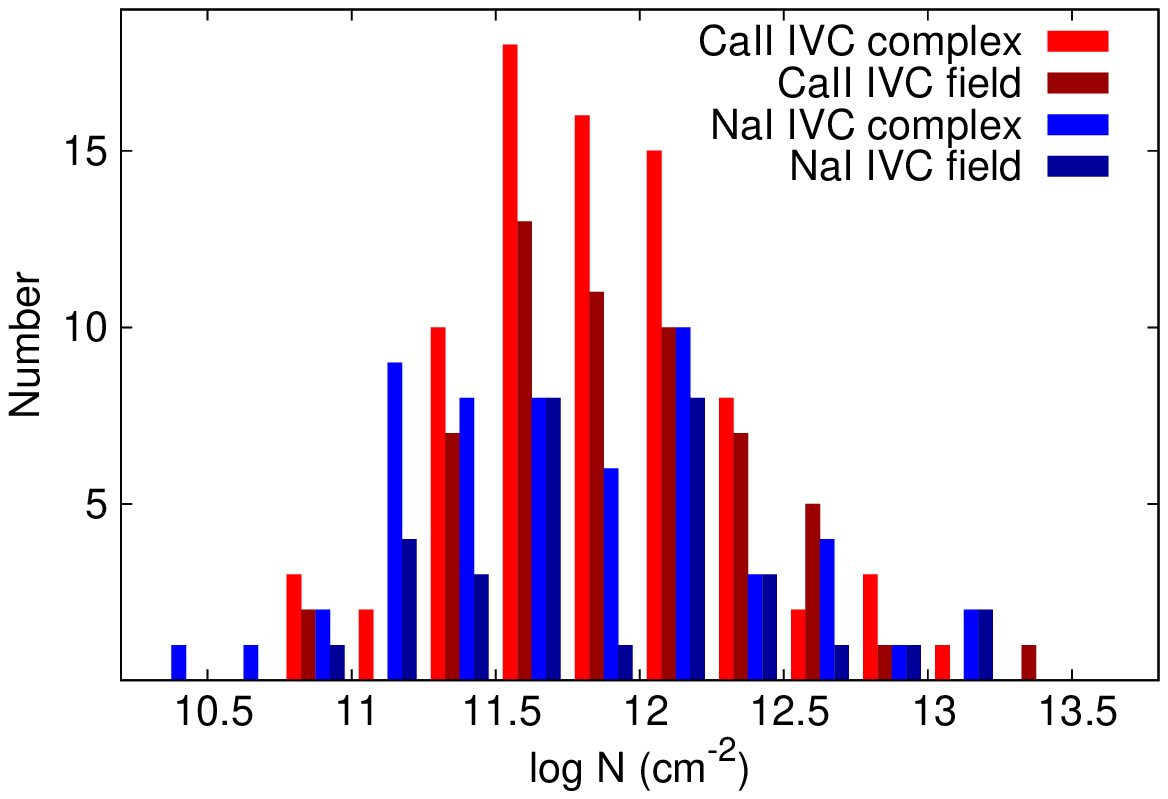}\\[0ex]
\includegraphics[width=0.48\textwidth,clip=,bb=60 60 395 291]{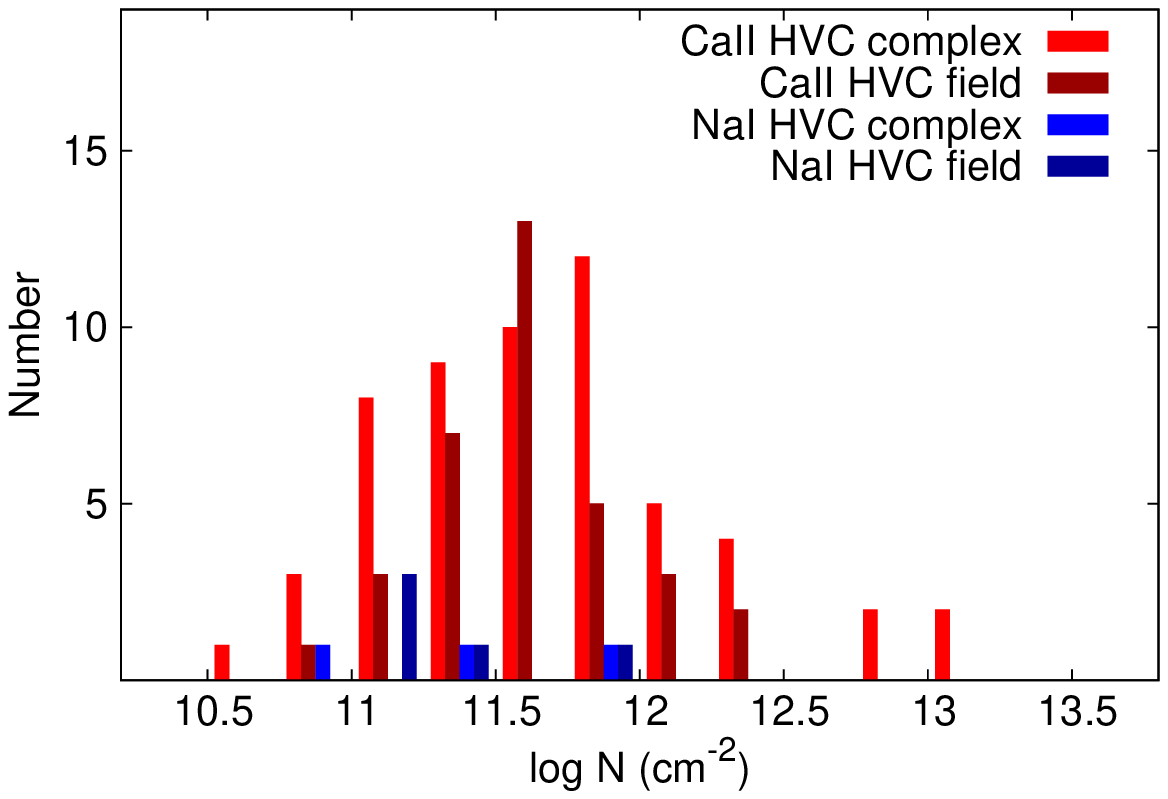}
\caption{Column density histograms for components (not) associated to an IVC/HVC complex.}
\label{fig_num_vs_association}
\end{figure}

To study whether there is a fundamental difference in the properties of the components that belong to a complex and those that do not (i.e., `field' absorbers), we show in Fig.\,\ref{fig_num_vs_association} column density histograms of these two populations. We find no significant difference with respect to the shape of the distributions, except that for HVC \ion{Ca}{ii} components several of the associated detections have column densities below $\log(N_\ion{Ca}{ii}/\mathrm{cm}^{-2})\approx11$. This is not be observed for `field' components. However, number counts are rather low in this column density regime. When looking at the \ion{H}{i} emission associated with the absorption, we do not detect any of the ``field'' HVC components. For the components that are associated with an HVC complex about 25\% show \ion{H}{i}. For intermediate-velocity features we see no significant deviation from the overall fraction $p(\ion{H}{i}\,\vert\,\ion{Ca}{ii})$ and $p(\ion{H}{i}\,\vert\,\ion{Na}{i})$, respectively.

The majority of the absorbers (30 IVCs and 35 HVCs) is associated with the Magellanic System (MS). In a few cases it is not clear whether they should be attributed to the MS or the anti-center shell (AC shell), which have intermediate velocities. For 11 of the HVC and 17 of the IVC groups we have corresponding \ion{H}{i} detections. Two sight lines (HE\,0251$-$5550 and PKS\,0506$-$61) are possibly associated with the Large Magellanic Cloud (LMC). From the ten absorption components with the highest deviation velocities of $\vert v_\mathrm{dev}\vert > 250$\,km$^{-1}$ in our sample, seven are likely associated with the Magellanic System (the corresponding \ion{H}{i} data show no emission lines). The remaining three components cannot be associated with a complex.

The directions towards QSO\,B1448$-$232 and PKS\,1508$-$05 indicate a possible association with the HVC complex L. While for QSO\,B1448$-$232 the LSR velocities are in agreement with the velocity range observed for complex L \citep[$v_\mathrm{lsr}=-188 \ldots -91$\,km\,s$^{-1}$][]{wakker04}, the absorption system in the direction of PKS\,1508$-$05 shows low negative velocities $v_\mathrm{lsr} \approx -17 \ldots -43$\,km\,s$^{-1}$. Such an intermediate-velocity part has not been observed previously in complex L. Eight intermediate- and two high-velocity groups are likely associated with the anti-center shell (AC-shell), nine with complex Galactic Center negative (GCN), and two with Galactic Center positive (GCP). 

17 IVC groups are associated with the intermediate-velocity Spur (IV-Spur). For 11 of these groups we have \ion{H}{i} data showing corresponding emission in 10 cases. The sight-line without \ion{H}{i} emission (3C\,281) passes the outer (clumpy) region of the IV-Spur. 

The intermediate-velocity complex PP-Arch hosts six of our absorption structures. Seven  are related to the intermediate-velocity Arch (IV Arch), though in the case of TON\,1480 the absorption component has a rather high deviation velocity of $v_\mathrm{dev}=-75$ km\,s$^{-1}$. Therefore, it is not entirely clear whether the latter is really part of the IV Arch. Two components in the direction of PKS\,1629$+$120 are possibly related to the IVC complex K and one absorption component in the direction of SDSS\,J014631.99$+$133506.3 is associated with the IVC complex ACHV. The sight lines towards seven QSOs pass the HVC complex WA and three pass complex WD. However, the absorption components are located at intermediate velocities.

\begin{figure*}[!t]
\centering
\includegraphics[width=0.8\textwidth,clip=,bb=103 84 550 375]{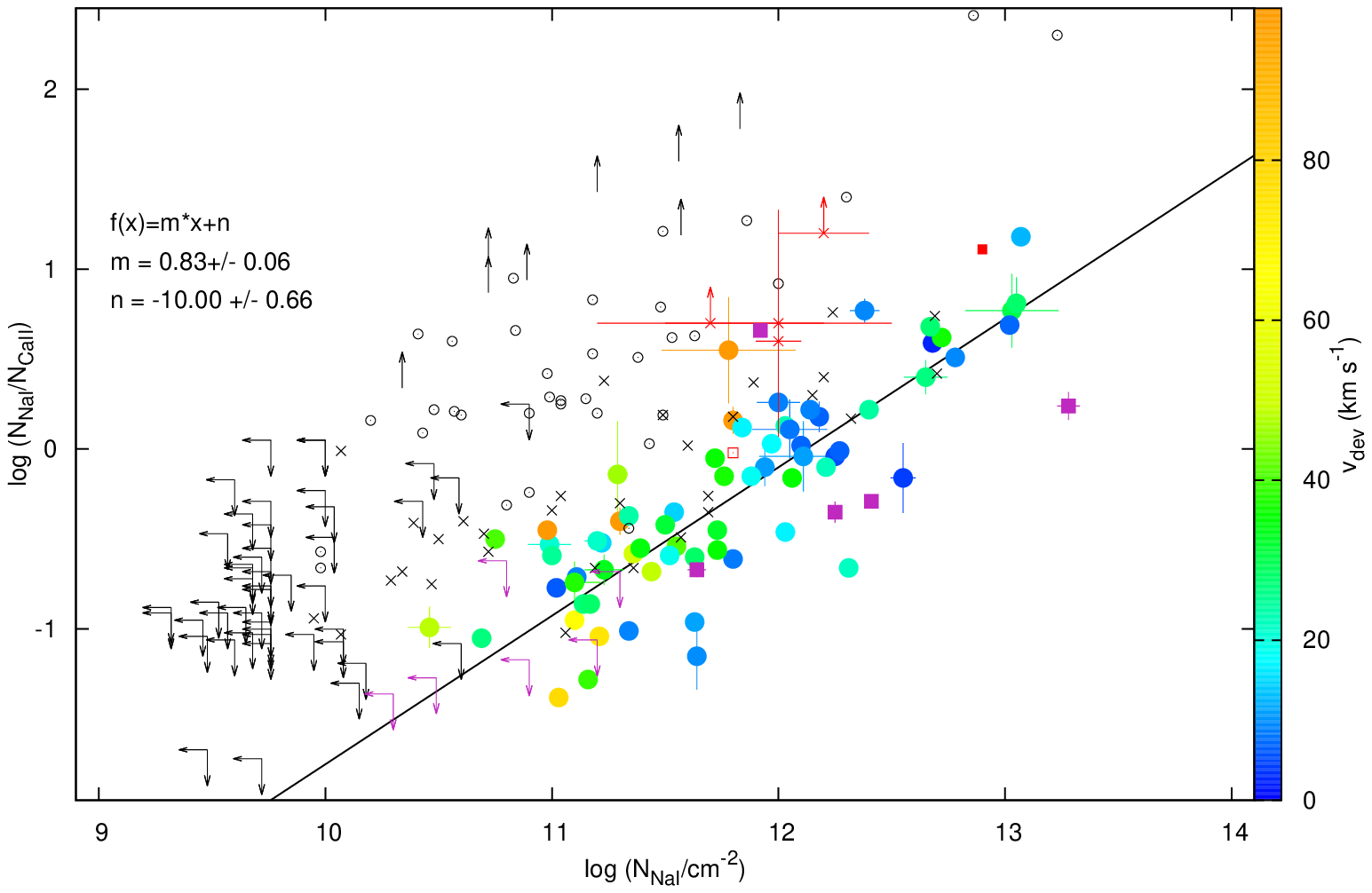}
\caption{\ion{Na}{i}/\ion{Ca}{ii} ratios as a function of $\log N_\ion{Na}{i}$. The red symbols show the DLA data ($1.06 \leq z \leq 1.18$) of \citet[][see their figure 5 for a detailed explanation of the different symbols]{kondoetal06}. Black circles and crosses show the data for the \ion{Ca}{ii} and \ion{Na}{i} absorbers in Milky Way \citep{vallerga93}  and in the LMC \citep{vladilo93}, respectively. 
Purple symbols show higher-redshift absorption lines extracted from our sample \citep{richteretal10}. Filled circles show the ratios found in our MW halo sample, with deviation velocities colour-coded. The black solid line represents a power-law fit with slope $\beta=0.8 \pm 0.1$.}
\label{figlogNNaI_logNCaII}
\end{figure*}

\subsection{The \ion{Na}{i}/\ion{Ca}{ii} ratio and dust depletion}\label{subsecdustdepletion}

As discussed earlier, the determination of absolute calcium and sodium abundances in the absorbers is afflicted with large systematic uncertainties related to the unknown amount of calcium/sodium being depleted onto dust grains, to beam smearing and to ionisation effects. 

To study the degree of dust depletion in the absorbers, the $N_\ion{Na}{i}$/$N_\ion{Ca}{ii}$ ratio has often been used as a probe \citep{welshetal90, bertinetal93, crawfordetal02, kondoetal06}, because sodium is hardly depleted onto grains while calcium can be strongly depleted \citep{savagesembach96}. 

Values of $N_\ion{Na}{i}/N_\ion{Ca}{ii} \leq 20$ are expected in diffuse, warm ($T \approx 10^3$\,K), low-density ($n_\mathrm{H} \approx 10$\,cm$^{-3}$) neutral and partly ionised gas, where much of the calcium remains in the gas phase \citep{crawfordetal02}. In contrast, values of  $N_\ion{Na}{i}/N_\ion{Ca}{ii} \geq 100$ are expected in cold ($T \approx 30$\,K), dense ($n_\mathrm{H} \geq 10^3$\,cm$^{-3}$) interstellar regions.

To compare the derived  $N_\ion{Na}{i}$/$N_\ion{Ca}{ii}$ ratios for the halo absorbers with absorbers located in different environments we display in Fig.\,\ref{figlogNNaI_logNCaII} column density ratios as a function of $\log N_\ion{Na}{i}$. Three other data sets were kindly provided by Sohei Kondo \citep[see also][]{kondoetal06}. The red symbols show damped Lyman $\alpha$ (DLA) data ($1.06 \leq z \leq 1.18$) obtained by \citet{kondoetal06} with the near-infrared instrument IRCS (Infrared Camera and Spectrograph) installed at the Subaru Telescope.  Black circles show the data for the Milky Way by \citet{vallerga93} who observed \ion{Ca}{ii} and \ion{Na}{i} absorbers towards 46 early type stars in the local ISM with distances between $d=20 \ldots 500$\,pc with the Coud\'{e} Echelle Spectrograph (CES) installed at the 3-m Shane telescope (Lick Observatory). Black crosses are values obtained for the LMC \citep{vladilo93}. For the latter the CES instrument was used to observe \ion{Ca}{ii} and \ion{Na}{i} absorbers towards 16 stars in a $30\arcmin \times 30\arcmin$ region of the LMC centred on SN\,1987A. Purple symbols show higher-redshift absorption lines extracted from our sample \citep[see][]{richteretal10}. The filled circles represent our $N_\ion{Na}{i}/N_\ion{Ca}{ii}$ ratios, with colour-coded deviation velocities. The black solid line represents a power-law fit of the form
\begin{equation}
\log(N_\ion{Na}{i}/N_\ion{Ca}{ii})=(0.8 \pm 0.1) \cdot \log (N_\ion{Na}{i}/\mathrm{cm}^{-2}) + (-10 \pm 1)
\end{equation}
through our data.

We find that the $N_\ion{Na}{i}$/$N_\ion{Ca}{ii}$ ratios in our sample are systematically smaller than those in the DLAs, in the LMC and in the Milky Way. This is not surprising since the halo absorbers are expected to contain less dust than gas clouds located in the discs of galaxies. Furthermore, apart from the dependency of the ratios on $\log N_\ion{Na}{i}$, there is a trend with deviation velocities. While low-velocity clouds reside mainly in the higher-column density regime, clouds with higher velocities are located in the lower left of the figure. This indicates that high-velocity clouds in our sample have less depletion of \ion{Ca}{ii} onto dust simply because the HVCs contain much less dust. The dust content gets larger with higher column densities such that the \ion{Na}{i}/\ion{Ca}{ii} ratio increases again. This trend is known as the ``Routly-Spitzer'' effect \citep{routly52}.

\subsection{\ion{Ca}{ii} and \ion{Na}{i} abundances}
\begin{figure*}
\centering
\includegraphics[width=0.8\textwidth,clip=,bb=97 109 556 375]{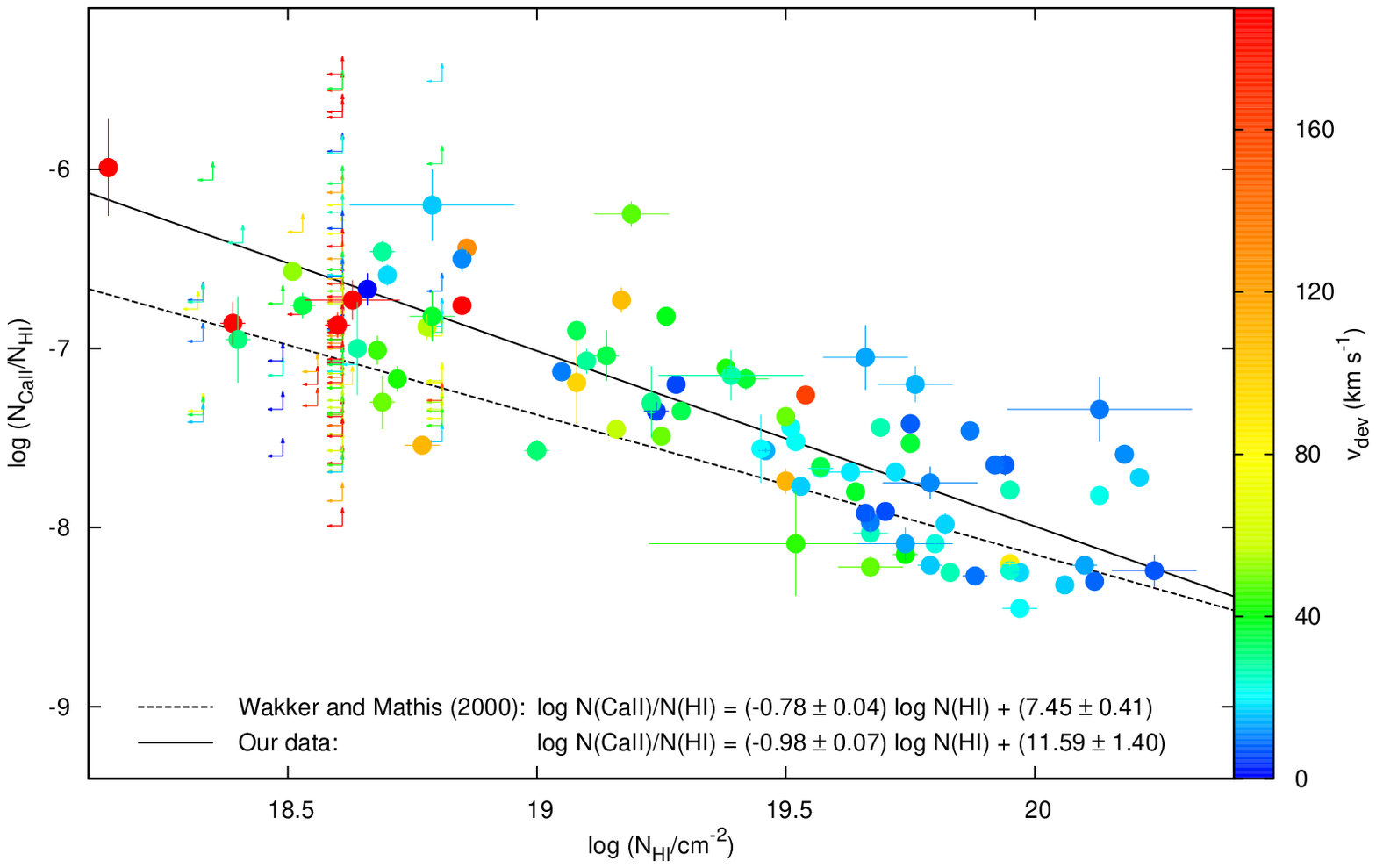}
\includegraphics[width=0.8\textwidth,clip=,bb=97 84 556 375]{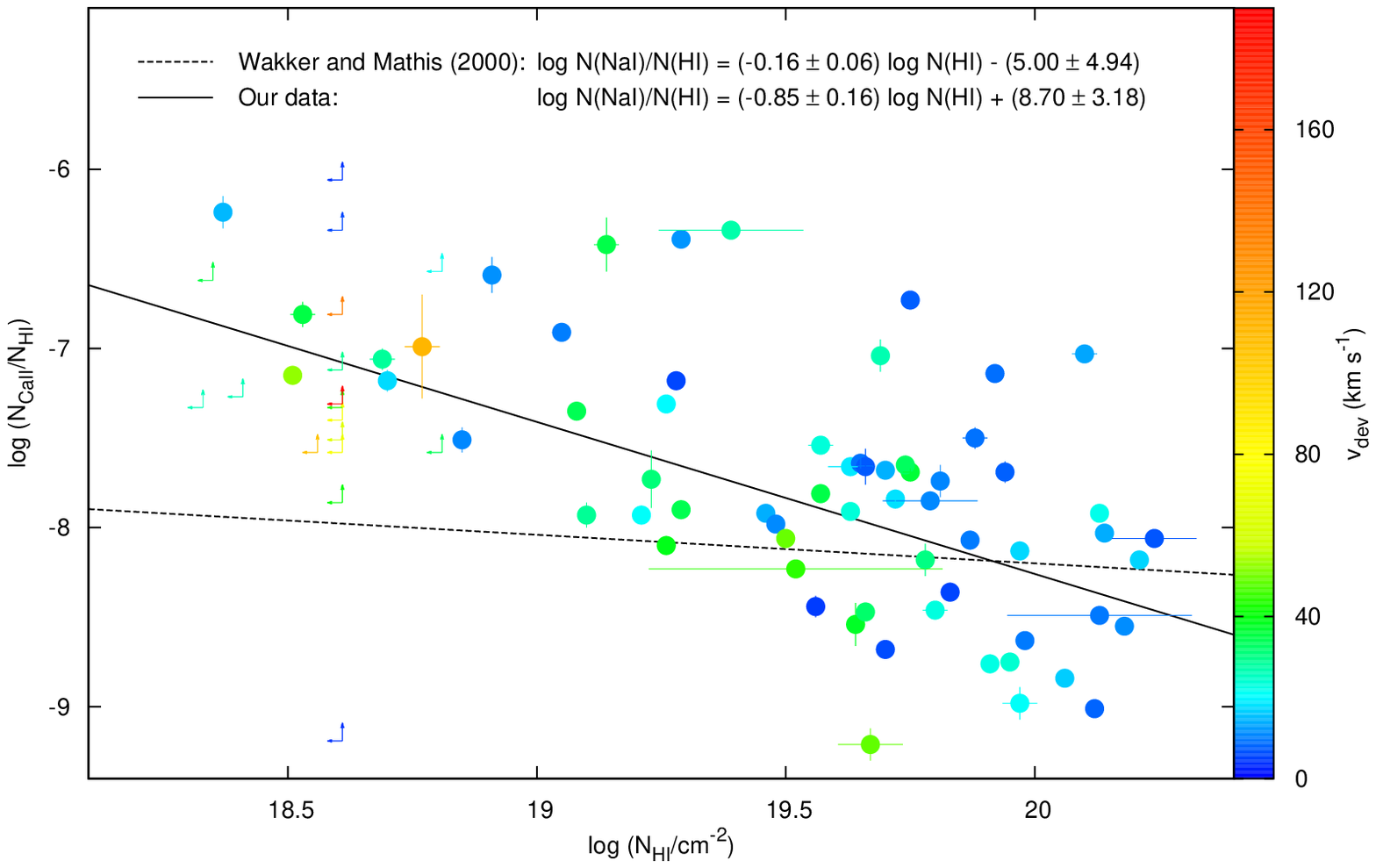}
\caption{\ion{Ca}{ii}/\ion{H}{i} and \ion{Na}{i}/\ion{H}{i} ratios calculated for our sample. Arrows show lower limits (associated with \ion{H}{i} non-detections). The colour of the data points marks their deviation velocity. The solid black line shows the best linear fit to the data, for comparison we have also plotted the relations found by \citet[][dashed line]{wakkermathis00}.}
\label{figabundances}
\end{figure*}

We detect \ion{H}{i} emission for many absorption components so that we can statistically investigate the \ion{Ca}{ii}/\ion{H}{i} and \ion{Na}{i}/\ion{H}{i} ratios (see Fig.\,\ref{figabundances}). For completeness we show the relation found by \citet[][dashed line]{wakkermathis00} along a best linear fit of our data (solid line), but note that the numbers are affected by systematic effects like beam-smearing or beam-filling and should be treated with caution. Beam-smearing is due to the finite angular resolution of the single-dish observations, which can lead to either over- or underestimation of the true \ion{H}{i} content along the QSO sight lines\footnote{Overestimation (beam-smearing problem) is due to a strong source of emission right next, i.e. within a beam size, to the QSO line of sight (but not much \ion{H}{i} being exactly on the sight line), while underestimation (beam-filling problem) is caused if an \ion{H}{i} feature is aligned with the QSO, but is much smaller than the single-dish beam.}. It is difficult to estimate the exact strength of both effects (which work against each other), even statistically, since the true angular and size distribution at small scales is not well-known. Previous studies with radio-synthesis telescopes \citep{richterwestmeierbruens05,benbekhtietal_09} have shown that the \ion{H}{i} features associated with the halo absorption systems are very clumpy on scales below 1\arcmin to 3\arcmin. Unfortunately, it turned out to be difficult to find a system where an \ion{H}{i} clump is exactly aligned with a QSO sight line, which would be required to have a sufficiently solid estimate of the local \ion{Ca}{ii}/\ion{H}{i} and \ion{Na}{i}/\ion{H}{i} ratios.

Nevertheless, Fig.\,\ref{figabundances} allows us at least to study some statistical trends. Arrows indicate lower limits of the \ion{Ca}{ii}/\ion{H}{i} and \ion{Na}{i}/\ion{H}{i} ratios (from the non-detection of hydrogen). The colour of the data points indicates their deviation velocity. Most of the higher-velocity absorbers have not been detected in \ion{H}{i}. The colour coding shows that the non-detections are mostly high-velocity absorbers; many of these obviously have low \ion{H}{i} column densities and/or small angular sizes so that they are not detected in \ion{H}{i} 21-cm emission.

Fig.\,\ref{figabundances} reveals similar trends for \ion{Ca}{ii} and \ion{Na}{i} (the ion ratios decrease with increasing \ion{H}{i} column densities). One important difference is that \ion{Ca}{ii} absorbers with higher radial velocities have generally higher \ion{Ca}{ii}/\ion{H}{i} ratios, while for \ion{Na}{i} such a trend is not as clearly visible. This indicates that the high-velocity absorbers have a lower dust content (i.e., less Calcium depletion and higher \ion{Ca}{ii}/\ion{H}{i} ratio) than low-velocity systems. Note that the solar abundance of \ion{Ca}{ii}/\ion{H}{i} is about $-5.7$ \citep{anders89}.

\subsection{Comparison with \ion{Ca}{ii} and \ion{Mg}{ii} absorption systems at redshifts $z>0$}
Our data shows that the Milky Way halo hosts a large number of low-column density \ion{Ca}{ii} and \ion{Na}{i} absorbers at intermediate- and high-velocities. If the Milky Way is a typical  spiral galaxy in the local Universe one would expect to find similar absorption systems around other low-redshift galaxies. Therefore, \citet{richteretal10} used 304 QSO spectra from the same UVES data set to search for \ion{Ca}{ii} absorbers at redshifts of $z\lesssim0.5$. They found 23 intervening \ion{Ca}{ii} absorbers, some of which have physical properties similar to those of the Milky Way halo structures. The detected column densities are in the range of $\log(N_\ion{Ca}{ii}/\mathrm{cm}^{-2})=11.25 \ldots 13.04$,  where the absorbers with $\log(N_\ion{Ca}{ii}/\mathrm{cm}^{-2}) > 11.5$ trace neutral and mildly ionised gas with $\log(N_\ion{H}{i}/\mathrm{cm}^{-2}) > 17.5$. One important result is that intervening \ion{Ca}{ii} absorbers outnumber damped Ly-$\alpha$ systems at low redshift by a factor of three, indicating that most of these systems trace low-column density gas in the outskirts of galaxies (i.e. they represent IVC/HVC analogues).

A good tracer for extra-planar gaseous structures around galaxies at higher redshifts is the \ion{Mg}{ii} doublet. There are two different populations of \ion{Mg}{ii} absorbers, the weak and the strong \ion{Mg}{ii} ones. While weak \ion{Mg}{ii} absorbers are thought to be located in the distant outskirts of galaxies, the strong \ion{Mg}{ii} systems are most likely located in the inner halo or even in the discs of galaxies \citep[e.g.,][]{petitjeanbergeron90, charltonandchurchill98,churchill99, dingcharltonchurchillpalma03}. A certain fraction of strong \ion{Mg}{ii} absorbers therefore is expected to represent the analogues of IVCs/HVCs. In the MW halo several \ion{Mg}{ii} absorbers could be associated with known IVC/HVC complexes \citep[e.g.,][]{savageetal00}.

We can now compare our CDD functions with the results of \citet{churchillvogtcharlton03} for \ion{Mg}{ii} and with those of \citet{richteretal10} for \ion{Ca}{ii}. In case of \ion{Ca}{ii} the power-law slope, $\beta_\ion{Ca}{ii}=-2.2$, is steeper than the value derived by \citet[][$\beta_\ion{Ca}{ii}=-1.7$]{richteretal10} and that obtained for \ion{Mg}{ii}, $\beta_\ion{Mg}{ii}=-1.59 \pm 0.05$ \citep{churchillvogtcharlton03}, inferred from strong \ion{Mg}{ii} absorbers at intermediate redshifts ($z=0.4-1.2$). As discussed in Section\,\ref{subseccdd}, this can be explained with the different absorber environments (disc+halo vs. halo) and the different dust-depletion levels.

\subsection{Small-scale structure}\label{subsec:substructure}
Since our QSO sample has a random sky distribution it is not very well-suited to study small-scale structure in Milky Way halo gas. Nevertheless, a few sight lines intersect the halo with relative angular separations of less than $\sim1\degr$. Cross-correlating all sight lines showing halo \ion{Ca}{ii} absorption with every other sight line in our sample having at least the same S/N level (to avoid selection effects), we identify 11 of such cases. Out of these, three sightlines show absorption components matching in radial velocity. For the pair QSO\,J1039$-$2719/QSO\,J1040$-$2727 (angular separation: $17\arcmin$) at a radial velocity of $v_\mathrm{lsr}=-17\,\mathrm{km\,s}^{-1}$ and $v_\mathrm{lsr}=-11\,\mathrm{km\,s}^{-1}$, respectively, we find consistent column densities of $\log (N_\ion{Ca}{ii}/\mathrm{cm}^{-2}) =12.5$ and 12.6. However, only the latter sight line shows also a high-velocity absorber at $v_\mathrm{lsr}=180\,\mathrm{km\,s}^{-1}$. The second (very close) pair 2QZ\,J225153.1$-$314620/2QZ\,J225154.8$-$314521 (angular distance: $1\arcmin$) has a match at $v_\mathrm{lsr}=-32\,\mathrm{km\,s}^{-1}$, but again (in the latter sight line) two components remain undetected. Finally, in the third case, even four sight lines are close to each other (2QZ\,J232046.7$-$294406, 2QZ\,J232114.2$-$294725, 2QZ\,J232059.3$-$295521, and 2QZ\,J232121.2$-$294351; angular distance: $\lesssim10\arcmin$). However, only the first two QSOs reveal one matching absorption component at $v_\mathrm{lsr}=-49\,\mathrm{km\,s}^{-1}$, which varies by 0.5\,dex in column density, i.e. $\log (N_\ion{Ca}{ii}/\mathrm{cm}^{-2}) =12.9$ vs. 12.4. Two IVC features in 2QZ\,J232046.7$-$294406 and an HVC in 2QZ\,J232114.2$-$294725 remain undetected in the other sight lines as well.

For all of the eight other pairs, the halo absorption components detected in one spectrum have no counterparts in the other (within a radial velocity of $\lesssim50\,\mathrm{km\,s}^{-1}$).

The situation is similar for \ion{Na}{i}. Seven pairs were found, four of them show one component (each) in one spectrum with a matching absorption line in the other spectrum, but only two of these four also have similar column densities. For all remaining components/pairs, column density variations of at least $\sim0.5\,$dex (or non-detection in one of the sight lines) occurs.

Although the total number of pairs is small, the few examples listed above suggest that the halo structures traced by \ion{Ca}{ii} and \ion{Na}{i} must be variable on angular scales of less than a fraction of a degree. This result is consistent with previous high-resolution radio observations \citep{richterwestmeierbruens05,benbekhtietal_09}.

\section{Summary}\label{secsummary}

With this study we have continued our project to systematically analyse the absorption properties of neutral gas in the Milky Way halo. We present results for a significantly enlarged data sample of 408 extragalactic sight lines through the Milky Way halo. Optical QSO spectra observed with VLT/UVES where combined with data from the new EBHIS and GASS surveys to search for \ion{Ca}{ii} and \ion{Na}{i} absorption and corresponding \ion{H}{i} emission features of neutral gas structures in the Milky Way halo.

The main results can be summarised as follows:

\begin{enumerate}
\item The filling factor (or covering fraction) of \ion{Ca}{ii}/\ion{Na}{i} absorbing gas in the halo is substantial. We detect \ion{Ca}{ii}/\ion{Na}{i} halo absorption in about 40\%/25\% of all sight lines. If one corrects for selection effects by applying (column-density) thresholds, the filling factors further increase by 10\%. Our large sample size allows us to separately study intermediate- and high-velocity absorbers (IVCs and HVCs). 

\item Associated \ion{H}{i} 21-cm emission is found for 50\% of all \ion{Ca}{ii} halo absorption components and for nearly all \ion{Na}{i} lines. These numbers reflect our \ion{H}{i} 21-cm detection limits for the absorbing halo structures. While the \ion{Na}{i} and \ion{H}{i} data probe a similar \ion{H}{i} column density regime down to few $10^{18}\,\mathrm{cm}^{-2}$, \ion{Ca}{ii} absorption arises also in gas at lower \ion{H}{i} column densities.

\item About half of all halo absorption components appear to be associated with known IVC and HVC complexes. Interestingly, for none of the high-velocity ``field'' components (i.e., halo absorbers that are not connected to a known HVC complex) \ion{H}{i} emission was detected, while for the halo absorbers related to known HVCs about 25\% have \ion{H}{i} counterparts. For IVCs such a phenomenon is not observed.

\item The observed \ion{Ca}{ii} and \ion{Na}{i} column density distribution functions follow a power-law $f(N)=CN^{\beta}$, with slopes $\beta_\ion{Ca}{ii}=-2.2\pm0.2$ and $\beta_\ion{Na}{i} =-1.4\pm0.1$. The former is steeper than what is found in extra-galactic absorber samples, which may be explained by the different absorber environments (disc vs. halo) and dust-depletion effects. Using the correlation between \ion{Ca}{ii}, \ion{Na}{i} and \ion{H}{i} we can convert the column densities to \ion{H}{i} column densities. It turns out that the (converted) distributions ($\beta_{\ion{H}{i}[\ion{Ca}{ii}]}=-1.26 \pm 0.02$ and $\beta_{\ion{H}{i}[\ion{Na}{i}]}=-1.44 \pm 0.06$) are roughly consistent with slopes derived for Galactic IVCs/HVCs based on \ion{H}{i} 21-cm surveys.

\item The $N_\ion{Na}{i}$/$N_\ion{Ca}{ii}$ ratios in the halo absorbers, that can be used as a probe for the dust content in the gas are systematically smaller than those in DLAs, in the LMC,  and in the Milky Way disc. As expected, depletion of \ion{Ca}{ii} onto dust grains is less important in the high-velocity absorbers, where metallicities and densities are expected to be lower than in gas close to the disc. 

\item Most of the \ion{Ca}{ii} and \ion{Na}{i} absorbers show single- or double-component absorption, with $b$-values of $<7\,\mathrm{km\,s}^{-1}$. Nevertheless, $b$-values are generally too high to be purely thermal, suggesting that other broadening mechanisms (unresolved subcomponent structure, turbulence, macroscopic gas motions) contribute to observed line widths.

\end{enumerate}

\begin{acknowledgements} 
The authors thank the Deutsche Forschungsgemeinschaft (DFG) for financial support under the research grant KE757/9-1 and KE757/7-1.
MTM thanks the Australian Research Council for a QEII Research Fellowship (DP0877998).
The work is based on observations with the 100-m telescope of the MPIfR (Max-Planck-Institut f\"{u}r Radioastronomie) at Effelsberg. We thank Sohei Kondo for providing the data sets used in Fig.\,\ref{figlogNNaI_logNCaII}. We are also grateful to the referee, Bart Wakker, for many useful comments and his suggestions to improve our manuscript.
\end{acknowledgements} 
\bibliographystyle{aa} 
\bibliography{references} 

\begin{thebibliography}{83}
\expandafter\ifx\csname natexlab\endcsname\relax\def\natexlab#1{#1}\fi

\bibitem[{{Anders} \& {Grevesse}(1989)}]{anders89}
{Anders}, E. \& {Grevesse}, N. 1989, \gca, 53, 197

\bibitem[{{Ben Bekhti} {et~al.}(2008){Ben Bekhti}, {Richter}, {Westmeier}, \&
  {Murphy}}]{benbekhtietal08}
{Ben Bekhti}, N., {Richter}, P., {Westmeier}, T., \& {Murphy}, M.~T. 2008,
  \aap, 487, 583

\bibitem[{{Ben Bekhti} {et~al.}(2009){Ben Bekhti}, {Richter}, {Winkel}, {Kenn},
  \& {Westmeier}}]{benbekhtietal_09}
{Ben Bekhti}, N., {Richter}, P., {Winkel}, B., {Kenn}, F., \& {Westmeier}, T.
  2009, \aap, 503, 483

\bibitem[{{Bertin} {et~al.}(1993){Bertin}, {Lallement}, {Ferlet}, \&
  {Vidal-Madjar}}]{bertinetal93}
{Bertin}, P., {Lallement}, R., {Ferlet}, R., \& {Vidal-Madjar}, A. 1993, \aap,
  278, 549

\bibitem[{{Blitz} {et~al.}(1999){Blitz}, {Spergel}, {Teuben}, {Hartmann}, \&
  {Burton}}]{blitz99}
{Blitz}, L., {Spergel}, D.~N., {Teuben}, P.~J., {Hartmann}, D., \& {Burton},
  W.~B. 1999, \apj, 514, 818

\bibitem[{{Boomsma} {et~al.}(2008){Boomsma}, {Oosterloo}, {Fraternali}, {van
  der Hulst}, \& {Sancisi}}]{boomsmaetal08}
{Boomsma}, R., {Oosterloo}, T.~A., {Fraternali}, F., {van der Hulst}, J.~M., \&
  {Sancisi}, R. 2008, \aap, 490, 555

\bibitem[{{Bouch{\'e}} {et~al.}(2012){Bouch{\'e}}, {Murphy}, {P{\'e}roux},
  {Contini}, {Martin}, {Forster Schreiber}, {Genzel}, {Lutz}, {Gillessen},
  {Tacconi}, {Davies}, \& {Eisenhauer}}]{bouche2012}
{Bouch{\'e}}, N., {Murphy}, M.~T., {P{\'e}roux}, C., {et~al.} 2012, \mnras,
  419, 2

\bibitem[{{Bouch{\'e}} {et~al.}(2006){Bouch{\'e}}, {Murphy}, {P{\'e}roux},
  {Csabai}, \& {Wild}}]{boucheetal06}
{Bouch{\'e}}, N., {Murphy}, M.~T., {P{\'e}roux}, C., {Csabai}, I., \& {Wild},
  V. 2006, \mnras, 813

\bibitem[{{Br{\" u}ns} {et~al.}(2005){Br{\" u}ns}, {Kerp}, {Staveley-Smith},
  {Mebold}, {Putman}, {Haynes}, {Kalberla}, {Muller}, \&
  {Filipovic}}]{Bruenskerp05}
{Br{\" u}ns}, C., {Kerp}, J., {Staveley-Smith}, L., {et~al.} 2005, \aap, 432,
  45

\bibitem[{{Braun} \& {Burton}(1999)}]{braunburton99}
{Braun}, R. \& {Burton}, W.~B. 1999, \aap, 341, 437

\bibitem[{{Charlton} \& {Churchill}(1998)}]{charltonandchurchill98}
{Charlton}, J.~C. \& {Churchill}, C.~W. 1998, \apj, 499, 181

\bibitem[{{Charlton} {et~al.}(2000){Charlton}, {Churchill}, \&
  {Rigby}}]{charltonetal00}
{Charlton}, J.~C., {Churchill}, C.~W., \& {Rigby}, J.~R. 2000, \apj, 544, 702

\bibitem[{{Churchill} {et~al.}(1999){Churchill}, {Rigby}, {Charlton}, \&
  {Vogt}}]{churchill99}
{Churchill}, C.~W., {Rigby}, J.~R., {Charlton}, J.~C., \& {Vogt}, S.~S. 1999,
  \apjs, 120, 51

\bibitem[{{Churchill} {et~al.}(2003){Churchill}, {Vogt}, \&
  {Charlton}}]{churchillvogtcharlton03}
{Churchill}, C.~W., {Vogt}, S.~S., \& {Charlton}, J.~C. 2003, \aj, 125, 98

\bibitem[{{Crawford} {et~al.}(2002){Crawford}, {Lallement}, {Price}, {Sfeir},
  {Wakker}, \& {Welsh}}]{crawfordetal02}
{Crawford}, I.~A., {Lallement}, R., {Price}, R.~J., {et~al.} 2002, \mnras, 337,
  720

\bibitem[{{de Boer}(2004)}]{deboer04}
{de Boer}, K.~S. 2004, in Astrophysics and Space Science Library, Vol. 312,
  High Velocity Clouds, ed. {H.~van Woerden, B.~P.~Wakker, U.~J.~Schwarz, \&
  K.~S.~de Boer }, 227

\bibitem[{{de Heij} {et~al.}(2002){de Heij}, {Braun}, \&
  {Burton}}]{deheijbraunburton02}
{de Heij}, V., {Braun}, R., \& {Burton}, W.~B. 2002, \aap, 391, 159

\bibitem[{{Dekker} {et~al.}(2000){Dekker}, {D'Odorico}, {Kaufer}, {Delabre}, \&
  {Kotzlowski}}]{dekkeretal.00}
{Dekker}, H., {D'Odorico}, S., {Kaufer}, A., {Delabre}, B., \& {Kotzlowski}, H.
  2000, in Proc. SPIE Vol. 4008, p. 534-545, Optical and IR Telescope
  Instrumentation and Detectors, Masanori Iye; Alan F. Moorwood; Eds., ed.
  M.~{Iye} \& A.~F. {Moorwood}, 534--545

\bibitem[{{Ding} {et~al.}(2003{\natexlab{a}}){Ding}, {Charlton}, {Bond},
  {Zonak}, \& {Churchill}}]{dingetal03}
{Ding}, J., {Charlton}, J.~C., {Bond}, N.~A., {Zonak}, S.~G., \& {Churchill},
  C.~W. 2003{\natexlab{a}}, \apj, 587, 551

\bibitem[{{Ding} {et~al.}(2003{\natexlab{b}}){Ding}, {Charlton}, {Churchill},
  \& {Palma}}]{dingcharltonchurchillpalma03}
{Ding}, J., {Charlton}, J.~C., {Churchill}, C.~W., \& {Palma}, C.
  2003{\natexlab{b}}, \apj, 590, 746

\bibitem[{{Fontana} \& {Ballester}(1995)}]{fontanaandbalester95}
{Fontana}, A. \& {Ballester}, P. 1995, The Messenger, 80, 37

\bibitem[{{Fox} {et~al.}(2006){Fox}, {Savage}, \& {Wakker}}]{foxetal06}
{Fox}, A.~J., {Savage}, B.~D., \& {Wakker}, B.~P. 2006, \apjs, 165, 229

\bibitem[{{Fraternali} {et~al.}(2007){Fraternali}, {Binney}, {Oosterloo}, \&
  {Sancisi}}]{fraternalietal_07}
{Fraternali}, F., {Binney}, J., {Oosterloo}, T., \& {Sancisi}, R. 2007, New
  Astronomy Review, 51, 95

\bibitem[{{Fraternali} {et~al.}(2004){Fraternali}, {Oosterloo}, {Boomsma},
  {Swaters}, \& {Sancisi}}]{Fraternali04}
{Fraternali}, F., {Oosterloo}, T., {Boomsma}, R., {Swaters}, R., \& {Sancisi},
  R. 2004, in IAU Symposium, Vol. 217, Recycling Intergalactic and Interstellar
  Matter, ed. {P.-A.~Duc, J.~Braine, \& E.~Brinks}, 136

\bibitem[{{Fraternali} {et~al.}(2001){Fraternali}, {Oosterloo}, {Sancisi}, \&
  {van Moorsel}}]{Fraternalietal_01}
{Fraternali}, F., {Oosterloo}, T., {Sancisi}, R., \& {van Moorsel}, G. 2001, in
  Astronomical Society of the Pacific Conference Series, Vol. 240, Gas and
  Galaxy Evolution, ed. {J.~E.~Hibbard, M.~Rupen, \& J.~H.~van Gorkom}, 286

\bibitem[{{Hill} {et~al.}(2009){Hill}, {Haffner}, \& {Reynolds}}]{hill09}
{Hill}, A.~S., {Haffner}, L.~M., \& {Reynolds}, R.~J. 2009, \apj, 703, 1832

\bibitem[{{Kalberla}(2003)}]{kalberla03}
{Kalberla}, P.~M.~W. 2003, \apj, 588, 805

\bibitem[{{Kalberla} {et~al.}(2005){Kalberla}, {Burton}, {Hartmann}, {Arnal},
  {Bajaja}, {Morras}, \& {P{\"o}ppel}}]{kalberlaetal05}
{Kalberla}, P.~M.~W., {Burton}, W.~B., {Hartmann}, D., {et~al.} 2005, \aap,
  440, 775

\bibitem[{{Kalberla} {et~al.}(2007){Kalberla}, {Dedes}, {Kerp}, \&
  {Haud}}]{kalberlaetal07}
{Kalberla}, P.~M.~W., {Dedes}, L., {Kerp}, J., \& {Haud}, U. 2007, \aap, 469,
  511

\bibitem[{{Kalberla} {et~al.}(2010){Kalberla}, {McClure-Griffiths}, {Pisano},
  {Calabretta}, {Ford}, {Lockman}, {Staveley-Smith}, {Kerp}, {Winkel},
  {Murphy}, \& {Newton-McGee}}]{kalberla10}
{Kalberla}, P.~M.~W., {McClure-Griffiths}, N.~M., {Pisano}, D.~J., {et~al.}
  2010, \aap, 521, A17

\bibitem[{{Kalberla} {et~al.}(1980){Kalberla}, {Mebold}, \&
  {Reich}}]{kalberla80}
{Kalberla}, P.~M.~W., {Mebold}, U., \& {Reich}, W. 1980, \aap, 82, 275

\bibitem[{{Kaufmann} {et~al.}(2006){Kaufmann}, {Mayer}, {Wadsley}, {Stadel}, \&
  {Moore}}]{kaufmann06}
{Kaufmann}, T., {Mayer}, L., {Wadsley}, J., {Stadel}, J., \& {Moore}, B. 2006,
  \mnras, 370, 1612

\bibitem[{{Kerp} {et~al.}(1999){Kerp}, {Burton}, {Egger}, {Freyberg},
  {Hartmann}, {Kalberla}, {Mebold}, \& {Pietz}}]{kerp99}
{Kerp}, J., {Burton}, W.~B., {Egger}, R., {et~al.} 1999, \aap, 342, 213

\bibitem[{{Kerp} {et~al.}(2011){Kerp}, {Winkel}, {Ben Bekhti}, {Fl{\"o}er}, \&
  {Kalberla}}]{Kerp11}
{Kerp}, J., {Winkel}, B., {Ben Bekhti}, N., {Fl{\"o}er}, L., \& {Kalberla},
  P.~M.~W. 2011, Astronomische Nachrichten, 332, 637

\bibitem[{{Kondo} {et~al.}(2006){Kondo}, {Kobayashi}, {Minowa}, {Tsujimoto},
  {Churchill}, {Takato}, {Iye}, {Kamata}, {Terada}, {Pyo}, {Takami}, {Hayano},
  {Kanzawa}, {Saint-Jacques}, {G{\"a}essler}, {Oya}, {Nedachi}, \&
  {Tokunaga}}]{kondoetal06}
{Kondo}, S., {Kobayashi}, N., {Minowa}, Y., {et~al.} 2006, \apj, 643, 667

\bibitem[{{Lockman} {et~al.}(2002){Lockman}, {Murphy}, {Petty-Powell}, \&
  {Urick}}]{lockman02}
{Lockman}, F.~J., {Murphy}, E.~M., {Petty-Powell}, S., \& {Urick}, V.~J. 2002,
  \apjs, 140, 331

\bibitem[{{Lu} {et~al.}(1994){Lu}, {Savage}, \& {Sembach}}]{Luetal_94}
{Lu}, L., {Savage}, B.~D., \& {Sembach}, K.~R. 1994, \apjl, 437, L119

\bibitem[{{Majewski}(2004)}]{majewski04}
{Majewski}, S. 2004, in Astronomical Society of the Pacific Conference Series,
  Vol. 327, Satellites and Tidal Streams, ed. F.~{Prada}, D.~{Martinez
  Delgado}, \& T.~J. {Mahoney}, 63

\bibitem[{{Marasco} \& {Fraternali}(2011)}]{marasco11}
{Marasco}, A. \& {Fraternali}, F. 2011, \aap, 525, A134

\bibitem[{{Masiero} {et~al.}(2005){Masiero}, {Charlton}, {Ding}, {Churchill},
  \& {Kacprzak}}]{masieroetal05}
{Masiero}, J.~R., {Charlton}, J.~C., {Ding}, J., {Churchill}, C.~W., \&
  {Kacprzak}, G. 2005, \apj, 623, 57

\bibitem[{{McClure-Griffiths} {et~al.}(2009){McClure-Griffiths}, {Pisano},
  {Calabretta}, {Ford}, {Lockman}, {Staveley-Smith}, {Kalberla}, {Bailin},
  {Dedes}, {Janowiecki}, {Gibson}, {Murphy}, {Nakanishi}, \&
  {Newton-McGee}}]{McClure_Griffithsetal09}
{McClure-Griffiths}, N.~M., {Pisano}, D.~J., {Calabretta}, M.~R., {et~al.}
  2009, \apjs, 181, 398

\bibitem[{{Muller} {et~al.}(1963){Muller}, {Oort}, \&
  {Raimond}}]{mulleroortraimond63}
{Muller}, C.~A., {Oort}, J.~H., \& {Raimond}, E. 1963, C. R. Acad. Sc. Paris,
  257, 1661

\bibitem[{{Oosterloo} {et~al.}(2007){Oosterloo}, {Fraternali}, \&
  {Sancisi}}]{Oosterlooetal_07}
{Oosterloo}, T., {Fraternali}, F., \& {Sancisi}, R. 2007, \aj, 134, 1019

\bibitem[{{Petitjean} \& {Bergeron}(1990)}]{petitjeanbergeron90}
{Petitjean}, P. \& {Bergeron}, J. 1990, \aap, 231, 309

\bibitem[{{Petitjean} {et~al.}(1993){Petitjean}, {Webb}, {Rauch}, {Carswell},
  \& {Lanzetta}}]{petitjean93}
{Petitjean}, P., {Webb}, J.~K., {Rauch}, M., {Carswell}, R.~F., \& {Lanzetta},
  K. 1993, \mnras, 262, 499

\bibitem[{{Prochter} {et~al.}(2006){Prochter}, {Prochaska}, \&
  {Burles}}]{prochteretal06}
{Prochter}, G.~E., {Prochaska}, J.~X., \& {Burles}, S.~M. 2006, \apj, 639, 766

\bibitem[{{Putman} {et~al.}(2002){Putman}, {de Heij}, {Staveley-Smith},
  {Braun}, {Freeman}, {Gibson}, {Burton}, {Barnes}, {Banks}, {Bhathal}, {de
  Blok}, {Boyce}, {Disney}, {Drinkwater}, {Ekers}, {Henning}, {Jerjen},
  {Kilborn}, {Knezek}, {Koribalski}, {Malin}, {Marquarding}, {Minchin},
  {Mould}, {Oosterloo}, {Price}, {Ryder}, {Sadler}, {Stewart}, {Stootman},
  {Webster}, \& {Wright}}]{Putmanetal02}
{Putman}, M.~E., {de Heij}, V., {Staveley-Smith}, L., {et~al.} 2002, \aj, 123,
  873

\bibitem[{{Putman} {et~al.}(2003){Putman}, {Staveley-Smith}, {Freeman},
  {Gibson}, \& {Barnes}}]{putmanetal_03}
{Putman}, M.~E., {Staveley-Smith}, L., {Freeman}, K.~C., {Gibson}, B.~K., \&
  {Barnes}, D.~G. 2003, \apj, 586, 170

\bibitem[{{Richter}(2006)}]{richterp06}
{Richter}, P. 2006, in Reviews in Modern Astronomy, Vol.~19, Reviews in Modern
  Astronomy, ed. S.~{Roeser}, 31

\bibitem[{{Richter} {et~al.}(2009){Richter}, {Charlton}, {Fangano}, {Bekhti},
  \& {Masiero}}]{richteretal09}
{Richter}, P., {Charlton}, J.~C., {Fangano}, A.~P.~M., {Bekhti}, N.~B., \&
  {Masiero}, J.~R. 2009, \apj, 695, 1631

\bibitem[{{Richter} {et~al.}(1999){Richter}, {de Boer}, {Widmann},
  {Kappelmann}, {Gringel}, {Grewing}, \& {Barnstedt}}]{richterdeboeretal99}
{Richter}, P., {de Boer}, K.~S., {Widmann}, H., {et~al.} 1999, \nat, 402, 386

\bibitem[{{Richter} {et~al.}(2011){Richter}, {Krause}, {Fechner}, {Charlton},
  \& {Murphy}}]{richteretal10}
{Richter}, P., {Krause}, F., {Fechner}, C., {Charlton}, J.~C., \& {Murphy},
  M.~T. 2011, \aap, 528, A12

\bibitem[{{Richter} {et~al.}(2003{\natexlab{a}}){Richter}, {Savage}, {Sembach},
  {Tripp}, \& {Jenkins}}]{richtersavagesembachetal03}
{Richter}, P., {Savage}, B.~D., {Sembach}, K.~R., {Tripp}, T.~M., \& {Jenkins},
  E.~B. 2003{\natexlab{a}}, in Astrophysics and Space Science Library, Vol.
  281, The IGM/Galaxy Connection. The Distribution of Baryons at z=0, ed. J.~L.
  {Rosenberg} \& M.~E. {Putman}, 85

\bibitem[{{Richter} {et~al.}(2001{\natexlab{a}}){Richter}, {Sembach}, {Wakker},
  \& {Savage}}]{richtersembachetal01}
{Richter}, P., {Sembach}, K.~R., {Wakker}, B.~P., \& {Savage}, B.~D.
  2001{\natexlab{a}}, \apjl, 562, L181

\bibitem[{{Richter} {et~al.}(2001{\natexlab{b}}){Richter}, {Sembach}, {Wakker},
  {Savage}, {Tripp}, {Murphy}, {Kalberla}, \&
  {Jenkins}}]{richtersembachetal01b}
{Richter}, P., {Sembach}, K.~R., {Wakker}, B.~P., {et~al.} 2001{\natexlab{b}},
  \apj, 559, 318

\bibitem[{{Richter} {et~al.}(2003{\natexlab{b}}){Richter}, {Wakker}, {Savage},
  \& {Sembach}}]{richterwakkersavagesembach03}
{Richter}, P., {Wakker}, B.~P., {Savage}, B.~D., \& {Sembach}, K.~R.
  2003{\natexlab{b}}, \apj, 586, 230

\bibitem[{{Richter} {et~al.}(2005){Richter}, {Westmeier}, \&
  {Br{\"u}ns}}]{richterwestmeierbruens05}
{Richter}, P., {Westmeier}, T., \& {Br{\"u}ns}, C. 2005, \aap, 442, L49

\bibitem[{{Routly} \& {Spitzer}(1952)}]{routly52}
{Routly}, P.~M. \& {Spitzer}, Jr., L. 1952, \apj, 115, 227

\bibitem[{{Rudie} {et~al.}(2012){Rudie}, {Steidel}, {Trainor}, {Rakic},
  {Bogosavljevic}, {Pettini}, {Reddy}, {Shapley}, {Erb}, \& {Law}}]{rudie2012}
{Rudie}, G.~C., {Steidel}, C.~C., {Trainor}, R.~F., {et~al.} 2012, ArXiv
  e-prints

\bibitem[{{Sancisi} {et~al.}(2008){Sancisi}, {Fraternali}, {Oosterloo}, \& {van
  der Hulst}}]{sancisi08}
{Sancisi}, R., {Fraternali}, F., {Oosterloo}, T., \& {van der Hulst}, T. 2008,
  \aapr, 15, 189

\bibitem[{{Savage} \& {Massa}(1987)}]{savage_massa87}
{Savage}, B.~D. \& {Massa}, D. 1987, \apj, 314, 380

\bibitem[{{Savage} \& {Sembach}(1996)}]{savagesembach96}
{Savage}, B.~D. \& {Sembach}, K.~R. 1996, \araa, 34, 279

\bibitem[{{Savage} {et~al.}(2000){Savage}, {Wakker}, {Jannuzi}, {Bahcall},
  {Bergeron}, {Boksenberg}, {Hartig}, {Kirhakos}, {Murphy}, {Sargent},
  {Schneider}, {Turnshek}, \& {Wolfe}}]{savageetal00}
{Savage}, B.~D., {Wakker}, B., {Jannuzi}, B.~T., {et~al.} 2000, \apjs, 129, 563

\bibitem[{{Sembach} {et~al.}(1991){Sembach}, {Savage}, \&
  {Massa}}]{Sembachetal91}
{Sembach}, K.~R., {Savage}, B.~D., \& {Massa}, D. 1991, \apj, 372, 81

\bibitem[{{Sembach} {et~al.}(2003){Sembach}, {Wakker}, {Savage}, {Richter},
  {Meade}, {Shull}, {Jenkins}, {Sonneborn}, \& {Moos}}]{sembach03}
{Sembach}, K.~R., {Wakker}, B.~P., {Savage}, B.~D., {et~al.} 2003, \apjs, 146,
  165

\bibitem[{{Shapiro} \& {Field}(1976)}]{shapirofield76}
{Shapiro}, P.~R. \& {Field}, G.~B. 1976, \apj, 205, 762

\bibitem[{{Steidel} {et~al.}(2010){Steidel}, {Erb}, {Shapley}, {Pettini},
  {Reddy}, {Bogosavljevi{\'c}}, {Rudie}, \& {Rakic}}]{steidel2010}
{Steidel}, C.~C., {Erb}, D.~K., {Shapley}, A.~E., {et~al.} 2010, \apj, 717, 289

\bibitem[{{Thom} {et~al.}(2006){Thom}, {Putman}, {Gibson}, {Christlieb},
  {Flynn}, {Beers}, {Wilhelm}, \& {Lee}}]{thom2006}
{Thom}, C., {Putman}, M.~E., {Gibson}, B.~K., {et~al.} 2006, \apjl, 638, L97

\bibitem[{{Vallerga} {et~al.}(1993){Vallerga}, {Vedder}, {Craig}, \&
  {Welsh}}]{vallerga93}
{Vallerga}, J.~V., {Vedder}, P.~W., {Craig}, N., \& {Welsh}, B.~Y. 1993, \apj,
  411, 729

\bibitem[{{van Woerden} {et~al.}(1999){van Woerden}, {Schwarz}, {Peletier},
  {Wakker}, \& {Kalberla}}]{vanWoerden99}
{van Woerden}, H., {Schwarz}, U.~J., {Peletier}, R.~F., {Wakker}, B.~P., \&
  {Kalberla}, P.~M.~W. 1999, \nat, 400, 138

\bibitem[{{Vladilo} {et~al.}(1993){Vladilo}, {Molaro}, {Monai}, {D'Odorico},
  {Ferlet}, {Madjar}, \& {Dennefeld}}]{vladilo93}
{Vladilo}, G., {Molaro}, P., {Monai}, S., {et~al.} 1993, \aap, 274, 37

\bibitem[{{Wakker}(1991)}]{wakker91}
{Wakker}, B.~P. 1991, \aap, 250, 499

\bibitem[{{Wakker}(2001)}]{wakker01}
{Wakker}, B.~P. 2001, \apjs, 136, 463

\bibitem[{{Wakker}(2004)}]{wakker04}
{Wakker}, B.~P. 2004, in Astrophysics and Space Science Library, Vol. 312, High
  Velocity Clouds, ed. {H.~van Woerden, B.~P.~Wakker, U.~J.~Schwarz, \&
  K.~S.~de Boer }, 25

\bibitem[{{Wakker}(2006)}]{wakker06}
{Wakker}, B.~P. 2006, \apjs, 163, 282

\bibitem[{{Wakker} {et~al.}(1999){Wakker}, {Howk}, {Savage}, {van Woerden},
  {Tufte}, {Schwarz}, {Benjamin}, {Reynolds}, {Peletier}, \&
  {Kalberla}}]{wakkeretal99}
{Wakker}, B.~P., {Howk}, J.~C., {Savage}, B.~D., {et~al.} 1999, \nat, 402, 388

\bibitem[{{Wakker} \& {Mathis}(2000)}]{wakkermathis00}
{Wakker}, B.~P. \& {Mathis}, J.~S. 2000, \apjl, 544, L107

\bibitem[{{Wakker} {et~al.}(2007){Wakker}, {York}, {Howk}, {Barentine},
  {Wilhelm}, {Peletier}, {van Woerden}, {Beers}, {Ivezi{\'c}}, {Richter}, \&
  {Schwarz}}]{wakker_york_howketal07}
{Wakker}, B.~P., {York}, D.~G., {Howk}, J.~C., {et~al.} 2007, \apjl, 670, L113

\bibitem[{{Wakker} {et~al.}(2008){Wakker}, {York}, {Wilhelm}, {Barentine},
  {Richter}, {Beers}, {Ivezi{\'c}}, \& {Howk}}]{wakkeryorkwilhelmetal08}
{Wakker}, B.~P., {York}, D.~G., {Wilhelm}, R., {et~al.} 2008, \apj, 672, 298

\bibitem[{{Welsh} {et~al.}(1990){Welsh}, {Vedder}, \& {Vallerga}}]{welshetal90}
{Welsh}, B.~Y., {Vedder}, P.~W., \& {Vallerga}, J.~V. 1990, \apj, 358, 473

\bibitem[{{Westmeier}(2007)}]{westmeier07}
{Westmeier}, T. 2007, PhD thesis, {Rheinische
  Friedrich-Wilhelms-Universit\"{a}t Bonn}

\bibitem[{{Winkel} {et~al.}(2011){Winkel}, {Ben Bekhti}, {Darmst{\"a}dter},
  {Fl{\"o}er}, {Kerp}, \& {Richter}}]{winkel11}
{Winkel}, B., {Ben Bekhti}, N., {Darmst{\"a}dter}, V., {et~al.} 2011, \aap,
  533, A105

\bibitem[{{Winkel} {et~al.}(2010){Winkel}, {Kalberla}, {Kerp}, \&
  {Fl{\"o}er}}]{winkel10a}
{Winkel}, B., {Kalberla}, P.~M.~W., {Kerp}, J., \& {Fl{\"o}er}, L. 2010, \apjs,
  188, 488

\end{thebibliography}

\clearpage
% \onecolumn

\longtabL{1}{
\begin{landscape}
\begin{longtable}{lllllllllllll}
\caption{QSO sight lines along which halo \ion{Ca}{ii} and \ion{Na}{i} was detected. For many of them, a follow-up search for \ion{H}{i} emission was performed. The table contains galactic coordinates, $l$ and $b$, radial velocity, $v_\mathrm{lsr}$, column densities, and $\log N$, $b$-values of all identified components. If an association to a known IVC/HVC complex is likely, the name of the complex is given in the last column. Primary QSO names are according to the SIMBAD database (\texttt{http://simbad.u-strasbg.fr/simbad/}). In some cases the SIMBAD identifiers cannot be found in the NASA/IPAC extragalactic database (NED, \texttt{http://ned.ipac.caltech.edu/}). In these cases the NED identifier is given in parentheses. }\label{qsotable}\\
\hline
\hline
QSO & R.A.  & Dec. & l  & b & $v_\mathrm{lsr}$ & $\log N_\ion{Ca}{ii}$ & $b_\ion{Ca}{ii}$ & $\log N_\ion{Na}{i}$ & $b_\ion{Na}{i}$ &$\log N_\ion{H}{i}$ & $b_\ion{H}{i}$ &Complex\\
& [hhmmss.s]  & [ddmmss] & [deg]  & [deg] & $[\mathrm{km\,s}^{-1}]$ & $[\mathrm{cm}^{-2}]$ & $[\mathrm{km\,s}^{-1}]$ & $[\mathrm{cm}^{-2}]$ & $[\mathrm{km\,s}^{-1}]$ &$[\mathrm{cm}^{-2}]$ & $[\mathrm{km\,s}^{-1}]$& \\
\hline
\endfirsthead
\caption{Continued.} \\
\hline
QSO & R.A.  & Dec. & l  & b & $v_\mathrm{lsr}$ & $\log N_\ion{Ca}{ii}$ & $b_\ion{Ca}{ii}$ & $\log N_\ion{Na}{i}$ & $b_\ion{Na}{i}$ &$\log N_\ion{H}{i}$ & $b_\ion{H}{i}$ &Complex \\
& [hhmmss.s]  & [ddmmss] & [deg]  & [deg] & $[\mathrm{km\,s}^{-1}]$ & $[\mathrm{cm}^{-2}]$ & $[\mathrm{km\,s}^{-1}]$ & $[\mathrm{cm}^{-2}]$ & $[\mathrm{km\,s}^{-1}]$ &$[\mathrm{cm}^{-2}]$ & $[\mathrm{km\,s}^{-1}]$& \\
\hline
\endhead
\hline
\multicolumn{7}{l}{n/a \textit{data not available}\qquad --- \textit{non-detection}}
\endfoot
\hline
\multicolumn{7}{l}{n/a \textit{data not available}\qquad --- \textit{non-detection}}
% \tablefoot{n/a \textit{data not available} --- \textit{non-detection}}
\endlastfoot
HE\,0001$-$2340 & 00 03 45.0 & $-$23 23 55 & 49.39 & $-$78.60 & $-$126 & 11.89 & 6.0 & --- & --- & --- & --- & MS \\ 
&  &  &  &  & $-$112 & 11.76 & 6.0 & --- & --- & 19.50 & 17.5 & MS \\ 
&  &  &  &  & $-$98 & 11.89 & 6.0 & --- & --- & 19.08 & 24.6 & MS \\ 
&  &  &  &  & 14 & 11.53 & 4.1 & --- & --- & --- & --- &  \\ 
QSO\,B0002$-$422 & 00 04 48.2 & $-$41 57 28 & 332.68 & $-$72.37 & 89 & 11.75 & 5.7 & --- & --- & 19.95 & 18.8 & MS \\ 

{\small\it([HB89]\,0002$-$422)}  \\ 
SDSS\,J000815.33$-$095854.0 & 00 08 15.3 & $-$09 58 54 & 90.19 & $-$70.06 & $-$202 & 11.62 & 6.0 & --- & --- & --- & --- &  \\ 
&  &  &  &  & $-$48 & 11.55 & 6.0 & --- & --- & 18.72 & 8.1 & MS \\ 
2QZ\,J000852.7$-$290044 & 00 08 52.7 & $-$29 00 44 & 19.11 & $-$80.43 & $-$36 & 12.53 & 7.0 & --- & --- & --- & --- & MS \\ 
FBQS\,J0040$-$0146 & 00 40 57.6 & $-$01 46 32 & 116.84 & $-$64.52 & $-$177 & 12.02 & 4.1 & --- & --- & --- & --- & MS \\ 
LBQS\,0042$-$2930 & 00 45 08.5 & $-$29 14 32 & 335.84 & $-$87.47 & $-$28 & 12.23 & 3.4 & 11.63 & 2.0 & 18.69 & 14.8 &  \\ 
LBQS\,0049$-$2820 & 00 51 27.2 & $-$28 04 35 & 302.73 & $-$89.05 & $-$118 & 11.61 & 3.2 & --- & --- & --- & --- & MS \\ 
&  &  &  &  & $-$171 & 11.97 & 8.0 & --- & --- & --- & --- & MS \\ 
&  &  &  &  & $-$186 & 11.24 & 1.1 & --- & --- & --- & --- & MS \\ 
&  &  &  &  & $-$205 & 12.15 & 3.0 & --- & --- & 18.14 & 6.1 & MS \\ 
LBQS\,0056$-$0009 & 00 59 05.5 & +00 06 52 & 127.11 & $-$62.70 & 17 & 11.99 & 1.1 & --- & --- & --- & --- & MS \\ 
QSO\,J0103$+$1316 & 01 03 11.3 & +13 16 18 & 127.34 & $-$49.50 & $-$351 & 11.72 & 4.0 & n/a & n/a & --- & --- & MS \\ 

{\small\it([HB89]\,0100$+$130)}\\
UM\,669 & 01 05 16.8 & $-$18 46 42 & 144.52 & $-$81.06 & 106 & 11.43 & 2.0 & 10.98 & 3.0 & --- & --- & MS \\ 
&  &  &  &  & 167 & 11.24 & 3.2 & --- & --- & --- & --- & MS \\ 
&  &  &  &  & 192 & 11.36 & 1.0 & --- & --- & --- & --- & MS \\ 
UM\,086 & 01 08 21.8 & +06 23 28 & 130.51 & $-$56.23 & $-$26 & 11.43 & 4.4 & n/a & n/a & n/a & n/a &  \\ 
QSO\,B0109$-$353 & 01 11 43.6 & $-$35 03 01 & 275.47 & $-$80.97 & $-$108 & 12.44 & 7.0 & --- & --- & 19.17 & 22.3 & MS \\ 

{\small\it([HB89]\,0109$-$353)}&  &  &  &  & 79 & 12.41 & 6.9 & 11.03 & 6.0 & --- & --- &  \\ 
&  &  &  &  & $-$162 & 12.28 & 12.6 & --- & --- & 19.54 & 13.9 & MS \\ 
QSO\,B0112$-$30 & 01 15 04.7 & $-$30 25 14 & 246.77 & $-$83.86 & $-$5 & 11.89 & 2.0 & --- & --- & 19.24 & 3.6 & MS \\ 

{\small\it([VCV96]\,0112$-$30)}&  &  &  &  & $-$13 & 11.89 & 2.0 & 13.07 & 0.5 & 20.10 & 15.3 & MS \\ 
3C\,37 & 01 18 18.5 & +02 58 06 & 136.12 & $-$59.21 & $-$17 & 12.09 & 4.5 & 12.68 & 2.1 & n/a & n/a & MS \\ 
PKS\,0119$-$04 & 01 22 27.9 & $-$04 21 27 & 142.30 & $-$66.06 & $-$48 & 11.59 & 1.6 & --- & --- & 19.74 & 14.1 & MS \\ 
QSO\,B0122$-$379 & 01 24 17.4 & $-$37 44 23 & 271.91 & $-$77.34 & 40 & 11.72 & 3.9 & --- & --- & --- & --- & MS \\ 

{\small\it([HB89]\,0122$-$379)} \\ 
FBQS\,J0125$-$0018 & 01 25 17.2 & $-$00 18 29  & 141.18 & $-$61.97 & $-$302 & 12.90 & 19.5 & --- & --- & --- & --- & MS \\ 
FBQS\,J0125$-$0005 & 01 25 28.8 & $-$00 05 56 & 141.16 & $-$61.76 & $-$54 & 11.43 & 0.8 & 11.29 & 0.6 & 19.52 & 16.9 & MS \\ 
2MASSi\,J0129449$-$403345 & 01 29 44.9 & $-$40 33 45 & 274.88 & $-$74.41 & $-$33 & 11.54 & 4.8 & --- & --- & --- & --- & MS \\ 
PKS\,0130$-$17 & 01 32 43.5 & $-$16 54 49 & 168.12 & $-$76.02 & 8 & 11.82 & 7.0 & 11.11 & 0.7 & 20.12 & 13.4 & MS \\

[ICS96]\,013312.9$-$412216 & 01 35 23.2 & $-$41 06 57 & 272.86 & $-$73.35 & 285 & 13.14 & 0.7 & --- & --- & --- & --- & MS \\ 
SDSS\,J013901.40$-$082443.9 & 01 39 01.4 & $-$08 24 44 & 156.17 & $-$68.16 & $-$100 & 12.18 & 2.0 & --- & --- & --- & --- & MS \\ 
PKS\,0139$-$09 & 01 41 25.8 & $-$09 28 44 & 159.05 & $-$68.77 & $-$26 & 11.58 & 3.0 & --- & --- & 19.83 & 10.4 &  \\ 
&  &  &  &  & $-$16 & 11.58 & 2.0 & --- & --- & 19.79 & 5.2 &  \\ 
SDSS\,J014214.74$+$002324.2 & 01 42 14.8 & +00 23 24 & 148.93 & $-$59.89 & 5 & n/a & n/a & 11.12 & 3.6 & 19.56 & 4.0 & MS \\ 
&  &  &  &  & 11 & n/a & n/a & 12.32 & 0.9 & 18.91 & 3.6 & MS \\ 
SDSS\,J014631.99$+$133506.3 & 01 46 32.0 & +13 35 06 & 142.83 & $-$47.15 & $-$104 & 11.36 & 10.4 & n/a & n/a & n/a & n/a & ACHV \\ 
HE\,0151$-$4326 & 01 53 27.2 & $-$43 11 38 & 268.92 & $-$69.61 & 134 & 12.42 & 6.6 & --- & --- & 18.86 & 14.4 & MS \\ 
3C\,57 & 02 01 57.2 & $-$11 32 33 & 173.08 & $-$67.26 & $-$101 & 11.04 & 1.9 & --- & --- & --- & --- & MS \\ 
PKS\,0202$-$17 & 02 04 57.7 & $-$17 01 20 & 185.99 & $-$70.23 & $-$11 & 12.04 & 0.8 & 11.94 & 0.8 & 19.79 & 5.5 & MS \\ 
QSO\,B0216$+$0803 & 02 18 57.3 & +08 17 28 & 156.92 & $-$48.72 & 11 & n/a & n/a & 12.07 & 1.3 & 19.81 & 8.7 &  \\ 

{\small\it([HB89]\,0216$+$080)}\\
2QZ\,J022620.4$-$285751 & 02 26 20.5 & $-$28 57 51 & 223.51 & $-$69.02 & $-$29 & 12.03 & 4.5 & 11.17 & 2.1 & 19.10 & 8.3 & MS \\ 
PKS\,0232$-$04 & 02 35 07.4 & $-$04 02 06 & 174.46 & $-$56.16 & 40 & 11.39 & 0.6 & --- & --- & --- & --- &  \\ 
PKS\,0244$-$128 & 02 46 58.5 & $-$12 36 30 & 190.42 & $-$59.32 & $-$214 & 10.97 & 1.0 & --- & --- & --- & --- & AC shell \\ 
&  &  &  &  & $-$205 & 10.62 & 1.0 & --- & --- & --- & --- & AC shell \\ 
&  &  &  &  & $-$194 & 11.23 & 1.0 & --- & --- & --- & --- & AC shell \\ 
HE\,0251$-$5550 & 02 52 40.1 & $-$55 38 32 & 273.87 & $-$54.10 & 214 & 11.53 & 6.0 & --- & --- & 18.39 & 8.6 & LMC \\ 
&  &  &  &  &  & --- & --- & --- & --- & 19.22 & 13.9 &  \\ 
SDSS\,J025607.24$+$011038.6 & 02 56 07.3 & +01 10 39 & 174.65 & $-$48.76 & $-$11 & 12.00 & 0.7 & 12.18 & 3.0 & 20.24 & 3.8 &  \\ 
2MASSi\,J0302113$-$314030 & 03 02 11.3 & $-$31 40 30 & 229.96 & $-$61.27 & $-$209 & 12.93 & 10.0 & --- & --- & --- & --- & MS \\ 
2QZ\,J030249.6$-$321600 & 03 02 49.7 & $-$32 16 01 & 231.18 & $-$61.12 & $-$66 & 11.41 & 0.7 & --- & --- & --- & --- & MS \\ 
&  &  &  &  & 136 & 11.18 & 0.6 & --- & --- & --- & --- & MS \\ 
&  &  &  &  & 163 & 11.18 & 1.1 & --- & --- & --- & --- & MS \\ 
&  &  &  &  & 189 & 11.42 & 6.0 & --- & --- & --- & --- & MS \\ 
UM\,680 & 03 10 06.0 & $-$19 21 25 & 206.67 & $-$57.33 & $-$9 & --- & --- & 12.01 & 1.3 & 19.65 & 2.1 &  \\ 
UM\,681 & 03 10 09.0 & $-$19 22 08 & 206.70 & $-$57.32 & $-$8 & 11.94 & 6.8 & 12.05 & 0.9 & n/a & n/a &  \\ 
SDSS\,J031856.61$-$060037.6 & 03 18 56.6 & $-$06 00 38 & 188.62 & $-$49.09 & $-$19 & 12.02 & 1.6 & n/a & n/a & --- & --- & AC shell \\ 
PKS\,0328$-$272 & 03 30 32.6 & $-$27 04 40 & 222.21 & $-$54.74 & 387 & 11.70 & 1.2 & 11.30 & 1.7 & --- & --- & MS \\ 
PKS\,0330$-$450 & 03 32 44.1 & $-$44 55 57 & 252.79 & $-$53.36 & 60 & 10.96 & 1.8 & --- & --- & --- & --- & MS \\ 
&  &  &  &  & 71 & 11.13 & 1.0 & --- & --- & --- & --- &  \\ 
3C\,95 & 03 51 28.5 & $-$14 29 09 & 205.48 & $-$46.33 & $-$23 & 11.71 & 3.2 & 11.34 & 4.0 & 19.80 & 10.3 & AC shell \\ 
3C\,94 & 03 52 30.6 & $-$07 11 02 & 196.56 & $-$42.72 & $-$18 & 12.03 & 6.0 & 11.88 & 1.6 & 19.72 & 12.5 & AC shell \\ 
PKS\,0402$-$362 & 04 03 53.8 & $-$36 05 02 & 237.74 & $-$48.48 & $-$72 & 11.73 & 1.9 & --- & --- & --- & --- &  \\ 
PKS\,0420$-$01 & 04 23 15.8 & $-$01 20 33 & 195.29 & $-$33.14 & $-$198 & 11.55 & 0.4 & --- & --- & --- & --- & AC shell \\ 
HE\,0421$-$2624 & 04 23 54.0 & $-$26 18 01 & 224.67 & $-$42.92 & 177 & 11.87 & 5.4 & --- & --- & --- & --- & MS \\ 
PKS\,0422$-$380 & 04 24 42.2 & $-$37 56 21 & 240.65 & $-$44.40 & $-$122 & 11.68 & 0.6 & --- & --- & --- & --- &  \\ 
CTS\,0436 & 04 26 44.5 & $-$52 08 20 & 260.26 & $-$43.03 & $-$11 & --- & --- & 11.50 & 0.9 & 19.48 & 14.0 &  \\ 
QSO\,B0438$-$166 & 04 40 26.5 & $-$16 32 34 & 214.14 & $-$36.25 & $-$14 & n/a & n/a & 12.11 & 1.5 & 20.14 & 12.2 & AC shell \\ 

{\small\it([HB89]\,0438$-$166)}\\ 
QSO\,B0449$-$1645 & 04 52 13.6 & $-$16 40 12 & 215.56 & $-$33.68 & $-$10 & --- & --- & 11.35 & 5.8 & 19.98 & 10.4 & AC shell \\ 
{\small\it(H\,0449$-$1645)}\\
QSO\,B0450$-$1310B & 04 53 12.8 & $-$13 05 46 & 211.75 & $-$32.07 & $-$20 & n/a & n/a & 11.95 & 2.4 & 19.26 & 6.8 & AC shell \\ 
{\small\it(H\,0450$-$1310)}&  &  &  &  & $-$5 & n/a & n/a & 11.47 & 5.1 & 19.83 & 2.3 & AC shell \\ 
QSO\,B0458$-$0203 & 05 01 12.8 & $-$01 59 14 & 201.45 & $-$25.30 & $-$6 & 12.08 & 4.0 & 12.10 & 2.2 & 19.28 & 2.6 & AC shell \\ 

{\small\it([HB89]\,0458$-$0203)}  \\ 
PKS\,0506$-$61 & 05 06 44.0 & $-$61 09 41 & 270.55 & $-$36.07 & 297 & 12.09 & 6.5 & --- & --- & 18.85 & 13.8 & LMC \\ 
HE\,0515$-$4414 & 05 17 07.6 & $-$44 10 55 & 249.61 & $-$34.97 & $-$58 & 11.28 & 0.6 & --- & --- & --- & --- &  \\ 
&  &  &  &  & $-$41 & 11.25 & 3.8 & 10.75 & 2.7 & --- & --- &  \\ 
&  &  &  &  & $-$17 & 10.92 & 1.0 & --- & --- & --- & --- &  \\ 
&  &  &  &  & $-$4 & 11.72 & 2.8 & 9.42 & 1.9 & --- & --- &  \\ 
HE\,0517$-$3649 & 05 19 39.8 & $-$36 46 13 & 240.82 & $-$33.43 & $-$115 & 11.45 & 0.7 & --- & --- & --- & --- &  \\ 
HS\,0810$+$2554 & 08 13 31.3 & +25 45 03 & 196.88 & 28.63 & $-$23 & n/a & n/a & 11.15 & 5.0 & 19.91 & 36.9 &  \\ 
QSO\,B0827$+$2421 & 08 30 52.1 & +24 10 60 & 200.02 & 31.88 & $-$21 & n/a & n/a & 11.28 & 3.0 & 19.21 & 7.6 & IV arch \\ 
{\small\it(FBQS\,J083052.0$+$241059)}\\
PKS\,0839$+$18 & 08 42 05.1 & +18 35 41 & 207.28 & 32.48 & $-$22 & 12.13 & 1.3 & --- & --- & n/a & n/a &  \\ 
&  &  &  &  & $-$12 & 12.15 & 2.4 & 12.11 & 1.2 & n/a & n/a &  \\ 
&  &  &  &  & 97 & 11.56 & 4.0 & --- & --- & n/a & n/a &  \\ 
PKS\,0906$+$01 & 09 09 10.1 & +01 21 36 & 228.94 & 30.92 & $-$27 & 11.32 & 0.6 & --- & --- & n/a & n/a & IV spur \\ 
&  &  &  &  & $-$16 & 11.60 & 0.8 & --- & --- & n/a & n/a & IV spur \\ 
SDSS\,J091127.61$+$055054.0 & 09 11 27.6 & +05 50 54 & 224.68 & 33.67 & $-$26 & 11.74 & 0.8 & 10.69 & 0.6 & n/a & n/a & IV spur \\ 
PKS\,0922$+$14 & 09 25 07.3 & +14 44 26 & 216.53 & 40.61 & $-$211 & 11.33 & 0.6 & --- & --- & n/a & n/a &  \\ 
&  &  &  &  & $-$84 & 11.87 & 0.5 & --- & --- & n/a & n/a &  \\ 
&  &  &  &  & $-$30 & 12.26 & 1.0 & 13.03 & 1.6 & n/a & n/a & IV arch \\ 
&  &  &  &  & $-$23 & 12.18 & 1.0 & 12.40 & 1.0 & n/a & n/a & IV arch \\ 
HE\,0926$-$0201 & 09 29 13.6 & $-$02 14 46 & 235.73 & 33.17 & 176 & 11.67 & 3.1 & --- & --- & 18.68 & 18.3 & WA \\ 
PKS\,0932$+$02 & 09 35 18.2 & +02 04 16 & 232.39 & 36.84 & 98 & 11.76 & 1.3 & --- & --- & n/a & n/a & WA/WB \\ 
SDSS\,J095352.69$+$080103.6 & 09 53 52.7 & +08 01 04 & 228.92 & 43.90 & $-$49 & 11.76 & 3.0 & --- & --- & 19.25 & 7.9 & IV spur \\ 
QSO\,B0952$+$179 & 09 54 56.8 & +17 43 31 & 216.46 & 48.36 & $-$30 & n/a & n/a & 11.60 & 8.8 & 19.78 & 11.0 & IV spur \\ 

{\small\it([HB89]\,0952$+$179)}\\
CTQ\,0460 & 10 39 09.5 & $-$23 13 26 & 267.40 & 30.37 & $-$23 & 11.90 & 6.8 & 12.03 & 1.3 & 19.57 & 1.9 &  \\ 
&  &  &  &  & 83 & 11.99 & 7.3 & --- & --- & 18.66 & 12.5 & WA/WB \\ 
QSO\,J1039$-$2719 & 10 39 21.8 & $-$27 19 16 & 270.03 & 26.98 & $-$11 & 12.59 & 8.0 & 11.63 & 6.0 & 20.18 & 9.4 &  \\ 

{\small\it([HB89]\,1037$-$270)}&  &  &  &  & 184 & 11.29 & 5.0 & --- & --- & --- & --- & WD \\ 
QSO\,J1040$-$2727 & 10 40 32.2 & $-$27 27 49 & 270.36 & 27.00 & $-$17 & 12.49 & 7.9 & 12.03 & 2.3 & 20.21 & 10.7 &  \\ 

{\small\it([HB89]\,1038$-$272)}\\
HE\,1043$-$1002 & 10 45 40.6 & $-$10 18 13 & 259.49 & 41.72 & 30 & 12.01 & 1.2 & n/a & n/a & --- & --- & IV WA \\ 
SDSS\,J104642.83$+$053107.0 & 10 46 42.8 & +05 31 07 & 243.36 & 53.30 & $-$16 & n/a & n/a & 11.61 & 1.2 & n/a & n/a & IV spur \\ 
&  &  &  &  & $-$8 & n/a & n/a & 11.52 & 1.9 & n/a & n/a &  \\ 
SDSS\,J104656.70$+$054150.3 & 10 46 56.7 & +05 41 50 & 243.19 & 53.45 & $-$27 & n/a & n/a & 10.90 & 3.4 & n/a & n/a & IV spur \\ 
QSO\,B1101$-$26 & 11 03 25.3 & $-$26 45 16 & 274.96 & 30.19 & $-$16 & 11.74 & 6.0 & 11.22 & 3.0 & 20.06 & 6.8 &  \\ 

{\small\it([HB89]\,1100$-$264)}&  &  &  &  & $-$27 & 11.34 & 3.8 & --- & --- & --- & --- &  \\ 
&  &  &  &  & 199 & 11.74 & 5.8 & --- & --- & --- & --- & WD \\ 
&  &  &  &  & 147 & 11.42 & 6.0 & --- & --- & --- & --- & WD \\ 
&  &  &  &  & 108 & 11.15 & 6.0 & --- & --- & --- & --- & WD \\ 
&  &  &  &  & 71 & 10.89 & 6.0 & --- & --- & --- & --- &  \\ 
SDSS\,J110729.03$+$004811.1 & 11 07 29.0 & +00 48 11 & 255.14 & 53.74 & $-$27 & 12.25 & 7.0 & 12.65 & 2.4 & 19.69 & 5.9 & IV spur \\ 
PKS\,1111$+$149 & 11 13 58.7 & +14 42 27 & 236.71 & 64.15 & $-$41 & 12.09 & 5.4 & 11.55 & 6.7 & n/a & n/a & IV spur \\ 
&  &  &  &  & $-$23 & 12.97 & 0.6 & 12.31 & 0.2 & n/a & n/a & IV spur \\ 
QSO\,B1119$+$183 & 11 22 29.7 & +18 05 26 & 232.07 & 67.62 & $-$38 & 11.84 & 1.8 & 11.10 & 2.0 & 19.64 & 13.5 & IV spur \\ 

{\small\it([HB89]\,1119$+$183)}\\
HE\,1126$-$2259 & 11 29 10.9 & $-$23 16 28 & 279.48 & 35.82 & $-$293 & 11.66 & 6.0 & --- & --- & --- & --- &  \\ 
SDSS\,J112932.64$+$020422.8 & 11 29 32.7 & +02 04 23 & 261.41 & 58.18 & 71 & 11.63 & 3.8 & --- & --- & --- & --- &  \\ 
&  &  &  &  & 85 & 11.42 & 2.0 & --- & --- & --- & --- &  \\ 
PKS\,1127$-$14 & 11 30 07.0 & $-$14 49 27 & 275.28 & 43.64 & 23 & 11.65 & 1.0 & --- & --- & 19.74 & 14.9 &  \\ 
&  &  &  &  & 30 & 11.55 & 1.3 & --- & --- & --- & --- &  \\ 
HS\,1140$+$2711\, & 11 42 54.3 & +26 54 58 & 209.81 & 74.75 & $-$43 & 12.27 & 7.7 & n/a & n/a & 19.38 & 3.2 & IV arch \\ 
&  &  &  &  &  & --- & --- & n/a & n/a & 20.05 & 15.2 &  \\ 
2MASSi\,J1159065$+$133738 & 11 59 06.5 & +13 37 38 & 258.19 & 71.79 & $-$33 & n/a & n/a & 12.09 & 10.5 & 19.74 & 5.5 & IV spur \\ 
SDSS\,J120342.24$+$102831.7 & 12 03 42.2 & +10 28 32 & 266.64 & 69.91 & $-$34 & 11.94 & 7.7 & 11.39 & 2.4 & 19.29 & 4.0 & IV spur \\ 
UM\,474 & 12 05 50.2 & +02 01 32 & 277.49 & 62.62 & 33 & 12.24 & 1.5 & 13.05 & 1.7 & 19.39 & 12.4 &  \\ 
&  &  &  &  & 41 & 12.29 & 9.9 & 11.73 & 3.9 & --- & --- & IV WA \\ 
LBQS\,1209$+$1046 & 12 11 41.6 & +10 30 17 & 271.66 & 70.92 & 76 & 11.54 & 1.7 & --- & --- & --- & --- &  \\ 
&  &  &  &  & $-$26 & 11.64 & 4.0 & --- & --- & 19.67 & 33.2 & IV spur \\ 
PKS\,1210$+$134 & 12 13 32.2 & +13 07 21 & 268.74 & 73.42 & $-$35 & 11.91 & 4.0 & 11.76 & 4.0 & 19.57 & 3.8 & IV spur \\ 
&  &  &  &  &  & --- & --- & --- & --- & 20.05 & 14.6 &  \\ 
TON\,1480 & 12 15 09.2 & +33 09 55 & 173.14 & 80.11 & $-$76 & 10.98 & 3.5 & --- & --- & --- & --- & IV arch \\ 
&  &  &  &  & $-$57 & 11.59 & 7.6 & --- & --- & --- & --- & IV arch \\ 
&  &  &  &  & $-$49 & 11.45 & 2.0 & 10.46 & 2.0 & 19.67 & 18.1 & IV arch \\ 
&  &  &  &  & $-$37 & 10.96 & 3.8 & --- & --- & --- & --- & IV arch \\ 
&  &  &  &  & $-$17 & 10.92 & 2.0 & --- & --- & --- & --- & IV arch \\ 
&  &  &  &  & $-$9 & 11.60 & 6.0 & --- & --- & --- & --- & IV arch \\ 
&  &  &  &  & 8 & 11.37 & 7.0 & --- & --- & --- & --- &  \\ 
&  &  &  &  & $-$27 & 11.59 & 3.8 & 11.00 & 2.0 & --- & --- & IV arch \\ 
LBQS\,1229$-$0207 & 12 32 00.0 & $-$02 24 05 & 293.16 & 60.10 & $-$21 & 11.52 & 1.9 & 10.99 & 0.2 & 19.97 & 12.4 & IV spur \\ 
&  &  &  &  & 55 & 11.94 & 3.3 & 11.36 & 1.4 & 18.51 & 14.5 &  \\ 
&  &  &  &  & 38 & 11.77 & 8.9 & 11.72 & 1.3 & 18.53 & 2.6 &  \\ 
&  &  &  &  & 22 & 11.71 & 3.9 & --- & --- & --- & --- &  \\ 
&  &  &  &  & 47 & --- & --- & 11.28 & 3.0 & --- & --- &  \\ 
LBQS\,1232$+$0815 & 12 34 37.6 & +07 58 41 & 290.41 & 70.44 & 88 & 11.62 & 1.8 & --- & --- & --- & --- &  \\ 
&  &  &  &  & 74 & 11.88 & 6.8 & --- & --- & --- & --- &  \\ 
PKS\,1243$-$072 & 12 46 04.2 & $-$07 30 47 & 300.59 & 55.33 & 24 & n/a & n/a & 11.72 & 2.0 & 19.63 & 19.2 &  \\ 
LBQS\,1246$-$0217 & 12 49 24.9 & $-$02 33 40 & 301.91 & 60.31 & $-$21 & 12.00 & 7.7 & --- & --- & 19.52 & 13.8 & IV spur \\ 
HE\,1249$-$0207 & 12 51 51.4 & $-$02 23 33 & 303.14 & 60.48 & 157 & 11.45 & 0.7 & --- & --- & --- & --- &  \\ 
NGC\,4748 & 12 52 12.4 & $-$13 24 54 & 303.22 & 49.46 & $-$20 & 11.72 & 4.6 & 11.84 & 2.6 & 19.97 & 11.2 &  \\ 
&  &  &  &  & 112 & 11.23 & 1.0 & 11.78 & 0.3 & 18.77 & 2.1 &  \\ 
&  &  &  &  & 118 & 10.76 & 1.0 & --- & --- & --- & --- &  \\ 
PKS\,1252$+$11 & 12 54 38.3 & +11 41 06 & 305.87 & 74.54 & $-$34 & 11.90 & 3.6 & 11.23 & 3.5 & --- & --- & IV spur \\ 
&  &  &  &  & $-$24 & 11.71 & 6.0 & 11.20 & 1.8 & 19.95 & 20.3 & IV spur \\ 
&  &  &  &  & 77 & 11.62 & 0.9 & --- & --- & --- & --- &  \\ 
PKS\,B1256$-$177 & 12 58 38.3 & $-$18 00 03 & 305.35 & 44.84 & $-$25 & 11.89 & 1.5 & --- & --- & 19.45 & 7.0 &  \\ 
HE\,1300$-$2431 & 13 03 00.1 & $-$24 47 12 & 306.26 & 38.01 & $-$28 & 12.07 & 9.4 & n/a & n/a & 19.51 & 11.1 &  \\ 
&  &  &  &  & 26 & 12.16 & 8.0 & n/a & n/a & 19.95 & 13.5 &  \\ 
3C\,281 & 13 07 53.9 & +06 42 14 & 314.51 & 69.20 & $-$36 & 12.84 & 0.6 & --- & --- & --- & --- & IV spur \\ 
&  &  &  &  & $-$45 & 11.38 & 0.5 & --- & --- & --- & --- & IV spur \\ 
2MASSi\,J1320299$-$052335 & 13 20 30.0 & $-$05 23 35 & 316.20 & 56.73 & $-$34 & --- & --- & 11.19 & 3.0 & 19.66 & 12.6 &  \\ 
PKS\,1327$-$206 & 13 30 07.7 & $-$20 56 16 & 314.94 & 41.03 & 22 & 12.31 & 6.9 & 12.21 & 2.4 & 20.13 & 4.6 & IV WA \\ 
&  &  &  &  & 35 & 12.10 & 1.7 & 12.72 & 2.0 & 19.14 & 5.2 & IV WA \\ 

[HB89]\,1331$+$170 & 13 33 35.8 & +16 49 04 & 348.51 & 75.81 & $-$26 & 12.00 & 4.9 & 11.14 & 2.6 & --- & --- & IV spur \\ 
&  &  &  &  & $-$11 & 12.35 & 11.6 & 11.34 & 5.5 & 18.85 & 2.7 &  \\ 
HE\,1341$-$1020 & 13 44 27.1 & $-$10 35 42 & 323.53 & 50.15 & $-$65 & 11.77 & 4.0 & --- & --- & --- & --- &  \\ 
&  &  &  &  & 56 & 11.71 & 3.6 & --- & --- & 19.16 & 13.2 &  \\ 
HE\,1347$-$2457 & 13 50 38.9 & $-$25 12 17 & 319.48 & 35.76 & $-$95 & 11.54 & 3.0 & --- & --- & --- & --- &  \\ 
PKS\,1349$-$439 & 13 52 56.5 & $-$44 12 40 & 314.41 & 17.29 & $-$106 & 12.05 & 3.7 & 11.10 & 2.0 & --- & --- &  \\ 
&  &  &  &  & $-$116 & 12.25 & 1.4 & 11.21 & 5.6 & --- & --- &  \\ 
&  &  &  &  & $-$124 & 11.74 & 2.9 & --- & --- & --- & --- &  \\ 
&  &  &  &  & $-$51 & 11.70 & 3.0 & --- & --- & 19.67 & 21.4 &  \\ 
PKS\,B1354$-$107 & 13 56 46.8 & $-$11 01 29 & 327.65 & 48.69 & $-$108 & 12.00 & 6.5 & n/a & n/a & --- & --- &  \\ 
PKS\,1354$+$19 & 13 57 04.4 & +19 19 06 & 8.99 & 73.04 & $-$38 & 11.97 & 1.0 & --- & --- & 18.79 & 4.6 &  \\ 
&  &  &  &  & $-$54 & 11.51 & 6.0 & --- & --- & --- & --- &  \\ 
QSO\,B1415$+$172 & 14 18 03.7 & +17 03 25 & 10.71 & 67.60 & $-$10 & 12.13 & 1.2 & --- & --- & --- & --- &  \\ 

{\small\it([HB89]\,1415$+$172)}&  &  &  &  & $-$17 & 13.30 & 0.6 & --- & --- & --- & --- &  \\ 
&  &  &  &  & $-$37 & 12.25 & 1.1 & --- & --- & 19.42 & 13.7 &  \\ 
PKS\,1420$-$27 & 14 22 49.2 & $-$27 27 56 & 326.67 & 31.16 & $-$74 & 11.32 & 3.5 & --- & --- & --- & --- &  \\ 
&  &  &  &  & $-$81 & 11.04 & 0.5 & --- & --- & --- & --- &  \\ 
&  &  &  &  & $-$47 & --- & --- & 11.49 & 1.1 & --- & --- &  \\ 
SDSS\,J142253.31$-$000148.9 & 14 22 53.3 & $-$00 01 49 & 345.66 & 55.07 & $-$12 & 12.79 & 1.5 & 11.64 & 3.2 & 20.13 & 19.7 &  \\ 
PKS\,1421$+$122 & 14 23 30.1 & +11 59 51 & 2.41 & 63.64 & $-$31 & 11.92 & 5.1 & 11.50 & 1.8 & 19.23 & 14.6 & GCN \\ 
&  &  &  &  & $-$155 & 11.52 & 1.1 & --- & --- & --- & --- &  \\ 
PKS\,1424$-$11 & 14 27 38.1 & $-$12 03 50 & 336.83 & 44.39 & $-$20 & 12.61 & 0.9 & --- & --- & 19.66 & 6.4 &  \\ 
&  &  &  &  & $-$11 & 12.71 & 1.4 & 12.55 & 1.3 & --- & --- &  \\ 
MRK\,1383 & 14 29 06.6 & +01 17 06 & 349.22 & 55.13 & $-$19 & 11.84 & 3.0 & --- & --- & 19.82 & 13.4 &  \\ 
QSO\,B1429$-$0053B & 14 32 29.0 & $-$01 06 14 & 347.75 & 52.74 & $-$338 & 11.90 & 1.1 & --- & --- & --- & --- &  \\ 
{\small\it(LBQS\,1429$-$0053B)}&  &  &  &  & $-$27 & 12.70 & 0.8 & --- & --- & --- & --- &  \\ 
&  &  &  &  & 265 & 12.18 & 1.3 & --- & --- & --- & --- &  \\ 
HE\,1434$-$1600 & 14 36 50.1 & $-$16 13 26 & 336.55 & 39.67 & 76 & 11.39 & 2.7 & --- & --- & --- & --- &  \\ 
LBQS\,1444$+$0126 & 14 46 53.0 & +01 13 56 & 354.71 & 52.10 & 16 & 11.76 & 4.0 & --- & --- & 19.53 & 7.9 &  \\ 
&  &  &  &  & $-$48 & 11.46 & 4.0 & --- & --- & --- & --- &  \\ 
QSO\,B1448$-$232 & 14 51 02.5 & $-$23 29 31 & 335.44 & 31.72 & $-$157 & 11.48 & 5.3 & --- & --- & --- & --- & L \\ 

{\small\it([HB89]\,1448$-$232)}&  &  &  &  & $-$150 & 11.64 & 1.4 & 11.80 & 2.1 & --- & --- & L \\ 
&  &  &  &  & $-$130 & 11.62 & 8.6 & --- & --- & --- & --- & L \\ 
&  &  &  &  & $-$99 & 10.93 & 1.5 & --- & --- & --- & --- & L \\ 
&  &  &  &  & 36 & 12.22 & 9.4 & 12.06 & 6.1 & 19.75 & 15.0 &  \\ 
PKS\,1508$-$05 & 15 10 53.6 & $-$05 43 07 & 353.91 & 42.94 & $-$17 & 12.56 & 0.6 & --- & --- & 19.76 & 14.7 & L low vel. \\ 
&  &  &  &  & $-$27 & 11.76 & 1.4 & --- & --- & --- & --- &  \\ 
&  &  &  &  & $-$37 & 14.01 & 0.5 & --- & --- & --- & --- &  \\ 
&  &  &  &  & $-$43 & 11.52 & 1.3 & --- & --- & --- & --- &  \\ 
&  &  &  &  & 8 & 12.29 & 5.0 & 12.25 & 2.6 & 19.94 & 5.9 &  \\ 
SDSS\,J151352.52$+$085555.7 & 15 13 52.0 & +08 55 55 & 11.30 & 51.78 & $-$37 & n/a & n/a & 11.37 & 2.8 & n/a & n/a & K \\ 
QSO\,J1621$-$0042 & 16 21 16.9 & $-$00 42 51 & 12.91 & 32.45 & $-$6 & 12.28 & 6.7 & 12.27 & 5.9 & --- & --- &  \\ 
{\small\it(SDSS\,J1621$-$0042)}\\
PKS\,1629$+$120 & 16 31 45.2 & +11 56 03 & 27.85 & 36.35 & $-$40 & n/a & n/a & 11.53 & 6.0 & n/a & n/a & K \\ 
&  &  &  &  & $-$15 & n/a & n/a & 12.14 & 3.0 & n/a & n/a & K \\ 
FBQS\,J2107$-$0620 & 21 07 57.7 & $-$06 20 11 & 43.67 & $-$33.14 & $-$19 & 11.94 & 4.6 & 11.97 & 3.2 & 19.63 & 5.3 & PP arch \\ 
PKS\,2115$-$30 & 21 18 10.6 & $-$30 19 12 & 15.74 & $-$43.56 & 22 & 12.59 & 0.7 & --- & --- & 18.79 & 8.8 &  \\ 
LBQS\,2132$-$4321 & 21 36 06.2 & $-$43 08 17 & 357.59 & $-$47.67 & $-$237 & 11.76 & 6.2 & --- & --- & --- & --- & GCN \\ 
PKS\,2134$+$004 & 21 36 38.6 & +00 41 54 & 55.47 & $-$35.58 & 26 & 12.11 & 5.3 & 11.52 & 4.0 & 18.70 & 5.1 &  \\ 
&  &  &  &  & 58 & 11.39 & 1.8 & --- & --- & 18.69 & 13.0 &  \\ 
&  &  &  &  & 70 & 11.47 & 4.7 & --- & --- & --- & --- & GCP \\ 
PKS\,2149$-$306 & 21 51 55.5 & $-$30 27 54 & 17.08 & $-$50.78 & 53 & 11.93 & 4.0 & --- & --- & --- & --- &  \\ 
FBQS\,J2155$-$0922 & 21 55 01.5 & $-$09 22 25 & 47.47 & $-$44.82 & $-$209 & 11.75 & 6.0 & --- & --- & --- & --- & GCN \\ 
&  &  &  &  & $-$166 & 11.13 & 1.1 & --- & --- & --- & --- & GCN \\ 
PKS\,2155$-$152 & 21 58 06.3 & $-$15 01 09 & 40.64 & $-$48.02 & $-$9 & 12.27 & 5.5 & 12.78 & 2.2 & 19.92 & 3.9 &  \\ 
PKS\,2204$-$54 & 22 07 43.7 & $-$53 46 34 & 339.90 & $-$49.93 & $-$243 & 11.53 & 0.8 & --- & --- & --- & --- & MS \\ 
&  &  &  &  & $-$33 & 11.64 & 6.0 & --- & --- & 18.64 & 10.5 & MS \\ 
SDSS\,J221511.93$-$004549.9 & 22 15 11.9 & $-$00 45 50 & 61.44 & $-$44.19 & $-$7 & 12.33 & 6.5 & 13.02 & 2.7 & 19.75 & 2.3 &  \\ 
HE\,2225$-$2258 & 22 27 56.9 & $-$22 43 02.6 & 32.63 & $-$57.28 & $-$118 & 12.11 & 7.0 & --- & --- & --- & --- & GCN \\ 
&  &  &  &  & $-$14 & 11.89 & 4.0 & 11.54 & 4.2 & 19.46 & 6.3 & GCP \\ 
&  &  &  &  & 8 & 11.79 & 5.4 & 11.02 & 3.6 & 19.70 & 12.7 &  \\ 
QSO\,B2225$-$404 & 22 28 27.0 & $-$40 09 59 & 359.81 & $-$57.79 & $-$122 & 12.48 & 0.6 & --- & --- & --- & --- & GCN \\ 

{\small\it([HB89]\,2225$-$404)}\\
PKS\,2227$-$08 & 22 29 40.1 & $-$08 32 54 & 55.22 & $-$51.70 & $-$30 & 11.93 & 4.2 & --- & --- & 19.23 & 16.2 &  \\ 
&  &  &  &  & $-$9 & 11.61 & 7.5 & 12.38 & 1.4 & 19.88 & 3.1 &  \\ 
2QZ\,J223951.8$-$294837 & 22 39 51.8 & $-$29 48 37 & 19.77 & $-$61.06 & $-$41 & 11.66 & 0.7 & --- & --- & --- & --- & GCN \\ 
PKS\,2243$-$123 & 22 46 18.2 & $-$12 06 51 & 53.87 & $-$57.07 & $-$11 & 12.41 & 6.8 & 11.80 & 3.7 & 19.87 & 3.9 & MS \\ 
2QZ\,J225153.1$-$314620 & 22 51 53.2 & $-$31 46 20 & 15.56 & $-$63.65 & $-$32 & 11.48 & 1.7 & n/a & n/a & --- & --- & GCN \\ 
2QZ\,J225154.8$-$314521 & 22 51 54.8 & $-$31 45 21 & 15.60 & $-$63.66 & $-$64 & 11.86 & 0.8 & n/a & n/a & --- & --- & MS \\ 
&  &  &  &  & $-$32 & 11.43 & 0.6 & n/a & n/a & 19.00 & 17.2 & MS \\ 
&  &  &  &  & $-$20 & 11.48 & 4.0 & n/a & n/a & --- & --- & MS \\ 
3C\,454.3 & 22 53 57.7 & +16 08 54 & 86.11 & $-$38.18 & $-$39 & 11.99 & 8.7 & 12.67 & 2.9 & n/a & n/a & PP arch \\ 
PKS\,2255$-$282 & 22 58 06.0 & $-$27 58 21 & 24.39 & $-$64.92 & $-$56 & 11.47 & 1.0 & --- & --- & --- & --- & GCN \\ 
&  &  &  &  & $-$31 & 11.45 & 1.7 & --- & --- & 18.40 & 3.5 & MS \\ 
QSO\,B2314$-$409 & 23 16 46.9 & $-$40 41 21 & 352.04 & $-$66.26 & $-$37 & 12.05 & 7.0 & --- & --- & --- & --- & MS \\ 

{\small\it([HB89]\,2314$-$409)}&  &  &  &  & $-$56 & 11.90 & 5.2 & --- & --- & 18.78 & 24.4 & MS \\ 
2QZ\,J232046.7$-$294406 & 23 20 46.7 & $-$29 44 06 & 20.13 & $-$69.93 & $-$49 & 12.44 & 2.0 & 11.16 & 0.6 & 19.26 & 10.0 & MS \\ 
&  &  &  &  & 165 & 13.05 & 1.8 & --- & --- & --- & --- & MS \\ 
2QZ\,J232114.2$-$294725 & 23 21 14.2 & $-$29 47 24 & 19.95 & $-$70.03 & $-$28 & 12.37 & 1.0 & n/a & n/a & --- & --- & MS \\ 
&  &  &  &  & $-$37 & 13.06 & 2.2 & n/a & n/a & --- & --- & MS \\ 
&  &  &  &  & $-$49 & 12.94 & 1.5 & n/a & n/a & 19.19 & 10.7 & MS \\ 
PKS\,2340$-$036 & 23 42 56.6 & $-$03 22 26 & 85.40 & $-$61.15 & $-$19 & --- & --- & 12.90 & 1.7 & 19.29 & 2.5 & MS \\ 
PSS\,J2344$+$0342 & 23 44 03.2 & +03 42 26 & 92.54 & $-$55.14 & $-$31 & n/a & n/a & 12.24 & 1.0 & --- & --- & PP arch \\ 
QSO\,J2346$+$1247 & 23 46 25.4 & +12 47 44 & 99.40 & $-$47.14 & $-$31 & --- & --- & 12.13 & 1.2 & 18.37 & 1.4 & PP arch \\ 

{\small\it([HB89]\,2343$+$125:BX0415)}\\
QSO\,B2343$+$125 & 23 46 28.2 & +12 49 00 & 99.33 & $-$47.06 & $-$68 & 12.12 & 3.4 & 11.44 & 1.2 & 19.50 & 8.9 & PP arch \\ 

{\small\it([HB89]\,2343$+$125)}&  &  &  &  & $-$53 & 12.18 & 8.0 & 11.73 & 0.9 & 19.08 & 2.2 & PP arch \\ 
&  &  &  &  & $-$40 & 11.93 & 4.0 & --- & --- & --- & --- & PP arch \\ 
&  &  &  &  & $-$29 & 11.92 & 4.0 & 12.14 & 1.2 & 19.05 & 6.6 & PP arch \\ 
QSO\,B2344$+$1229 & 23 46 32.8 & +12 45 40 & 99.33 & $-$47.12 & $-$30 & n/a & n/a & 12.02 & 1.4 & 19.70 & 10.1 & PP arch \\

{\small\it([VCV96]\,Q2344$+$1229)}\\
QSO\,B2345$+$006 & 23 48 19.2 & +00 57 18 & 91.99 & $-$58.07 & $-$253 & 11.73 & 1.3 & --- & --- & 18.60 & 12.7 & MS \\ 

{\small\it([HB89]\,2345$+$006B)}\\
QSO\,B2345$+$006A & 23 48 19.6 & +00 57 21 & 91.99 & $-$58.07 & $-$17 & 11.74 & 6.0 & 12.00 & 0.9 & 19.66 & 4.1 & MS \\ 

{\small\it([HB89]\,2345$+$006A)}&  &  &  &  & $-$292 & 11.12 & 12.9 & --- & --- & --- & --- & MS \\ 
&  &  &  &  & $-$277 & 11.90 & 8.4 & --- & --- & 18.63 & 13.1 & MS \\ 
HE\,2348$-$1444 & 23 51 29.9 & $-$14 27 48 & 72.14 & $-$71.15 & 91 & 11.43 & 1.0 & n/a & n/a & --- & --- & MS \\ 
NVSS\,J235953$-$124148 & 23 59 53.6 & $-$12 41 48 & 80.50 & $-$71.19 & $-$41 & 11.52 & 4.0 & --- & --- & --- & --- & MS \\ 
&  &  &  &  & 47 & 11.24 & 0.9 & --- & --- & --- & --- & MS
\end{longtable}
\end{landscape}
}
\end{document}